\documentclass[11pt,english]{article}
\usepackage[latin9]{inputenc}
\usepackage{color}
\usepackage{babel}
\usepackage{amsmath}
\usepackage{amssymb}
\usepackage{graphicx}
\usepackage{setspace}
\usepackage{esint}
\usepackage[bookmarks=false,
 breaklinks=false,pdfborder={0 0 1},backref=false,colorlinks=false]
 {hyperref}

\makeatletter
\@ifundefined{date}{}{\date{}}
\usepackage{babel}
\usepackage{latexsym}
\usepackage{bm}\usepackage{longtable}

\usepackage{xcolor}

\@addtoreset{equation}{section}

\makeatother

\begin{document}
\title{Holographic correlation functions of fermions in anisotropic plasma}
\maketitle
\begin{center}
Si-wen Li$^{a,}$\footnote{Email: siwenli@dlmu.edu.cn}, Yan-qing
Zhao$^{b,}$\footnote{Email: zhaoyanqing@hainnu.edu.cn} 
\par\end{center}

\begin{center}
\vspace{2mm}
 $^{a}$\emph{Department of Physics, School of Science, Dalian Maritime
University, }\\
 \emph{Dalian 116026, China}\\
 
\par\end{center}

\begin{center}
$^{b}$\emph{College of Physics and Electronic Engineering, Hainan
Normal University,}\\
 \emph{Haikou 571158, China} 
\par\end{center}

\vspace{12mm}

\begin{abstract}
By using the gauge-gravity duality, we study the holographic fermionic
correlation functions in strongly coupled anisotropic plasmas. Starting
from the isotropic AdS-Schwarzschild black brane background, we revisit
the prescription for computing the retarded Green's function of a
probe Dirac fermion and then generalize the formulas with respect
to the anisotropic geometries. The method is applied to three distinct
holographic models that capture different physical origins of anisotropy:
axion-induced, magnetic-field-induced and unquenched-flavor-induced.
Numerical results for the holographic correlation functions reveal
direction-dependent corrections, negative dips in the imaginary part,
Landau levels in the fermionic dispersion (magnetic field), and a
momentum-independent pseudogap indicating an incoherent metallic phase
(flavors). Our results complement and go beyond the hard thermal loop
approximation, providing non-perturbative insights into fermionic
excitations in strongly coupled anisotropic plasmas relevant for heavy-ion
collisions and certain condensed matter systems. 
\end{abstract}
\newpage{}

\tableofcontents{}

\section{Introduction}

Quantum chromodynamics (QCD) is the fundamental theory of the strong
interaction. Despite its success at high energies where asymptotic
freedom justifies perturbative methods, QCD becomes notoriously difficult
to solve in the low-energy regime, particularly for dense matter at
finite temperature. This difficulty stems from the strongly coupled
nature of quarks and gluons in the confinement and deconfinement phases,
where standard perturbative expansions break down. Consequently, the
investigation of strongly coupled and dense matter via gauge-gravity
duality, also known as the AdS/CFT correspondence, has emerged as
a powerful alternative for decades \cite{Aharony:1999ti,Maldacena:1997re,Witten:1998qj}.
This duality maps a strongly coupled quantum field theory (QFT) to
a weakly coupled classical gravity theory in a higher-dimensional
anti-de Sitter (AdS) spacetime, thereby providing a unique theoretical
tool to access non-perturbative phenomena.

Quark-gluon plasma (QGP), a state of matter created in heavy-ion collision
(HIC) experiments, is known to be both strongly coupled \cite{Shuryak:2003xe,Shuryak:2004cy}
and anisotropic \cite{Florkowski:2010cf,Martinez:2010sc,Martinez:2010sd,Ryblewski:2010bs}.
The anisotropy arises naturally due to the rapid expansion of the
fireball along the beam direction, leading to different pressure gradients
transverse and longitudinal to the collision axis. Moreover, off-central
collisions generate strong magnetic fields that further break rotational
symmetry. Therefore, constructing type IIB supergravity models that
capture anisotropic and strongly coupled QGP through gauge-gravity
duality is not only significant but also physically compelling \cite{Avila:2016mno,Banks:2016fab,Giataganas:2017koz}.
The most celebrated example of this correspondence is the duality
between the four-dimensional $\mathcal{N}=4$ $SU\left(N_{c}\right)$
super Yang-Mills theory on $N_{c}$ D3-branes and the type IIB super
string theory on $\mathrm{AdS_{5}}\times S^{5}$. This well-established
framework serves as the starting point for many holographic studies
of strongly coupled plasmas, including the anisotropic extensions
explored in this work.

On the other hand, fermions are the fundamental constituents of QFT
and nuclear physics. The study of fermionic self-energies, spectral
functions, and correlation functions has a long history, as these
quantities encode the response of the medium to external fermionic
probes. In weakly coupled systems, the fermionic self-energy can be
analyzed using perturbative QFT or characterized by the thermal mass
in a hot medium \cite{MLeBellac,JKapusta}. However, these weak-coupling
results may not hold in the strong-coupling regime, where non-perturbative
effects such as bound states, collective excitations, and large anomalous
dimensions dominate. Indeed, several studies have pointed out the
failure of perturbative expansions at strong coupling \cite{Harada:2008vk,Nakkagawa:2011ci}.
To address this gap, holographic methods have been extensively applied
to investigate strongly coupled fermionic systems, revealing novel
phenomena such as non-Fermi liquids, Fermi surfaces \cite{Lee:2008xf,Liu:2009dm,Iqbal:2009fd,Li:2023wyb,Li:2024apc,Li:2024jkd,Li:2025ahp,Li:2025zbj,Basu:2009qz,Fang:2014jka,Li:2026wqu}.
These studies, however, have primarily focused on isotropic backgrounds,
leaving the role of anisotropy largely unexplored.

Motivated by the above, in this work we focus on holographic fermionic
correlation functions, with a particular emphasis on anisotropy. Our
goal is to model fermions (for instance, the plasmino mode, which
is a collective fermionic excitation of the medium) in a setting that
more closely resembles the realistic QGP, where rotational symmetry
is explicitly broken. To achieve this, we first revisit the holographic
principle within the AdS/CFT dictionary, and then numerically demonstrate
the standard prescription for computing fermionic correlation functions
on the isotropic AdS-Schwarzschild black brane spacetime. Subsequently,
we generalize this method to incorporate anisotropic gravity backgrounds
and apply it to three distinct holographic models, each representing
a different physical origin of anisotropy.

Specifically, the three models employed in this work are: (i) anisotropy
induced by an axion field \cite{Banks:2016fab,Giataganas:2017koz,Mateos:2011ix,A2016mno,Mateos:2011tv,Cheng:2014qia,Cheng:2014sxa},
which corresponds to a theta term in the dual theory; (ii) anisotropy
induced by an external magnetic field \cite{DHoker:2009ixq,DHoker:2009mmn,Dudal:2015wfn,Zhao:2024ipr},
relevant for off-central HICs and magnetized plasmas like QGP; (iii)
anisotropy induced by unquenched flavors \cite{Penin:2017lqt,Garbayo:2022pqp,Hoyos:2020zeg}
which captures the backreaction of dynamical quarks. These three scenarios
cover the possible origins of anisotropy in dense strongly coupled
matter. We numerically evaluate the holographic fermionic correlation
functions in each model, obtaining a rich set of results, including
direction-dependent corrections to the Green function, anisotropic
dispersion curves, Landau levels under the external magnetic field
and an incoherent metallic phase induced by flavors. It is worth noting
that in thermal field theory, the plasmino mode, as a collective fermionic
excitation present at finite temperature or chemical potential, has
been extensively studied by using the Hard Thermal Loop (HTL) approximation
\cite{MLeBellac,Klimov:1981ka,Haque:2024gva}. The HTL method successfully
captures the dispersion relation and damping rate of plasminos in
the isotropic situation. However, its extension to strongly coupled
regimes is highly non-trivial, and becomes especially challenging
in the presence of anisotropy, where the HTL effective action loses
its simple Lorentz-invariant structure. Our holographic results therefore
complement and go beyond the HTL framework, providing non-perturbative
data for fermionic excitations in strongly coupled anisotropic plasmas.

The outline of this work is as follows. In Section 2, we revisit the
holographic principle for fermions and demonstrate numerically how
to evaluate the fermionic correlation function on the AdS-Schwarzschild
black brane background. In Section 3, we generalize the method to
include anisotropic gravity backgrounds, discuss the perturbative
expansion of the correlation function, and derive the structure of
the anisotropic Green function. In Sections 4, 5, and 6, we apply
our method to the models with anisotropies induced respectively by
axion, magnetic field, and flavors, presenting detailed numerical
results for the associated fermionic correlation functions. In Section
7, we provide a summary and discussion of our findings, including
their physical implications and connections to heavy-ion phenomenology
and condensed matter physics. Finally, in the Appendix, we list the
explicit forms of the metric functions used in the axion model presented
in Section 4.

\subsubsection*{Notation}

In this work, the indices of the bulk coordinates are denoted by the
capital letters $M,N,P...$ as $x^{M}$, while the corresponding indices
in the local Lorentz frame (tangent space) are denoted by the lowercase
letters $a,b,c...$. For a five-dimensional (5D) manifold, the indices
$M,N,P...$ run over 0, 1, ..., 4. The generators of the Clifford
algebra are denoted by $\gamma^{a},\Gamma^{M}$ and the vielbein $e^{a}_{M}$,
bulk metric $g_{MN}$ are related by 
\begin{equation}
\left\{ \gamma^{a},\gamma^{b}\right\} =2\eta^{ab},\left\{ \Gamma^{M},\Gamma^{N}\right\} =2g^{MN},\Gamma^{M}=e^{M}_{a}\gamma^{a},g_{MN}=e^{a}_{M}\eta_{ab}e^{b}_{N},
\end{equation}
where $\eta_{ab}$ refers to the Minkowskian metric. The spin connection
$\omega_{Mab}$ can be obtained by using the affine connection $\Gamma^{K}_{MN}$
as, 
\begin{equation}
\omega_{Mab}=\eta_{cb}\left(e^{N}_{a}\partial_{M}e^{c}_{N}-e^{N}_{a}\Gamma^{K}_{MN}e^{c}_{K}\right),
\end{equation}
where $\Gamma^{K}_{MN}$ is given by the metric as, 
\begin{equation}
\Gamma^{K}_{MN}=\frac{1}{2}g^{KL}\left(\partial_{M}g_{LN}+\partial_{N}g_{ML}-\partial_{L}g_{MN}\right).
\end{equation}
The indices denoted by using the Greek letters as $\mu,\nu...$run
from 0, 1, 2, 3. And the indices denoted by using lowercase letters
$i,j,k$ run over the spatial directions i.e. $i,j,k=1,2,3$. In this
work, the gamma matrices are chosen as $\gamma^{a}=\left(\gamma^{\mu},\gamma\right)$
as, 
\begin{equation}
\gamma^{\mu}=i\left(\begin{array}{cc}
0 & \sigma^{\mu}\\
\bar{\sigma}^{\mu} & 0
\end{array}\right),\gamma=\left(\begin{array}{cc}
1 & 0\\
0 & -1
\end{array}\right),
\end{equation}
where $\sigma^{\mu}=\left(1,-\tau^{i}\right),\bar{\sigma}^{\mu}=\left(1,\tau^{i}\right)$
defined by the Pauli matrices $\tau^{i}$'s. The fermionic spectral
function is defined as $\mathcal{S}\left(\omega,\vec{k}\right)=\mathrm{Im}G_{R}\left(\omega,\vec{k}\right)$
in our notation. The covariant derivative operator $D_{M}=\partial_{M}-ig_{f}A_{M}$
for fermions is defined through

\begin{equation}
\left(\nabla_{M}-ig_{f}A_{M}\right)\psi\equiv\left(D_{M}+\frac{1}{4}\omega_{Mab}\gamma^{ab}\right)\psi,
\end{equation}
where

\begin{equation}
\nabla_{M}=\partial_{M}+\frac{1}{4}\omega_{Mab}\gamma^{ab},\ \gamma^{ab}=\frac{1}{2}\left[\gamma^{a},\gamma^{b}\right].\label{eq:1.6}
\end{equation}
Here $g_{f}=g_{\mathrm{YM}}q_{f}$ is the effective coupling constant,
$g_{\mathrm{YM}}$ is the Yang-Mills coupling constant and $q_{f}$
is the $U(1)$ charge of the fermions. In this work, we will set $g_{f}=1$
during the numerical calculations for simplification since we do not
attempt to fit any experimental data.

\section{Revisit the fermionic correlation function on the AdS-Schwarzschild
black brane background}

In this section, we demonstrate the calculations and analyses in general
about the fermionic correlation function by using the holographic
principle in the AdS/CFT dictionary. We first apply this method to
the isotropic AdS-Schwarzschild black brane background and then generalize
it to the anisotropic gravity background in the other sections.

\subsection{The prescription for the fermionic correlation function}

We start from the holographic principle of the AdS/CFT, that is the
gravitational partition function $Z_{\mathrm{gravity}}$ in the bulk
is equal to the generating functional of the dual conformal field
theory (CFT) $Z_{\mathrm{CFT}}$ \cite{Aharony:1999ti,Witten:1998qj,Iqbal:2009fd}.
Specifically, for a spinor field $\psi$ in the $D+1$ dimensional
bulk, we can write down, 
\begin{equation}
Z_{\mathrm{CFT}}\left[\bar{\psi}_{0},\psi_{0}\right]=Z_{\mathrm{gravity}}\left[\bar{\psi},\psi\right]\big|_{\bar{\psi},\psi\rightarrow\bar{\psi}_{0},\psi_{0}},\label{eq:2.1}
\end{equation}
with 
\begin{align}
Z_{\mathrm{CFT}}\left[\bar{\psi_{0}},\psi_{0}\right] & =\left\langle \exp\left\{ \int_{\partial\mathcal{M}}\left(\bar{\eta}\psi_{0}+\bar{\psi}_{0}\eta\right)d^{D}x\right\} \right\rangle ,\nonumber \\
Z_{\mathrm{gravity}}\left[\bar{\psi},\psi\right] & =e^{-S^{ren}_{\mathrm{gravity}}},\nonumber \\
S^{ren}_{\mathrm{gravity}} & =\int_{\mathcal{M}}\mathcal{L}^{ren}_{\mathrm{gravity}}\left[\bar{\psi},\psi\right]\sqrt{-g}d^{D+1}x.
\end{align}
Note that $\mathcal{M}$ denotes the bulk space and $\partial\mathcal{M}=\left\{ u\rightarrow0|\mathcal{M}\right\} $
represents the holographic boundary of $\mathcal{M}$. The boundary
value of $\psi$ is denoted by $\psi_{0}=\psi|_{\partial\mathcal{M}}$
as the source of the boundary fermionic operator $\eta$, and $\psi_{0}$
must be obtained by solving the equation of motion associated to the
action $S^{ren}_{\mathrm{gravity}}$. Here $\mathcal{L}^{ren}_{\mathrm{gravity}}$
is the renormalized Lagrangian of the bulk field $\psi$. Then using
formula in quantum field theory (QFT), the one-point function, i.e.
average value of $\bar{\eta}$ can be written as 
\begin{align}
\left\langle \bar{\eta}\right\rangle  & =\frac{1}{Z_{\mathrm{CFT}}}\frac{\delta Z_{\mathrm{CFT}}}{\delta\psi_{0}}\big|_{\bar{\psi}_{0},\psi_{0}\rightarrow0}=\frac{1}{Z_{\mathrm{gravity}}}\frac{\delta Z_{\mathrm{gravity}}}{\delta\psi}\big|_{\bar{\psi},\psi\rightarrow\bar{\psi}_{0},\psi_{0}}\nonumber \\
 & =\lim_{u\rightarrow0}\frac{\delta S^{ren}_{\mathrm{gravity}}}{\delta\psi_{0}}\equiv i\int\frac{d^{4}k}{\left(2\pi\right)^{4}}e^{ik_{\mu}x^{\mu}}\bar{\Pi}_{0}\left(\omega,\vec{k}\right),\label{eq:2.3}
\end{align}
where the relation (\ref{eq:2.1}) has been imposed. Therefore, the
two-point correlation function $G_{R}$ of $\eta$ is obtained as,
\begin{equation}
\left\langle \bar{\eta}\left(y\right)\right\rangle =\int d^{4}x\bar{\psi}_{0}\left(x\right)\tilde{G}_{R}\left(y-x\right),\ \tilde{G}_{R}=-\gamma^{0}G^{\dagger}_{R}\gamma^{0},
\end{equation}
i.e. in the momentum space 
\begin{equation}
G_{R}\left(\omega,\vec{k}\right)=\lim_{u\rightarrow0}\Pi_{0}\chi^{-1}_{0},\ \psi_{0}\left(x\right)=\int\frac{d^{4}k}{\left(2\pi\right)^{4}}e^{ik_{\mu}x^{\mu}}\chi_{0}\left(\omega,\vec{k}\right).\label{eq:2.5}
\end{equation}
where, in the Fourier modes, $\omega,\vec{k}$ refers to the associated
frequency and 3-momentum. Altogether, it is possible to evaluate $\psi_{0},\Pi_{0}$
in order to evaluate the two-point correlation function $G_{R}$ of
$\eta$ by using the classical gravity action $S^{ren}_{\mathrm{gravity}}$
in holography.

\subsection{Application to the 5D AdS-Schwarzschild black brane background}

\subsubsection*{The Dirac equation}

As we will always focus on a four-dimensional dual field theory in
this work, let us apply the prescription to compute the fermionic
correlation function on the five-dimensional AdS-Schwarzschild black
brane background in this section as \cite{Lee:2008xf,Iqbal:2009fd,Li:2024apc,Li:2025zbj}.
In the other sections, our demonstration could be generalized easily
onto the anisotropic gravity background. In general, the static isotropic
metric can be written as, 
\begin{equation}
ds^{2}_{\left(5\mathrm{d}\right)}=g_{tt}dt^{2}+g_{xx}dx^{i}dx_{i}+g_{uu}du^{2},\label{eq:2.6}
\end{equation}
where $u$ is the radial coordinate and $g_{tt},g_{xx},g_{uu}$ depend
on $u$ only. Consider a probe fermion propagating in the bulk geometry
(\ref{eq:2.6}), its action is given by the Dirac action as, 
\begin{equation}
S_{\mathrm{bulk}}=i\int d^{4}xdu\sqrt{-g}\bar{\psi}\left(\Gamma^{M}\nabla_{M}-m\right)\psi,\label{eq:2.7}
\end{equation}
where the covariant derivative operator $\nabla_{M}$ is given in
(\ref{eq:1.6}). Varying the action (\ref{eq:2.7}) with respect to
$\bar{\psi}$, then inserting the following ansatz into the associated
equation of motion, 
\begin{equation}
\psi=\left(\begin{array}{c}
\psi_{R}\\
\psi_{L}
\end{array}\right)=\left(-gg^{-1}_{uu}\right)^{-1/4}\int\frac{d^{4}k}{\left(2\pi\right)^{4}}e^{ik_{\mu}x^{\mu}}\chi\left(u,k\right),
\end{equation}
and 
\begin{align}
\chi\left(u,k\right) & =\left[\begin{array}{c}
\chi_{R}\left(u,k\right)\\
\chi_{L}\left(u,k\right)
\end{array}\right],\chi_{R,L}=\left(\begin{array}{c}
\chi^{\left(1\right)}_{R,L}\\
\chi^{\left(2\right)}_{R,L}
\end{array}\right),k_{\mu}=\left(-\omega,k_{i}\right),\label{eq:2.9}
\end{align}
we can obtain the Dirac equation for spinor $\chi$ as, 
\begin{equation}
\left[\sqrt{\frac{g_{xx}}{g_{uu}}}\left(\gamma\partial_{u}-m\sqrt{g_{uu}}\right)+iK_{\mu}\gamma^{\mu}\right]\chi=0,\label{eq:2.10}
\end{equation}
or equivalently, 
\begin{align}
\sqrt{\frac{g_{xx}}{g_{uu}}}\left(\partial_{u}-m\sqrt{g_{uu}}\right)\chi_{R}-\sigma^{\mu}K_{\mu}\chi_{L} & =0,\nonumber \\
-\bar{\sigma}^{\mu}K_{\mu}\chi_{R}-\sqrt{\frac{g_{xx}}{g_{uu}}}\left(\partial_{u}+m\sqrt{g_{uu}}\right)\chi_{L} & =0,\label{eq:2.11}
\end{align}
where 
\begin{equation}
K_{\mu}=\left(\sqrt{-\frac{g_{xx}}{g_{tt}}}k_{0},k_{i}\right).
\end{equation}
To define the boundary value $\psi_{0}$, we need to solve respectively
the decoupled equations for $\chi_{L,R}$ near the boundary i.e. at
$u\rightarrow0$. The generic formulas of the decoupled equations
for $\chi_{L,R}$ can be derived as, 
\begin{align}
0= & -\sigma^{\mu}\partial_{u}S_{\mu}\sqrt{\frac{g_{xx}}{g_{uu}}}\left(\partial_{u}+m\sqrt{g_{uu}}\right)\chi_{L}-\sigma^{\mu}S_{\mu}\sqrt{\frac{g_{xx}}{g_{uu}}}\left[\partial^{2}_{u}+m\partial_{u}\left(\sqrt{g_{uu}}\right)+m\sqrt{g_{uu}}\partial_{u}\right]\chi_{L}\nonumber \\
 & +\sigma^{\mu}S_{\mu}m\sqrt{g_{xx}}\left(\partial_{u}+m\sqrt{g_{uu}}\right)\chi_{L}-\sigma^{\mu}K_{\mu}\chi_{L},\nonumber \\
0= & -\bar{\sigma}^{\mu}\partial_{u}S_{\mu}\sqrt{\frac{g_{xx}}{g_{uu}}}\left(\partial_{u}-m\sqrt{g_{uu}}\right)\chi_{R}-\bar{\sigma}^{\mu}S_{\mu}\sqrt{\frac{g_{xx}}{g_{uu}}}\left[\partial^{2}_{u}-m\partial_{u}\left(\sqrt{g_{uu}}\right)-m\sqrt{g_{uu}}\partial_{u}\right]\chi_{R}\nonumber \\
 & -\bar{\sigma}^{\mu}S_{\mu}m\sqrt{g_{xx}}\left(\partial_{u}-m\sqrt{g_{uu}}\right)\chi_{R}-\bar{\sigma}^{\mu}K_{\mu}\chi_{R},\label{eq:2.13}
\end{align}
where 
\begin{equation}
S_{\mu}=\frac{K_{\mu}}{K^{2}_{0}-K_{i}K^{i}}\sqrt{\frac{g_{xx}}{g_{uu}}}.
\end{equation}

\subsubsection*{The asymptotical solution}

By picking up the metric of the 5D AdS-Schwarzschild black brane background,
\begin{equation}
g_{tt}=-\frac{L^{2}}{u^{2}}f,g_{xx}=\frac{L^{2}}{u^{2}},g_{uu}=\frac{L^{2}}{u^{2}f},f=1-\frac{u^{4}}{u^{4}_{H}},\label{eq:2.15}
\end{equation}
the asymptotic solution at $u\rightarrow0$ can be found as, 
\begin{equation}
\chi_{L}=Au^{-mL}+Bu^{mL+1},\chi_{R}=Cu^{1-mL}+Du^{mL},\label{eq:2.16}
\end{equation}
where $A,B,C,D$ are constant Weyl spinors depending on 4-momentum
$k_{\mu}$. Hence, for the case $mL>0$, the boundary value $\chi_{0}$
of $\chi$ is given by $Au^{-mL}$ since it is the most divergent
term in $\chi_{L,R}$. So we can define 
\begin{equation}
\chi_{0}=\lim_{u\rightarrow0}u^{mL}\chi=\left(\begin{array}{c}
0\\
A
\end{array}\right).\label{eq:2.17}
\end{equation}
Afterwards, by imposing the ansatz (\ref{eq:2.9}) into action (\ref{eq:2.7}),
let us focus on the Fourier mode $\tilde{S}_{\mathrm{bulk}}$ of the
action $S_{\mathrm{bulk}}$ as, 
\begin{align}
S_{\mathrm{bulk}} & =\int\frac{d^{4}k}{\left(2\pi\right)^{4}}\tilde{S}_{\mathrm{bulk}},\nonumber \\
\tilde{S}_{\mathrm{bulk}} & =i\int du\bar{\chi}\left(k,u\right)\left[\sqrt{\frac{g_{xx}}{g_{uu}}}\left(\gamma\partial_{u}-m\sqrt{g_{uu}}\right)+iK_{\mu}\gamma^{\mu}\right]\chi\left(k,u\right).
\end{align}
Thus, in momentum space, the action becomes, 
\begin{align}
\tilde{S}_{\mathrm{bulk}} & =i\left(\bar{\chi}\gamma\chi\right)|^{0}_{u_{H}}-i\int du\sqrt{\frac{g_{uu}}{g_{xx}}}\left[\sqrt{\frac{g_{xx}}{g_{uu}}}\left(\partial_{u}\bar{\chi}\gamma+m\sqrt{g_{uu}}\bar{\chi}\right)+iK_{\mu}\bar{\chi}\gamma^{\mu}\right]\chi.\label{eq:2.19}
\end{align}
Note that the second term in (\ref{eq:2.19}) vanishes since it is
nothing but the Dirac equation conjugated to (\ref{eq:2.10}). In
this sense, the boundary action for $\chi$ reads 
\begin{align}
S_{\mathrm{bdry}} & =i\int d^{4}x\left(\bar{\chi}\gamma\chi\right)|_{u\rightarrow0}\nonumber \\
 & =\int d^{4}x\left(\chi^{\dagger}_{R}\chi_{L}-\chi^{\dagger}_{L}\chi_{R}\right)|_{u\rightarrow0}\nonumber \\
 & =\int d^{4}x\left(C^{\dagger}Au^{1-2mL}+D^{\dagger}A-h.c.\right)|_{u\rightarrow0},
\end{align}
so the renormalized action $S^{ren}$ and its the counter term are
expected to be 
\begin{align}
S^{ren}= & S_{\mathrm{bdry}}+S_{\mathrm{CT}}=\int d^{4}xD^{\dagger}A,\nonumber \\
S_{\mathrm{CT}}= & -\int d^{4}xC^{\dagger}Au^{1-2mL}|_{u\rightarrow0}.
\end{align}

\subsubsection*{The structure of the Green function}

Accordingly, $\bar{\Pi}_{0}$ given in (\ref{eq:2.3}) is obtained
as, 
\begin{equation}
\bar{\Pi}_{0}=-iu^{-mL}\frac{\delta\tilde{S}_{\mathrm{bulk}}}{\delta\chi_{0}}=iu^{-mL}\left(\chi^{\dagger}_{L},-\chi^{\dagger}_{R}\right),
\end{equation}
which, by using (\ref{eq:2.5}), leads to a generic form of $G_{R}\left(\omega,\vec{k}\right)$
as, 
\begin{equation}
G_{R}\left(\omega,\vec{k}\right)=\lim_{u\rightarrow0}u^{-2mL}\Pi\chi^{-1}=\left(\begin{array}{cc}
0 & -DA^{-1}\\
AD^{-1} & 0
\end{array}\right).\label{eq:2.23}
\end{equation}
As the fermionic Green function is expected to take the form, 
\begin{equation}
G_{R}\left(\omega,\vec{k}\right)=\gamma^{\mu}\mathcal{W}_{\mu}\left(\omega,\vec{k}\right)+\mathcal{V}\left(\omega,\vec{k}\right)=\left(\begin{array}{cc}
\mathcal{V} & i\sigma^{\mu}\mathcal{W}_{\mu}\\
i\bar{\sigma}^{\mu}\mathcal{W}_{\mu} & \mathcal{V}
\end{array}\right)\equiv\left(\begin{array}{cc}
0 & \mathcal{G}_{R}\\
-\mathcal{G}^{-1}_{R} & 0
\end{array}\right),\label{eq:2.24}
\end{equation}
by comparing (\ref{eq:2.24}) with (\ref{eq:2.23}), one can find
\begin{figure}
\begin{centering}
\includegraphics[scale=0.39]{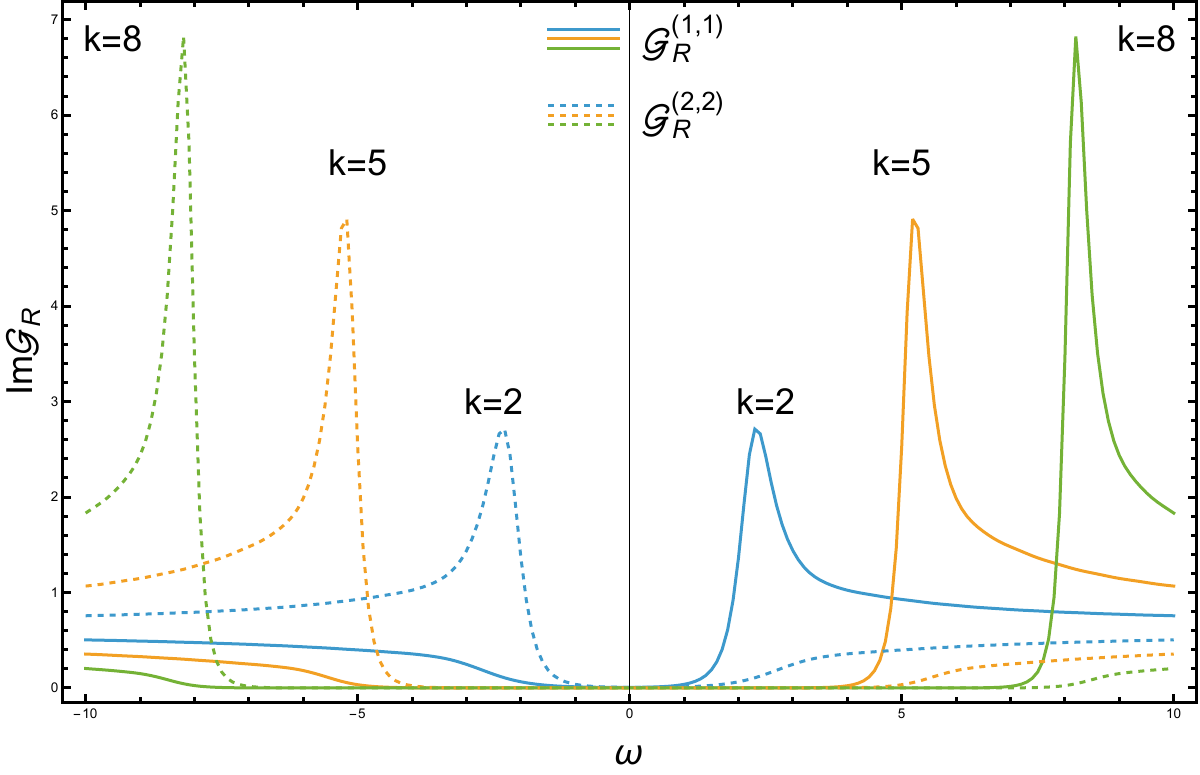}\includegraphics[scale=0.39]{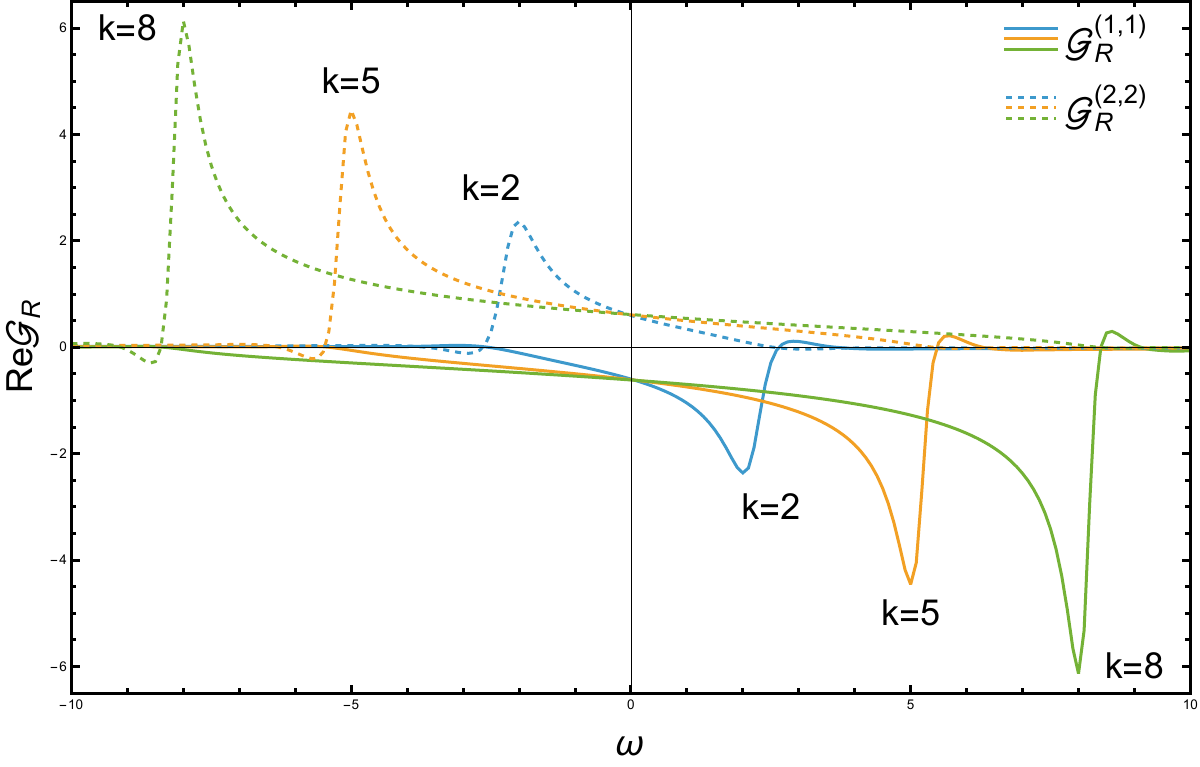} 
\par\end{centering}
\caption{The holographic fermionic Green function $\mathcal{G}_{R}$ on the
AdS-Schwarzschild black hole background as a function of $\omega$
with various momentum $\mathrm{k}$. The parameter is chosen as $T=\pi^{-1},u_{H}=1$.}
\label{fig:1} 
\end{figure}

\begin{align}
i\sigma^{\mu}\mathcal{W}_{\mu} & =-DA^{-1}=\mathcal{G}_{R},\nonumber \\
i\bar{\sigma}^{\mu}\mathcal{W}_{\mu} & =AD^{-1}=-\mathcal{G}^{-1}_{R},\nonumber \\
\left(i\sigma^{\mu}\mathcal{W}_{\mu}\right)\left(i\bar{\sigma}^{\nu}\mathcal{W}_{\nu}\right) & =-\mathcal{W}^{2}_{0}+\mathcal{W}^{2}_{i}=-1,\ \mathcal{V}=0.\label{eq:2.25}
\end{align}
Since the AdS-Schwarzschild black brane background is $SO\left(3\right)$
symmetric on the space $\mathbb{R}^{3}$ parameterized by $\left\{ x^{i}\right\} $,
it is possible to assume 
\begin{equation}
\mathcal{W}_{\mu}=\left\{ \mathcal{W}_{0},\mathcal{W}_{i}\right\} =\left\{ -i\mathcal{A},i\frac{\mathcal{B}}{\left|\vec{k}\right|}k_{i}\right\} ,
\end{equation}
where $\vec{k}$ refers to the spatial momentum in the Fourier modes.
Therefore, the relation between $\mathcal{W},\mathcal{A},\mathcal{B}$
is given as, 
\begin{equation}
\mathcal{G}_{R}=\mathcal{A}+\frac{\mathcal{B}}{\left|\vec{k}\right|}k_{i}\tau^{i},\mathcal{G}^{-1}_{R}=-\mathcal{A}+\frac{\mathcal{B}}{\left|\vec{k}\right|}k_{i}\tau^{i},\label{eq:2.27}
\end{equation}
where $\mathcal{A},\mathcal{B}$ need to be determined by solving
the Dirac equation (\ref{eq:2.11}). Due to the $SO\left(3\right)$
symmetry on the $\mathbb{R}^{3}$ parameterized by $\left\{ x^{i}\right\} $,
we can choose that the momentum is along $x^{3}$ which means $k_{\mu}=\left(-\omega,0,0,\mathrm{k}\right)$.
So the Green function $\mathcal{G}_{R}$ is diagonal as, 
\begin{equation}
\mathcal{G}_{R}=\left(\begin{array}{cc}
\mathcal{A}+\mathcal{B} & 0\\
0 & \mathcal{A}-\mathcal{B}
\end{array}\right)=\left(\begin{array}{cc}
\mathcal{G}^{\left(1,1\right)}_{R} & 0\\
0 & \mathcal{G}^{\left(2,2\right)}_{R}
\end{array}\right),\label{eq:2.28}
\end{equation}
then further defining the ratios as, 
\begin{equation}
\xi_{\left(\alpha\right)}=\left(-1\right)^{\alpha+1}\frac{\chi^{\left(\alpha\right)}_{R}}{\chi^{\left(\alpha\right)}_{L}},\ \alpha=1,2,\label{eq:2.29}
\end{equation}
the Green function is given by 
\begin{equation}
\mathcal{G}^{\left(\alpha,\alpha\right)}_{R}=\left(-1\right)^{\alpha}\lim_{u\rightarrow0}u^{-2mL}\xi_{\left(\alpha\right)}.\label{eq:2.30}
\end{equation}
The Dirac equation (\ref{eq:2.11}) yields the following flow equations
for $\xi_{\left(\alpha\right)}$ as, 
\begin{equation}
\sqrt{\frac{g_{xx}}{g_{uu}}}\xi^{\prime}_{\left(\alpha\right)}=\left(-1\right)^{\alpha+1}K_{0}-K_{3}+\left[K_{3}+\left(-1\right)^{\alpha+1}K_{0}\right]\xi^{2}_{\left(\alpha\right)}+2m\sqrt{g_{xx}}\xi_{\left(\alpha\right)},\label{eq:2.31}
\end{equation}
where $\xi^{\prime}_{\left(\alpha\right)}=\partial_{u}\xi_{\left(\alpha\right)}$.
For finite temperature $\pi^{-1}u^{-1}_{H}=T$, the in-falling boundary
conditions induce $\xi_{\left(1\right),\left(2\right)}|_{u=u_{H}}=\mp i$
as the boundary conditions for the ratios \cite{Liu:2009dm,Li:2024apc,Seo:2012bz,Seo:2013nva}.
Altogether, it is possible to evaluate numerically the ratios $\xi_{\left(\alpha\right)}$
with the in-falling boundary condition in order to obtain $\mathcal{G}^{\left(1,1\right)}_{R},\mathcal{G}^{\left(2,2\right)}_{R}$
by using (\ref{eq:2.30}). Afterwards, $\mathcal{A},\mathcal{B}$
can be obtained by using (\ref{eq:2.25}) and (\ref{eq:2.28}). Note
that according to (\ref{eq:2.31}), we can find $\mathcal{G}^{\left(1,1\right)}_{R}\left(\omega,\mathrm{k}\right)=\mathcal{G}^{\left(2,2\right)}_{R}\left(\omega,-\mathrm{k}\right),\mathcal{G}^{\left(1,1\right)*}_{R}\left(\omega,\mathrm{k}\right)=-\mathcal{G}^{\left(2,2\right)}_{R}\left(-\omega,\mathrm{k}\right)$.

\subsubsection*{The numerical calculations}

\begin{figure}
\begin{centering}
\includegraphics[width=6cm,totalheight=7cm,keepaspectratio]{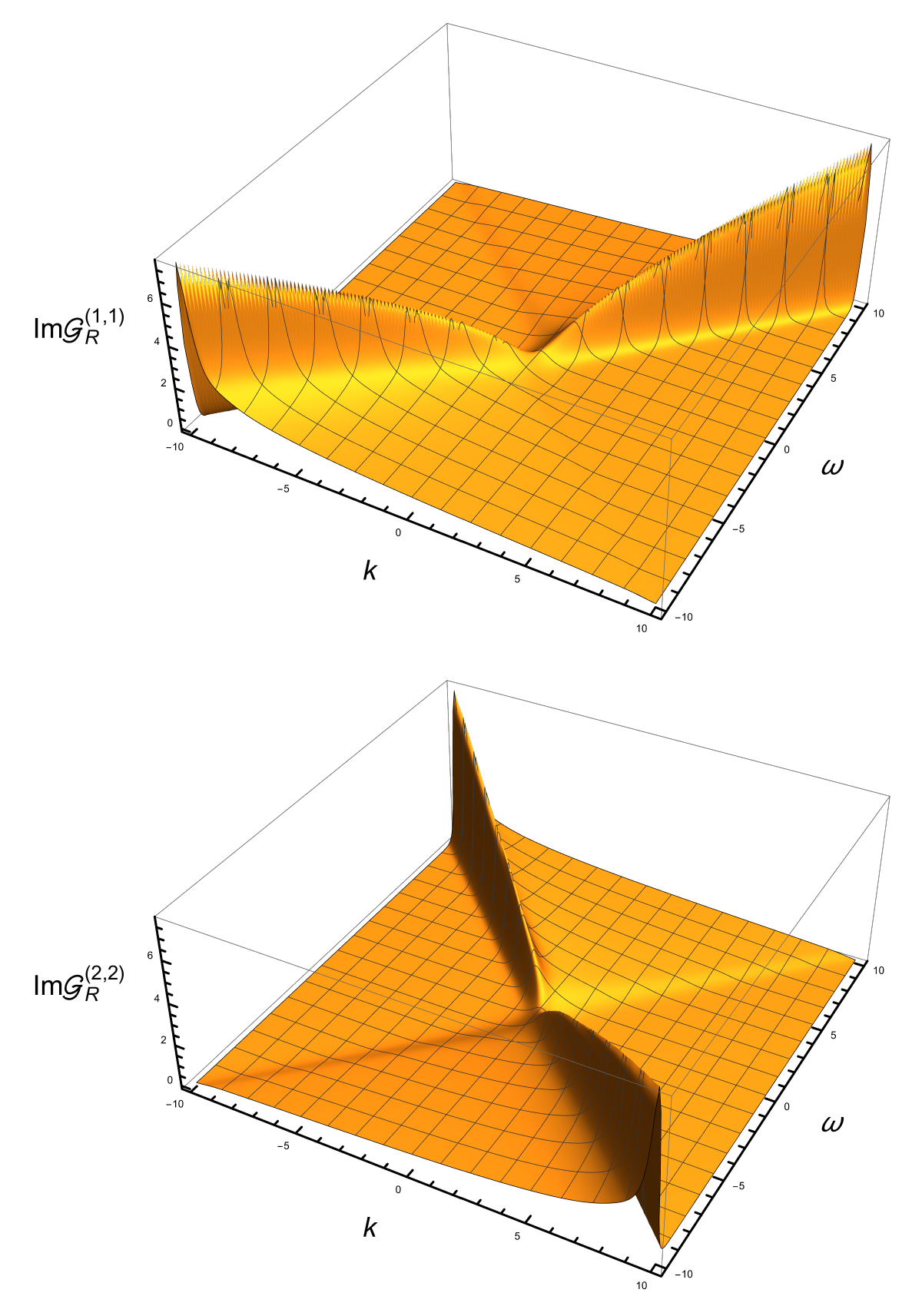}\includegraphics[width=11cm,totalheight=7cm,keepaspectratio]{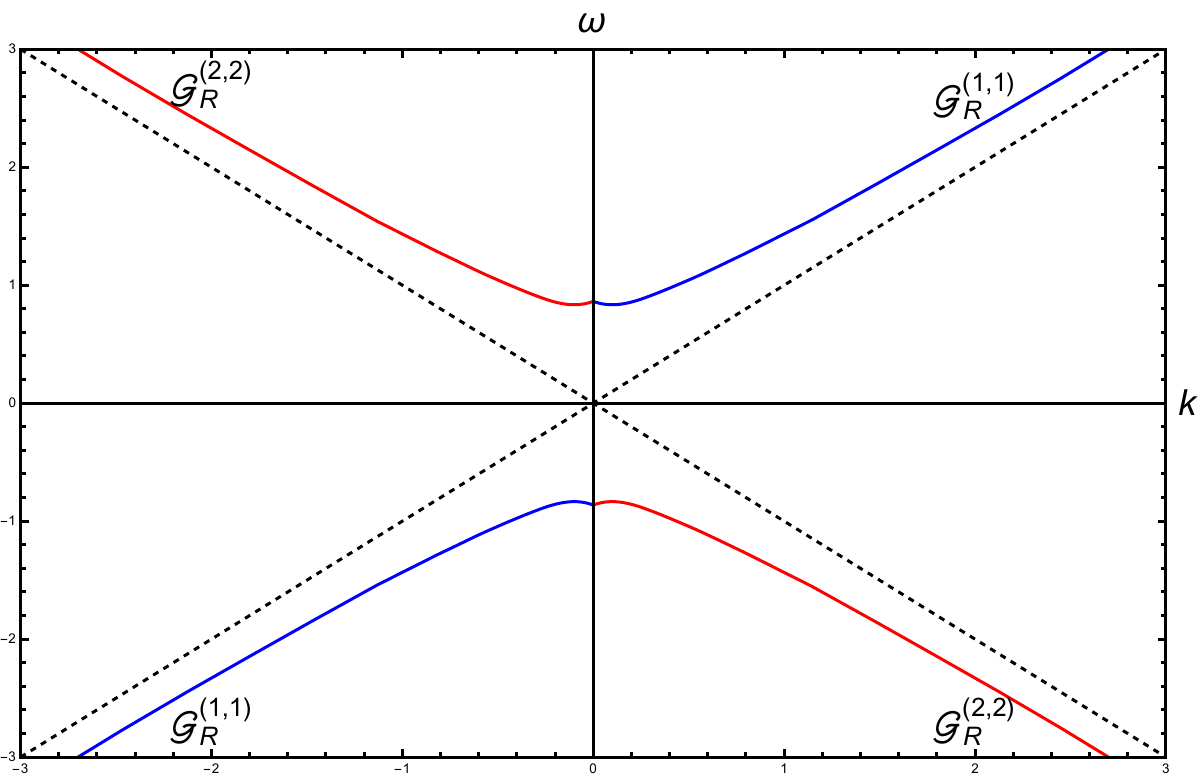} 
\par\end{centering}
\caption{The fermionic dispersion curve from the holographic Green function
with $T=\pi^{-1},u_{H}=L=1,m=0.01$. The horizontal and vertical axis
refer respectively to momentum $\mathrm{k}$ and frequency $\omega$
in the right figure.}
\label{fig:2} 
\end{figure}

Keeping the above content in hand, it is possible to compute numerically
the fermionic correlation function, and the results are displayed
in Figure \ref{fig:1} and \ref{fig:2}. 

The numerical calculation illustrates that the fermionic dispersion
curves are symmetric for $\omega\rightarrow-\omega$ since the chemical
potential is not turned on in this setup. In addition, for $\mathrm{k}>0$,
as the momentum increases, we can find $\omega$ on the dispersion
curves firstly decrease and then increase, approaching to the light
cone. The dispersion curves agree qualitatively with the hard thermal
loop approximation for fermion \cite{MLeBellac,Klimov:1981ka,Haque:2024gva}
describing the excitation of plasmino in the plasma, since the black
AdS is dual to the super Yang-Mills plasma in holography.

\section{The setup on the asymptotic AdS with anisotropy}

In this section, we will introduce some existing holographic backgrounds
with various anisotropies, then discuss a general method to compute
the fermionic Green function and the structure of the Green function.

\subsection{Dirac equation on the anisotropic background}

We consider the static anisotropic holographic backgrounds in this
work, which in general takes the following form, 
\begin{equation}
ds^{2}=g_{tt}dt^{2}+g_{xx}\left[\left(dx^{1}\right)^{2}+\left(dx^{2}\right)^{2}\right]+g_{zz}\left(dx^{3}\right)^{2}+g_{uu}du^{2},
\end{equation}
and $g_{tt},g_{xx},g_{zz},g_{uu}$ are functions of $u$. To compute
the holographic fermionic Green function, we introduce a probe Dirac
fermion whose action is 
\begin{equation}
S_{\mathrm{bulk}}=i\int d^{5}x\sqrt{-g}\bar{\psi}\left[\Gamma^{M}\left(\nabla_{M}-ig_{f}A_{M}\right)-m\right]\psi,\label{eq:3.2}
\end{equation}
where $\nabla_{M}$ is the covariant derivate operator for fermion
given in (\ref{eq:1.6}) and $A_{M}$ is the gauge field. The Dirac
operator is computed as, 
\begin{align}
\Gamma^{M}\nabla_{M}= & \frac{1}{\sqrt{-g_{tt}}}\gamma^{0}\partial_{0}+\frac{1}{\sqrt{g_{xx}}}\left(\gamma^{1}\partial_{1}+\gamma^{2}\partial_{2}\right)+\frac{1}{\sqrt{g_{zz}}}\gamma^{3}\partial_{3}+\frac{1}{\sqrt{g_{uu}}}\gamma\partial_{4}\nonumber \\
 & +\frac{g^{\prime}_{tt}g_{xx}+g_{tt}\left(2g^{\prime}_{xx}+g_{xx}g^{\prime}_{zz}g^{-1}_{zz}\right)}{4g_{tt}g_{xx}\sqrt{g_{uu}}}\gamma.
\end{align}
In general, the components of $A_{M}$ may depend on $x^{\mu}$, hence
we need to rewrite $\psi$ as,

\begin{equation}
\psi=\left(\begin{array}{c}
\psi_{R}\\
\psi_{L}
\end{array}\right)=\left(-gg^{-1}_{uu}\right)^{-1/4}\int\frac{d^{4}k}{\left(2\pi\right)^{4}}e^{ik_{\mu}x^{\mu}}\chi\left(u,k,x\right),
\end{equation}
so that the Dirac equation obtained by varying (\ref{eq:3.2}) becomes,
\begin{equation}
\sqrt{\frac{g_{xx}}{g_{uu}}}\left(\gamma\partial_{u}-m\sqrt{g_{uu}}\right)\chi+i\gamma^{\mu}\hat{K}_{\mu}\chi=0,\label{eq:3.5}
\end{equation}
where 
\begin{equation}
\hat{K}_{\mu}=\left\{ -\sqrt{\frac{g_{xx}}{-g_{tt}}}\left(i\partial_{0}+g_{f}A_{0}\right),-i\partial_{1}-g_{f}A_{1},-i\partial_{2}-g_{f}A_{2},-\sqrt{\frac{g_{xx}}{g_{zz}}}\left(i\partial_{3}+g_{f}A_{3}\right)\right\} .\label{eq:3.6}
\end{equation}
If all the components of $A_{M}$ do not depend on $x^{\mu}$, the
Dirac equation (\ref{eq:3.5}) can be simply obtained by replacing
$-i\partial_{\mu}\rightarrow k_{\mu}$ in (\ref{eq:3.6}).

\subsection{The structure of the anisotropic correlation function}

Since the anisotropic holographic background illustrates $g_{xx}\neq g_{zz}$,
the associated Green function should depend on the direction of the
momentum since the $SO\left(3\right)$ rotation symmetry is broken
down to $SO\left(2\right)$. Thus, according to (\ref{eq:2.27}),
the structure of the anisotropic correlation function is expected
to be 
\begin{align}
\mathcal{G}_{R} & =\mathcal{G}^{\perp}_{R}+\mathcal{G}^{\parallel}_{R},\nonumber \\
\mathcal{G}^{\perp}_{R} & =\mathcal{A}^{\perp}+\frac{\mathcal{B}^{\perp}}{\left|\vec{k}\right|}k_{\alpha}\tau^{\alpha},\alpha=1,2,\nonumber \\
\mathcal{G}^{\parallel}_{R} & =\mathcal{A}^{\parallel}+\frac{\mathcal{B}^{\parallel}}{\left|\vec{k}\right|}k_{3}\tau^{3}.
\end{align}
The vertical directions are defined as $x^{\perp}=\left(x^{1},x^{2}\right)$
and the parallel direction is defined as $x^{\parallel}=x^{3}$. While
$\mathcal{G}^{\parallel}_{R}$ is diagonal and its algorithm has been
demonstrated in Section 2, the vertical Green function $\mathcal{G}^{\perp}_{R}$
is off-diagonal. To obtain $\mathcal{G}^{\perp}_{R}$, we can follow
the steps, first, chose the momentum as $k_{\mu}=\left(-\omega,\mathrm{k},0,0\right)$.
Then write down the Green function in the representation of $\tau^{1}$.
In this sense, the Pauli matrix becomes 
\begin{equation}
\tau^{1}=\left(\begin{array}{cc}
1 & 0\\
0 & -1
\end{array}\right),\ \tau^{3}=\left(\begin{array}{cc}
0 & 1\\
1 & 0
\end{array}\right),
\end{equation}
so $\mathcal{G}^{\perp}_{R}$ becomes diagonal. Therefore the diagonal
Green function $\mathcal{G}^{\perp}_{R}$, in the representation of
$\tau^{1}$ $\mathcal{G}^{\perp}_{R}\big|_{\tau^{1}}$, can be obtained
by the following equation 
\begin{equation}
\sqrt{\frac{g_{xx}}{g_{uu}}}\xi^{\prime}_{\left(\alpha\right)}=\left(-1\right)^{\alpha+1}K_{0}-K_{1}+\left[K_{1}+\left(-1\right)^{\alpha+1}K_{0}\right]\xi^{2}_{\left(\alpha\right)}+2m\sqrt{g_{xx}}\xi_{\left(\alpha\right)}.\label{eq:3.9}
\end{equation}
and these equations can be computed by following the steps given in
Section 2. Note that, when the gauge field $A_{M}$ is taken into
account, $K_{\mu}$ should be replaced by $K_{\mu}\rightarrow\hat{K}_{\mu}$.
Afterwards, the Green function in the usual representation of $\tau^{3}$
can be obtained by the following unitary transformation: 
\begin{equation}
U^{\dagger}\mathcal{G}^{\perp}_{R}\big|_{\tau^{1}}U=\mathcal{G}^{\perp}_{R}\big|_{\tau^{3}},\ U=\frac{1}{\sqrt{2}}\left(\tau^{1}+\tau^{3}\right).
\end{equation}
In this work, we will use the above methods to evaluate the fermionic
correlation function in the various models.

\subsection{The perturbative method}

In our concerned gravity backgrounds, we will see some metrics take
the perturbative form as, 
\begin{equation}
g_{MN}=G_{MN}+\epsilon h_{MN}+\mathcal{O}\left(\epsilon^{2}\right),\epsilon\ll1,
\end{equation}
where $G_{MN}$ is the zeroth-order metric and $\epsilon$ is the
perturbative parameter. In order to find the corresponding perturbative
corrections to the associated Green function, we assume the ratios
defined in (\ref{eq:2.29}) can be decomposed as, 
\begin{equation}
\xi_{\left(\alpha\right)}=\Lambda_{\left(\alpha\right)}+\epsilon\lambda_{\left(\alpha\right)},\label{eq:3.12}
\end{equation}
where $\Lambda_{\left(\alpha\right)}$ is the zeroth-order solution
i.e. which is solved by using (\ref{eq:2.31}) with respect to $G_{MN}$.
Keeping these in hand, it is possible to derive the equations for
$\lambda_{\left(\alpha\right)}$. By the following series ($X_{1,2,...5}$
are the perturbative functions by the expansions with respect to $\epsilon$)
\begin{align}
\sqrt{\frac{g_{xx}}{g_{uu}}} & =\sqrt{\frac{g_{xx}}{g_{uu}}}\big|_{\epsilon=0}+\epsilon X_{1},\ K_{0}=K_{0}\big|_{\epsilon=0}+\epsilon X_{2},\nonumber \\
\sqrt{g_{xx}} & =\sqrt{g_{xx}}\big|_{\epsilon=0}+\epsilon X_{3},\ K_{3}=K_{3}\big|_{\epsilon=0}+\epsilon X_{4},\nonumber \\
K_{1} & =K_{1}\big|_{\epsilon=0}+\epsilon X_{5},
\end{align}
then imposing (\ref{eq:3.12}) onto (\ref{eq:2.31}) and (\ref{eq:3.9}),
the equations for $\lambda_{\left(\alpha\right)}$ are obtained as,
\begin{align}
\sqrt{\frac{G_{xx}}{G_{uu}}}\partial_{u}\lambda^{\parallel,\perp}_{\left(1\right)}= & X_{2}-X^{\parallel,\perp}+\left(X^{\parallel,\perp}+X_{2}\right)\left[\Lambda^{\parallel,\perp}_{\left(1\right)}\right]^{2}+2\lambda_{\left(1\right)}\left[\left(K^{\parallel,\perp}+K_{0}\right)\big|_{\epsilon=0}\right]\Lambda^{\parallel,\perp}_{\left(1\right)}\nonumber \\
 & +2m\left[X_{3}\Lambda^{\parallel,\perp}_{\left(1\right)}+\sqrt{G_{xx}}\lambda^{\parallel,\perp}_{\left(1\right)}\right]-X_{1}\partial_{u}\Lambda^{\parallel,\perp}_{\left(1\right)},\nonumber \\
\sqrt{\frac{G_{xx}}{G_{uu}}}\partial_{u}\lambda^{\parallel,\perp}_{\left(2\right)}= & -X_{2}-X^{\parallel,\perp}+\left(X^{\parallel,\perp}-X_{2}\right)\left[\Lambda^{\parallel,\perp}_{\left(2\right)}\right]^{2}+2\lambda_{\left(2\right)}\left[\left(K^{\parallel,\perp}-K_{0}\right)\big|_{\epsilon=0}\right]\Lambda^{\parallel,\perp}_{\left(2\right)}\nonumber \\
 & +2m\left[X_{3}\Lambda^{\parallel,\perp}_{\left(2\right)}+\sqrt{G_{xx}}\lambda^{\parallel,\perp}_{\left(2\right)}\right]-X_{1}\partial_{u}\Lambda^{\parallel,\perp}_{\left(2\right)},\label{eq:3.14}
\end{align}
where 
\begin{equation}
K^{\parallel}=K_{3},K^{\perp}=K_{1},X^{\parallel}=X_{4},X^{\perp}=X_{5}.
\end{equation}
Note that if the zeroth metric $G_{MN}$ is isotropic, the associated
zeroth-order solution satisfies $\Lambda^{\parallel}_{\left(\alpha\right)}=\Lambda^{\perp}_{\left(\alpha\right)}$.
If the anisotropic metric $g_{MN}$ has the same asymptotic behavior
as the AdS-Schwarzschild black brane metric near the horizon $u=u_{H}$,
it implies $\xi_{\left(\alpha\right)}\big|_{u=u_{H}}=\Lambda_{\left(\alpha\right)}\big|_{u=u_{H}}=\left(-1\right)^{\alpha}i$
so that it leads to $\lambda_{\left(\alpha\right)}\big|_{u=u_{H}}=0$.
This perturbative method will be applied to the asymptotic AdS backgrounds
which take the perturbative forms.

\section{The anisotropy induced by axion}

In this section, we consider the holographic model with anisotropy
induced by axion, then analyze the fermionic correlation functions
in this holographic system.

\subsection{The model}

The anisotropic gravity background induced by axion is proposed in
\cite{Banks:2016fab,Giataganas:2017koz,Mateos:2011ix,A2016mno,Mateos:2011tv,Cheng:2014qia,Cheng:2014sxa}.
This model is obtained by the system of D3-D7 intersection which describes
the dynamics of $N_{c}$ D3-branes with $N_{\mathrm{D7}}$ D7-branes
dissolved inhomogeneously in the ten-dimensional bulk spacetime in
the limit of $N_{c}\rightarrow\infty$ with fixed value of $N_{\mathrm{D7}}/N_{c}$.
The dynamics is described by the type IIB supergravity action in string
frame as\footnote{Action~(4.1) should be understood as the bosonic action restricted
to the fields relevant to the present background. In type IIB supergravity,
the self-dual condition $F_{5}=\ ^{\star}F_{5}$ does not follow from
varying this action and must be imposed separately at the level of
the equations of motion. See Ref.~\cite{Becker:2007} for further
details.},

\begin{equation}
S_{\mathrm{IIB}}=\frac{1}{2\kappa^{2}_{10}}\int d^{10}x\sqrt{-g}\left[e^{-2\phi}\left(\mathcal{R}+4\partial_{M}\phi\partial^{M}\phi\right)-\frac{1}{2}F^{2}_{1}-\frac{1}{4\cdot5!}F^{2}_{5}\right],\label{eq:4.1}
\end{equation}
which includes a dilaton $\phi$, a Ramond-Ramond zero-form $C_{0}$
as the axion field with its strength $F_{1}=dC_{0}$ and a Ramond-Ramond
four-form $C_{4}$ with its strength $F_{5}=dC_{4}$ in addition to
the ten-dimensional metric $g_{MN}$. This system can be reduced effectively
to a five-dimensional gravity theory as the axion-dilaton-Maxwell-gravity
and the action in Einstein frame takes form as \footnote{The $U\left(1\right)$ field $A_{M}$ originates from the ten-dimensional
metric, which provides the ansatz solution for the action (4.1), see
the details in \cite{Cheng:2014qia}. There should be a Chern-Simons
term $S=-\frac{N^{2}}{96\pi^{2}}\int A\wedge F\wedge F$ with $F=dA$
in the five-dimensional effective action (\ref{eq:4.2}) reduced from
the coincident $N$ D3-brane solution. However this CS term vanishes
since only one component of $A_{M}$ as $A=A_{0}dt$ is turned on
in this setup.} \cite{Giataganas:2017koz,Mateos:2011ix,Mateos:2011tv,Cheng:2014qia,Cheng:2014sxa},

\begin{equation}
S_{\mathrm{5D}}=\frac{1}{2\kappa^{2}}\int d^{5}x\sqrt{-g}\left[\mathcal{R}_{\left(\mathrm{5D}\right)}-\frac{2\Lambda}{L^{2}}-\frac{1}{2}\partial_{M}\phi\partial^{M}\phi-\frac{1}{2}e^{2\phi}\partial_{M}C_{0}\partial^{M}C_{0}-\frac{1}{4}F_{MN}F^{MN}\right]+S_{\mathrm{bdry}},\label{eq:4.2}
\end{equation}
where $\Lambda=-6$ is the cosmological constant, $F_{MN}=\partial_{M}A_{N}-\partial_{N}A_{M}$
is the $U\left(1\right)$ gauge field strength, $\kappa$ is the five-dimensional
gravity coupling constant and $S_{\mathrm{bdry}}$ is the gravity
boundary terms. The five-dimensional solution for the gravity system
in Einstein frame takes the form

\begin{align}
ds^{2} & =e^{-\frac{1}{2}\phi}\frac{L^{2}}{u^{2}}\left[-\mathcal{F}\mathcal{B}dt^{2}+\left(dx^{1}\right)^{2}+\left(dx^{2}\right)^{2}+\mathcal{H}\left(dx^{3}\right)^{2}+\frac{du^{2}}{\mathcal{F}}\right],\nonumber \\
A & =A_{0}dt,\ C_{0}=az,\ \mathcal{H}=e^{-\phi}\label{eq:4.3}
\end{align}
where $\mathcal{F},\mathcal{B},\phi,A_{0}$ are functions of the radial
coordinate $u$ only, and the parameters are given as,

\begin{equation}
L^{4}=4\pi g_{s}N_{c}l^{4}_{s}=\lambda l^{4}_{s},a=\frac{\lambda n_{\mathrm{D7}}}{4\pi N_{c}},n_{\mathrm{D7}}=\frac{dN_{\mathrm{D7}}}{dz}.
\end{equation}
Here $l_{s}$ is the length of string and $\lambda$ is the 't Hooft
coupling constant. Note that the field strength $a$ of the axion
dominates the density of the D7-branes along $z$ direction and the
anisotropy since the metric returns to the isotropic case without
D7-branes if $a=0$. The horizon and holographic boundary are located
at $u=u_{H}$ and $u=0$ respectively. While the functions $\mathcal{F},\mathcal{B},\phi,A_{0}$
in general are non-analytical, their asymptotical behaviors are required
as $\mathcal{F}\left(u_{H}\right)=0,$ and $\mathcal{F}\left(0\right)=\mathcal{B}\left(0\right)=1,\phi\left(0\right)=0$.
The Hawking temperature is given by

\begin{equation}
T=-\frac{\sqrt{\mathcal{B}}}{4\pi}\partial_{u}\mathcal{F}\big|_{u=u_{H}}.
\end{equation}
In particular, we consider the high temperature limit $T\gg a$ or
equivalently $u_{H}a\ll1$ in the dual plasma, in this case, the functions
$\mathcal{F},\mathcal{B},\phi,A_{0}$ can be written as series of
$u_{H}a$ \cite{Li:2026wqu,A2016mno,Mateos:2011tv,Cheng:2014qia,Cheng:2014sxa,Li:2022wwv}
which are given in the appendix. So it would be closer to the realistic
plasma. 
\begin{figure}[t]
\includegraphics[scale=0.39]{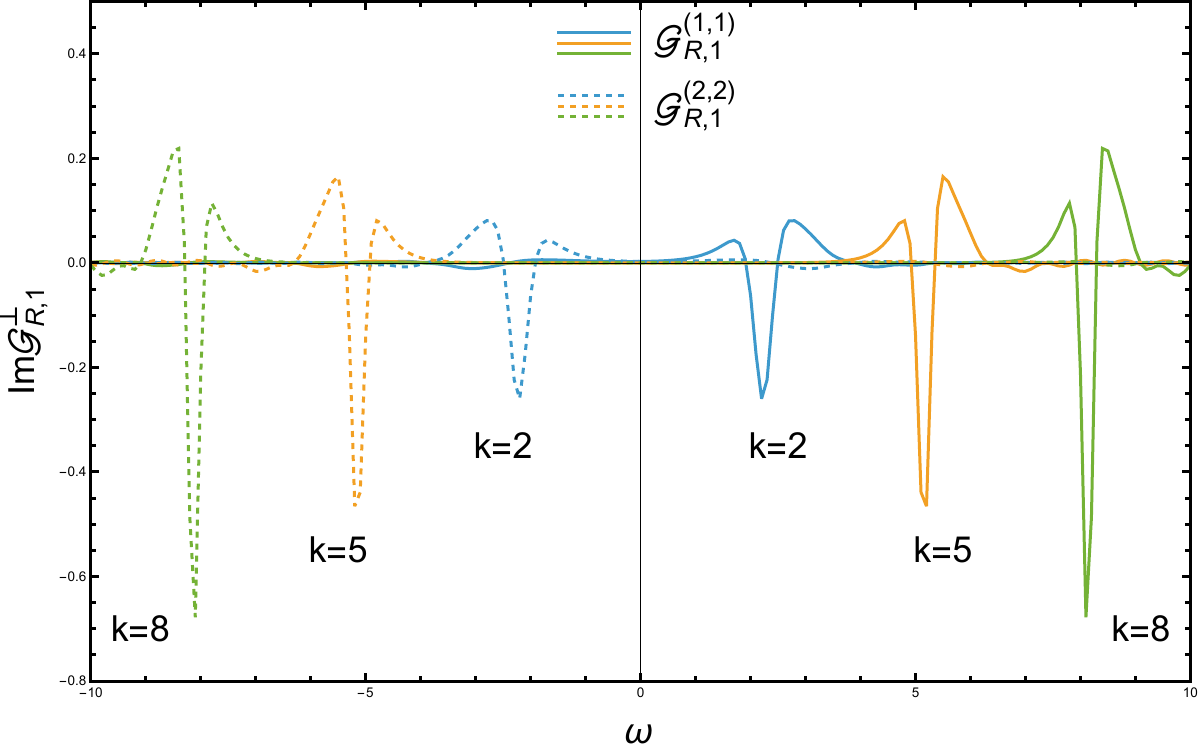}\includegraphics[scale=0.39]{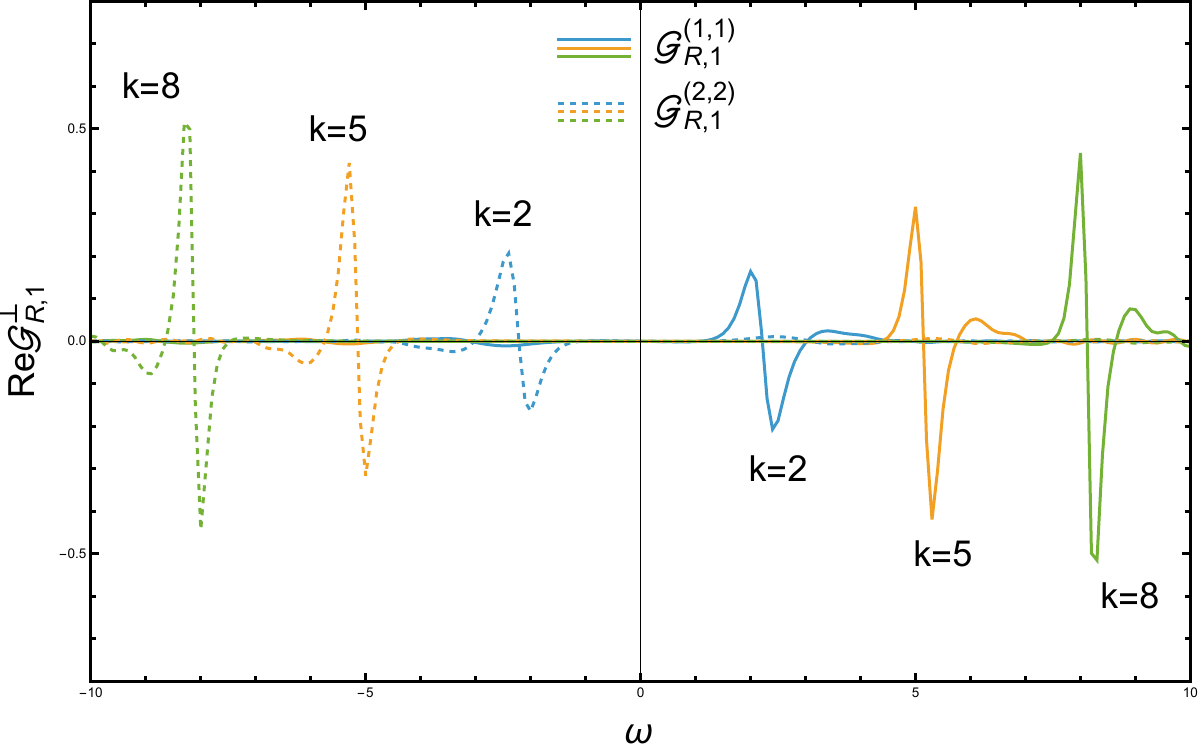}

\includegraphics[scale=0.39]{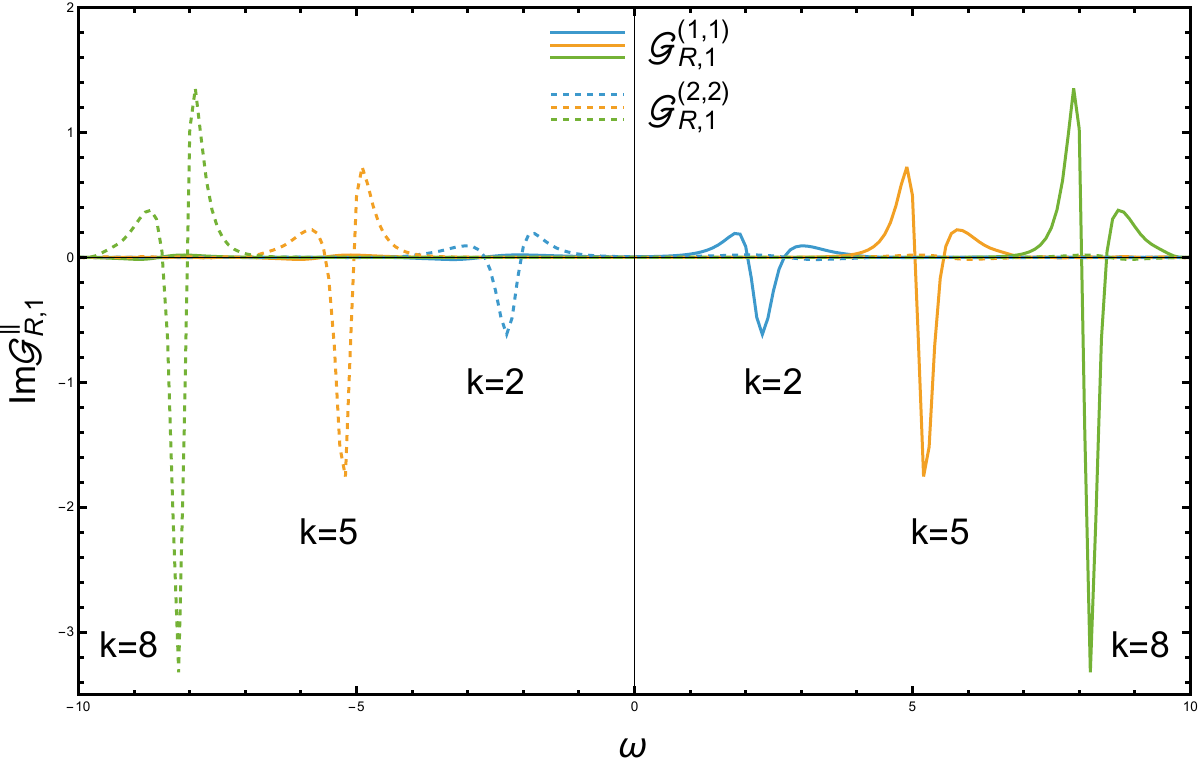}\includegraphics[scale=0.39]{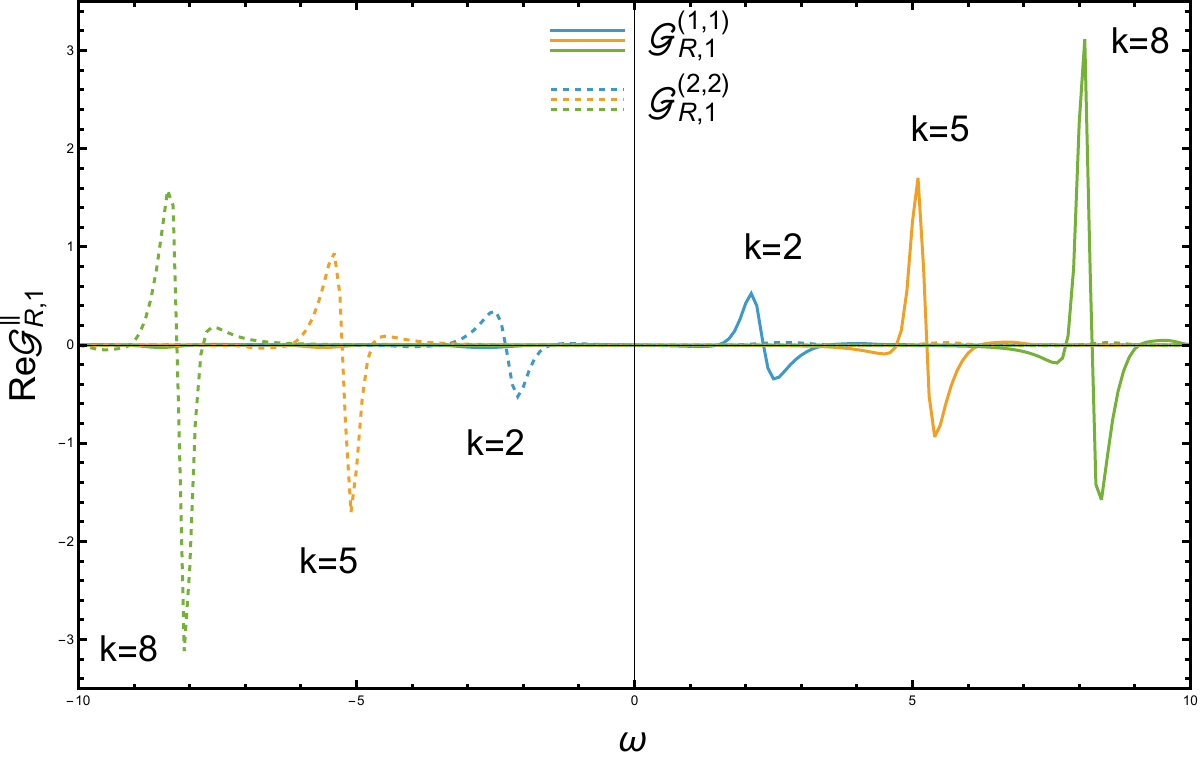}

\caption{Anisotropic perturbations $\mathcal{G}^{\left(\alpha,\alpha\right)}_{R,1}$
induced by axion to the fermionic correlation functions. The parameters
are chosen to be $u_{H}=L=1,m=0.01$.}
\label{fig:3} 
\end{figure}

\subsection{Numerical results}

Since the metric with anisotropy in this model can be written in the
perturbative form with respect to $u_{H}a$ (see the appendix), our
numerical results in this section are obtained by using the perturbative
methods discussed in Section 3.3. 
\begin{figure}[t]
\includegraphics[scale=0.39]{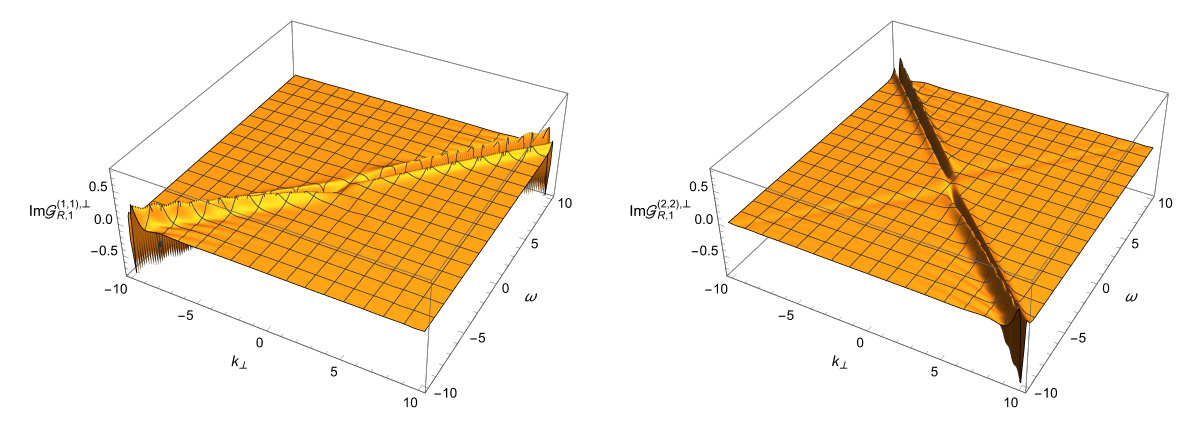}\includegraphics[scale=0.39]{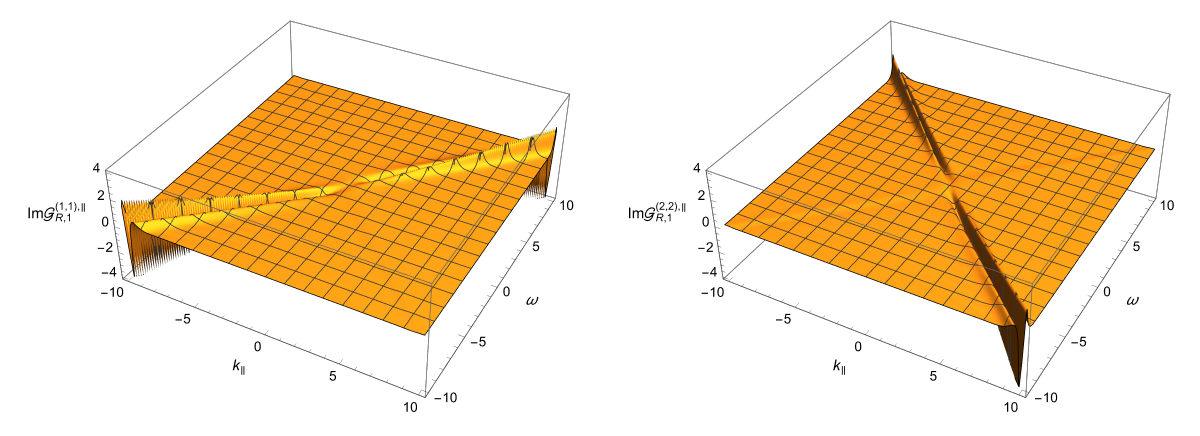}

\includegraphics[scale=0.39]{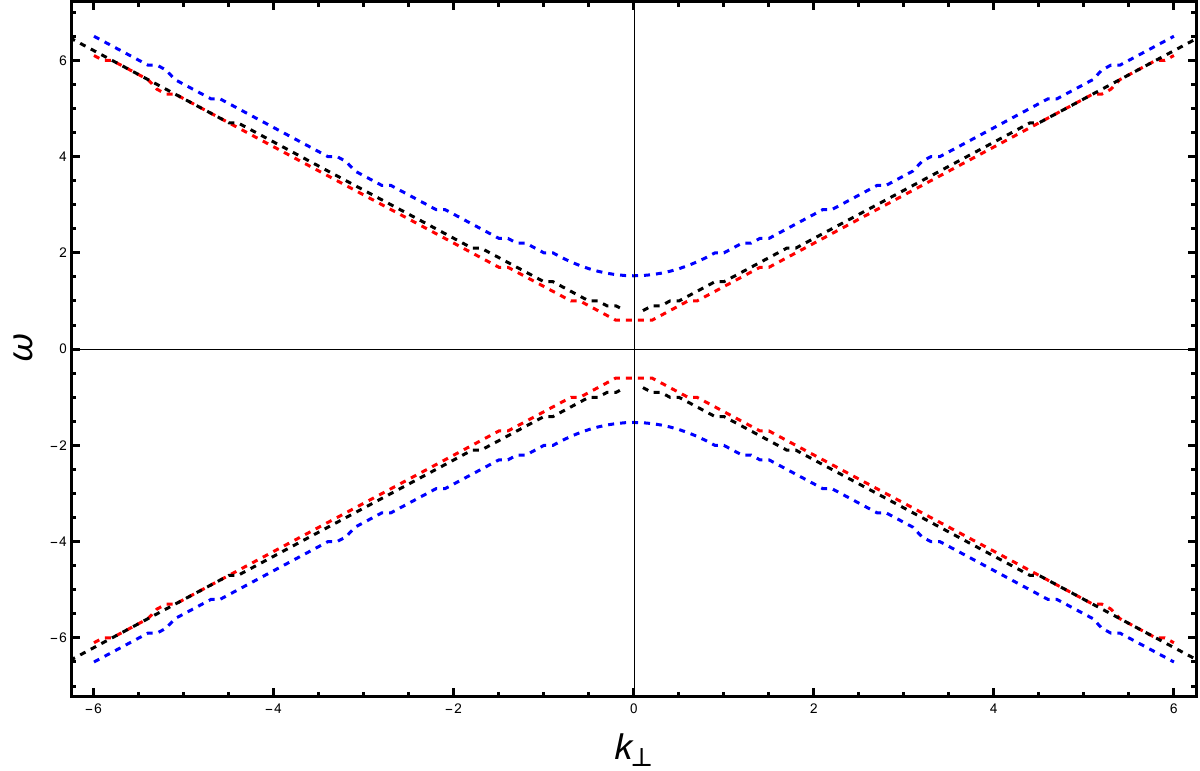}\includegraphics[scale=0.39]{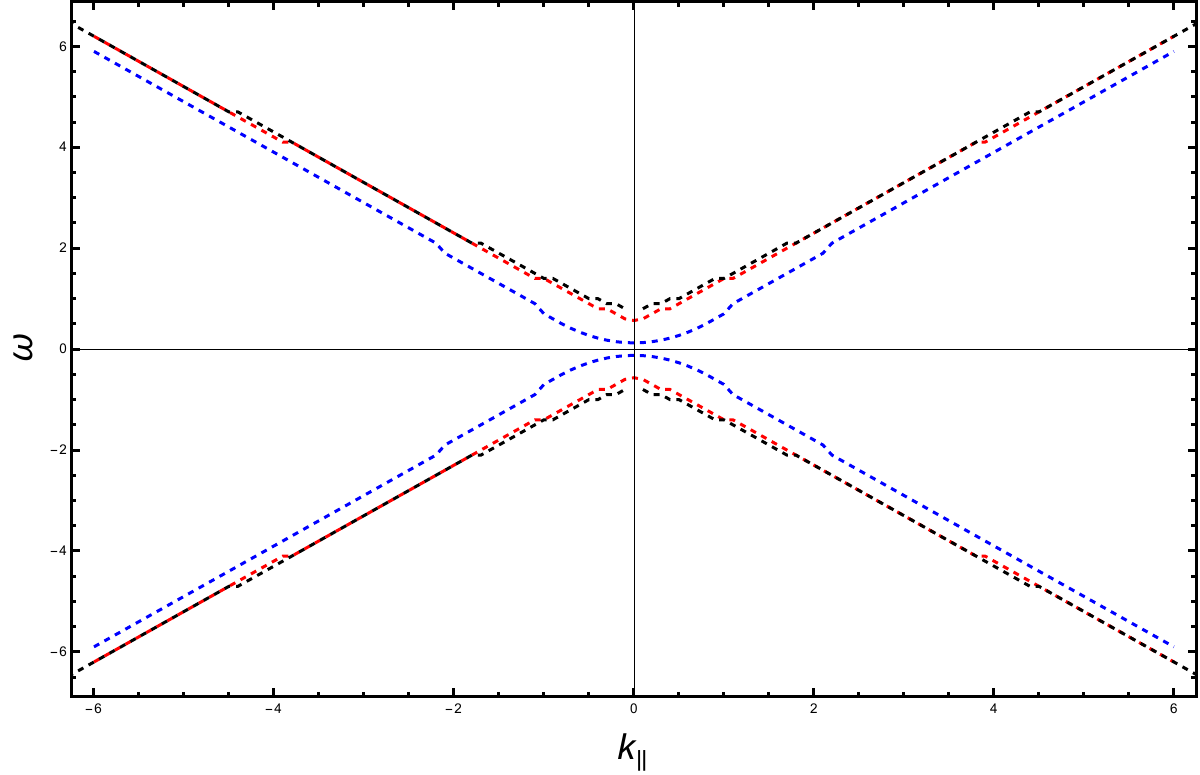}

\caption{The perturbative fermionic dispersion curve with respect to the axion
induced anisotropy. The parameters are set as $u_{H}=L=1,m=0.01$.
The blue and red dashed lines are derived from the peaks and dips,
respectively, in the perturbative part of the Green functions. The
black dashed line refers to the dispersion curve obtained from the
zeroth-order Green function.}
\label{fig:4} 
\end{figure}

\begin{figure}
\begin{centering}
\includegraphics[scale=0.39]{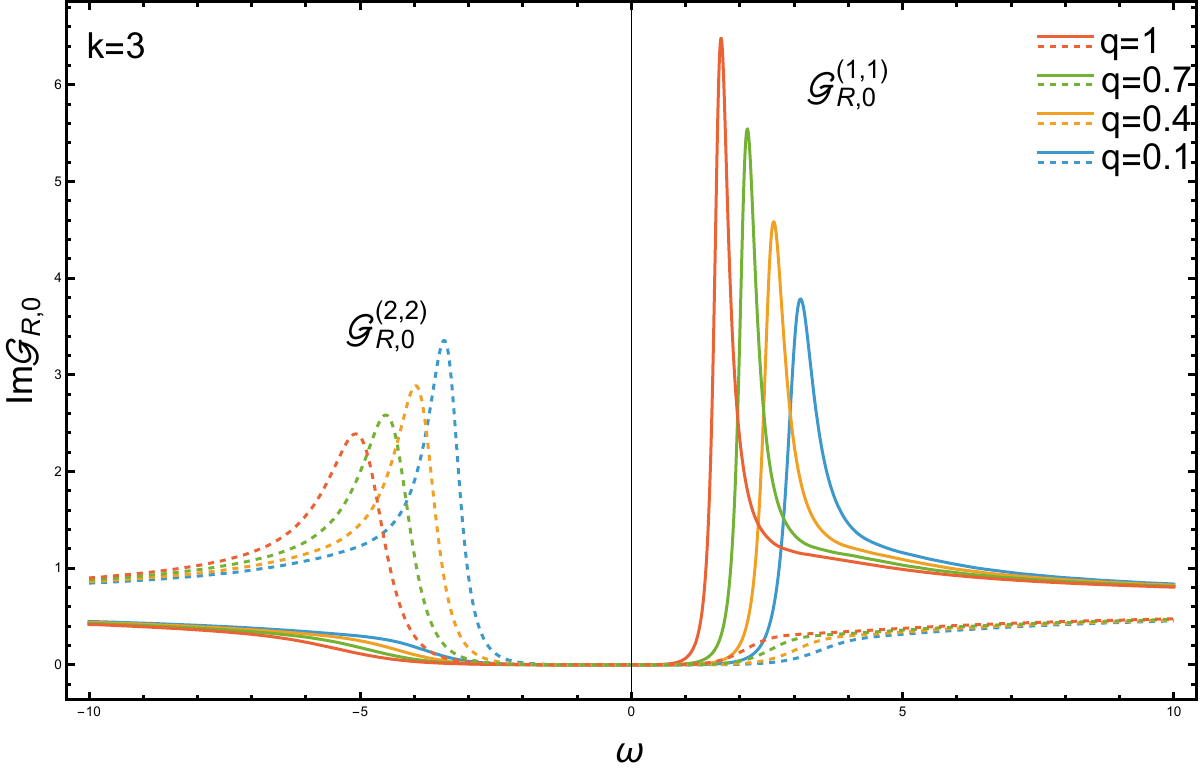}\includegraphics[scale=0.39]{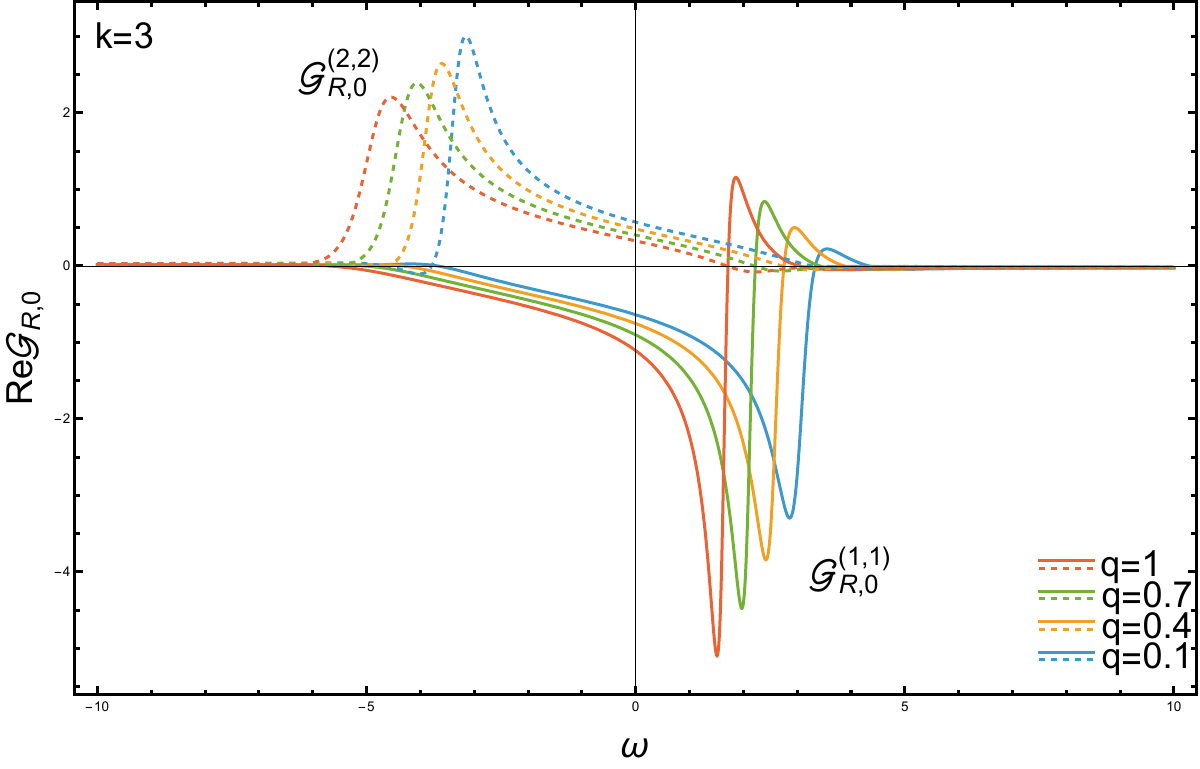} 
\par\end{centering}
\begin{centering}
\includegraphics[scale=0.39]{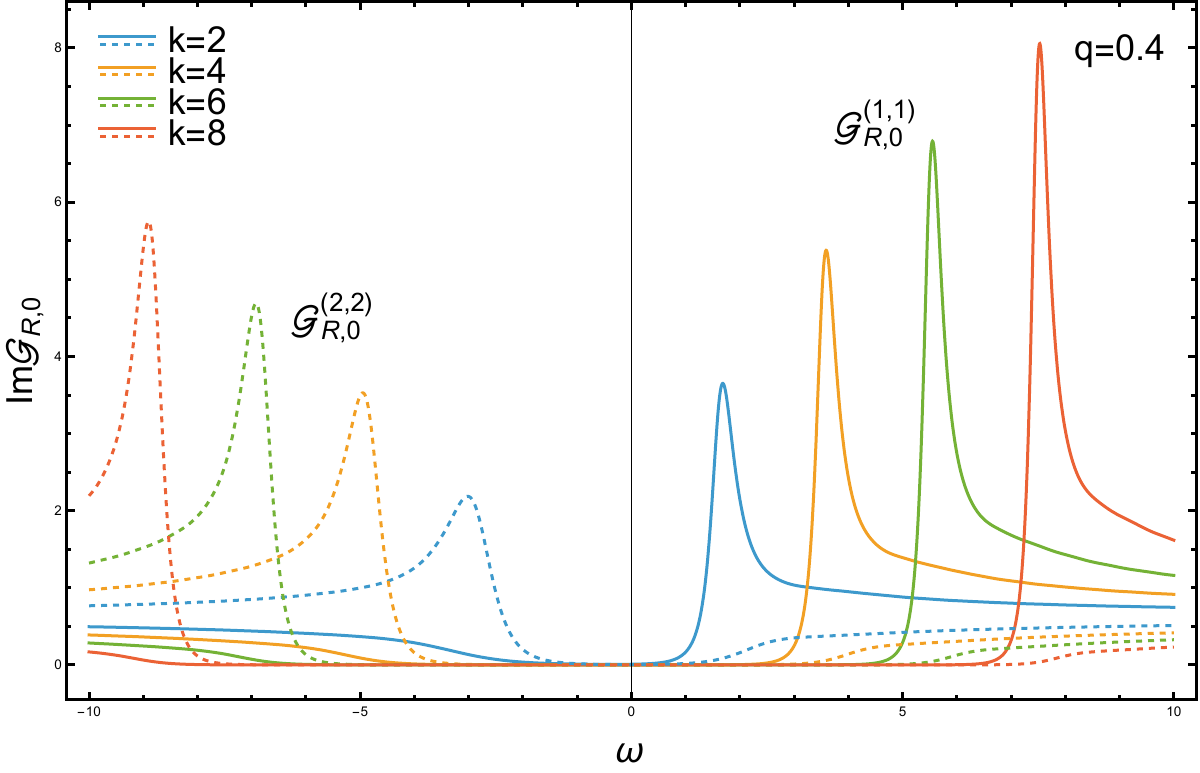}\includegraphics[scale=0.39]{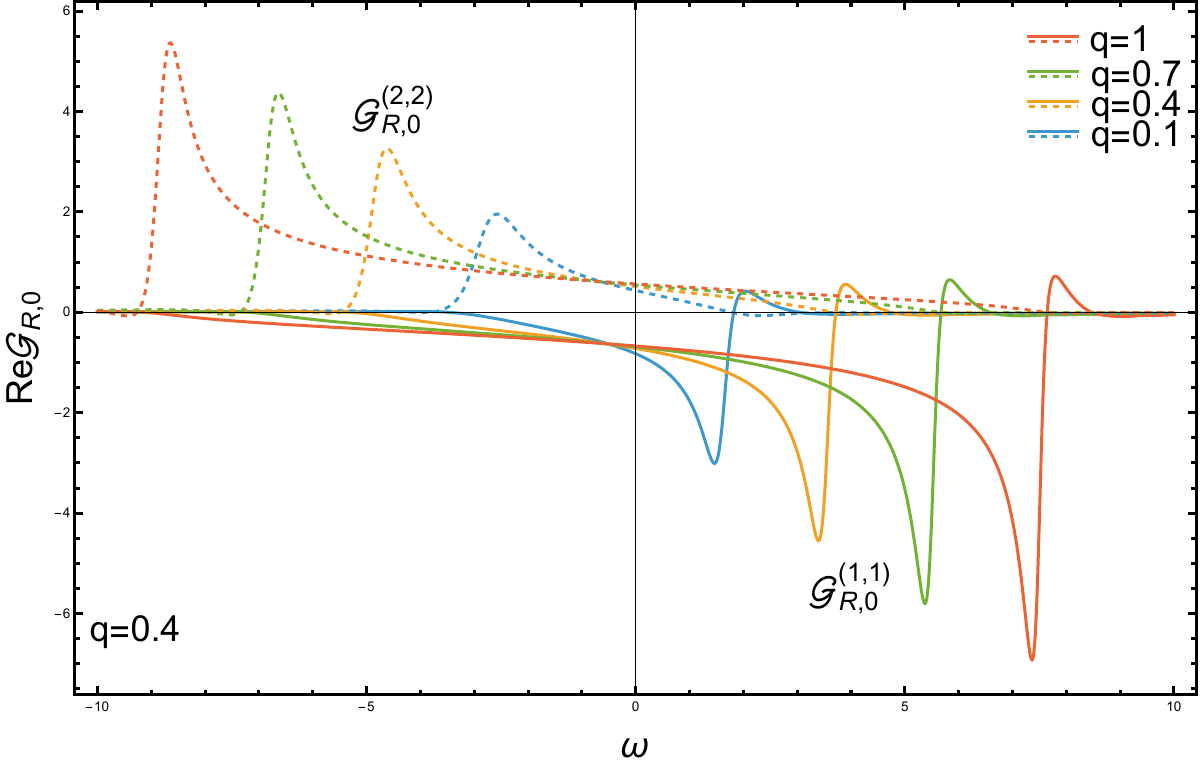} 
\par\end{centering}
\caption{The zeroth-order fermionic Green function $\mathcal{G}_{R,0}$ as
the isotropic correlation function with chemical potential. The parameters
are chosen as $u_{H}=g_{f}=L=1,m=0.01$. \textbf{Upper:} momentum
$\mathrm{k}$ is fixed. \textbf{Lower:} chemical potential is fixed
by fixing $q$.}
\label{fig:5} 
\end{figure}

\subsubsection*{Turning off the chemical potential}

\begin{figure}
\includegraphics[width=5.5cm,totalheight=7cm]{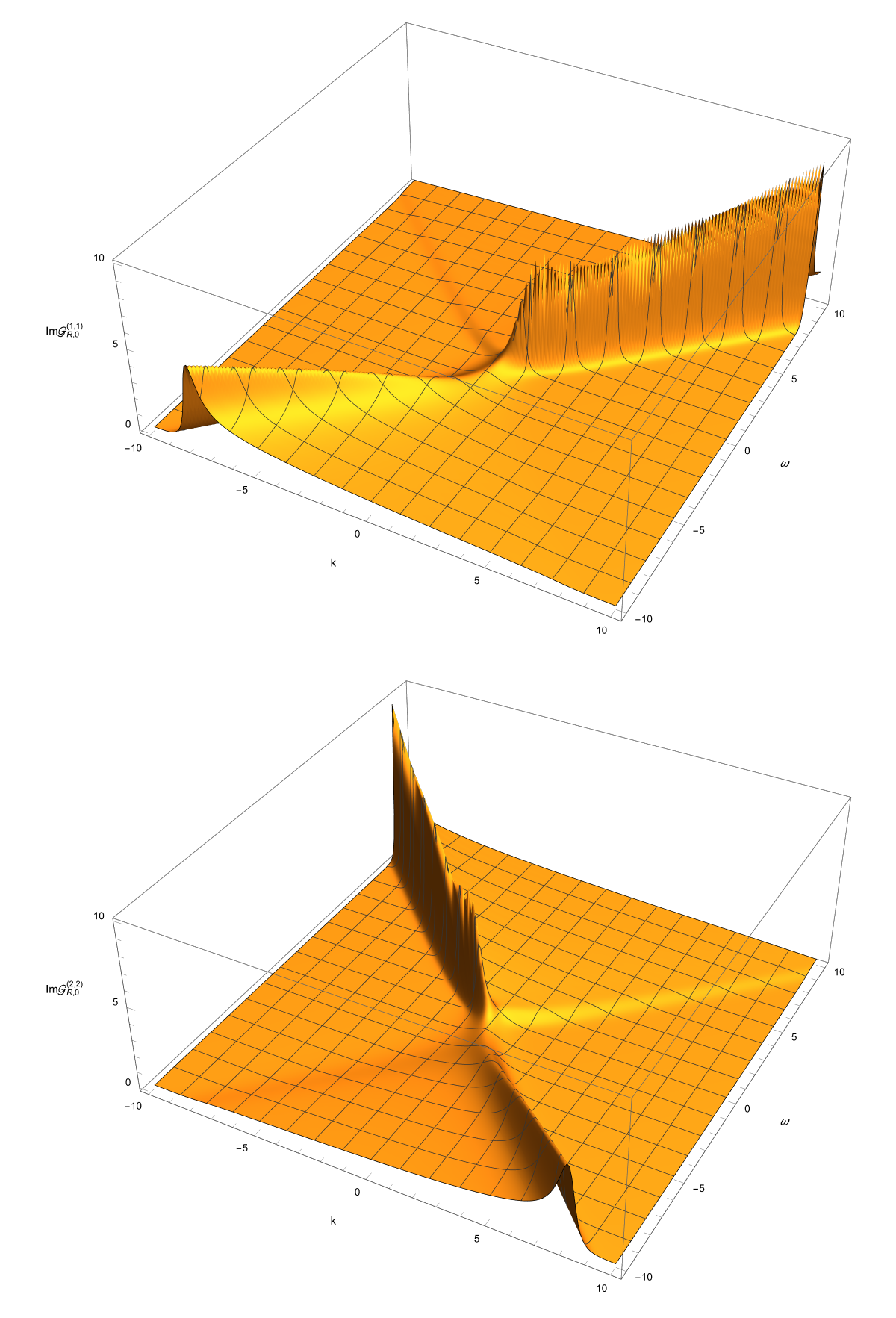}\includegraphics[width=10cm,totalheight=7cm]{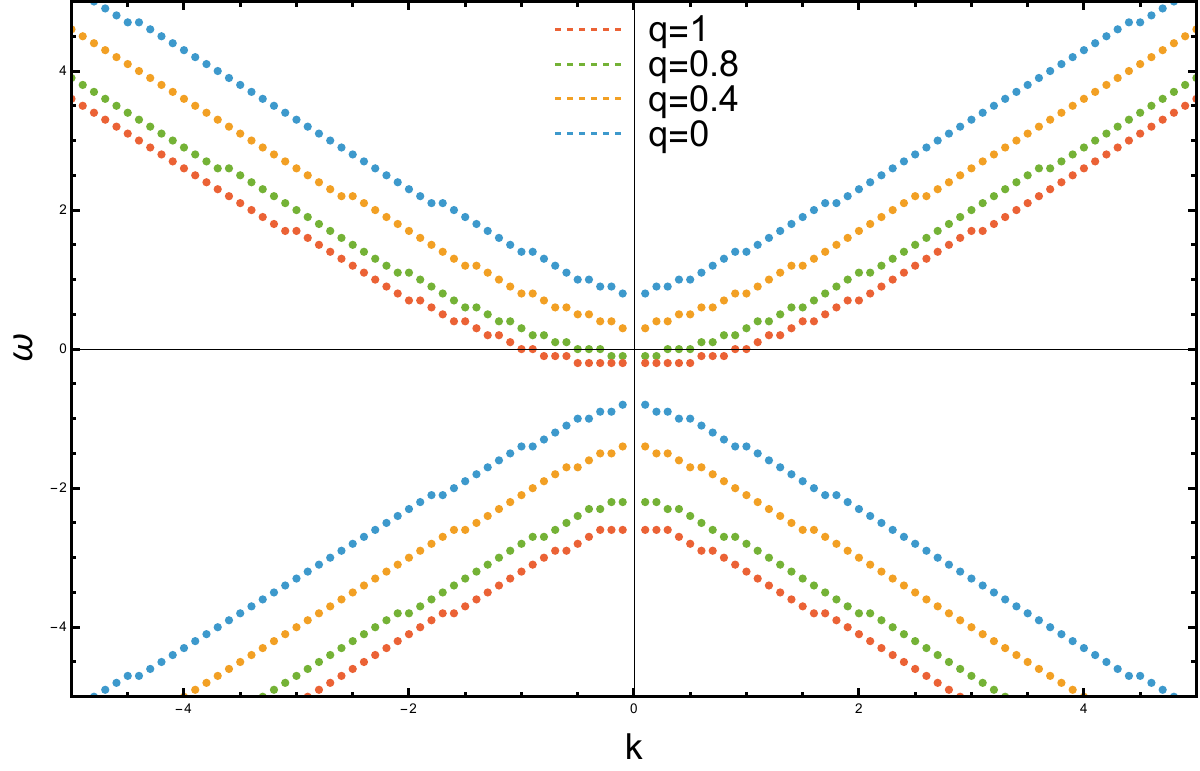}

\caption{The fermionic dispersion curve from the isotropic Green function with
chemical potential. The parameters are set as $u_{H}=g_{f}=L=1,m=0.01$.}
\label{fig:6} 
\end{figure}

The metric (\ref{eq:4.3}) takes the same asymptotic behavior as the
AdS-Schwarzschild metric at $u=u_{H}$ and $u=0$, hence the Dirac
field on bulk takes the same asymptotic solution as (\ref{eq:2.16}).
On the other hand, in the case without chemical potential i.e. $A_{0}=0$,
the metric (\ref{eq:4.3}) for $a=0$ returns to the metric of the
AdS-Schwarzschild black brane (\ref{eq:2.15}). Therefore, by choosing
the perturbative parameter in (\ref{eq:3.12}) as $\epsilon\rightarrow a^{2}u^{2}_{H}$,
the holographic Green function $\mathcal{G}^{\left(\alpha,\alpha\right)}_{R}$
can be written as, 
\begin{equation}
\mathcal{G}^{\left(\alpha,\alpha\right)}_{R}=\mathcal{G}^{\left(\alpha,\alpha\right)}_{R,0}+a^{2}u^{2}_{H}\mathcal{G}^{\left(\alpha,\alpha\right)}_{R,1}=\left(-1\right)^{\alpha}\lim_{u\rightarrow0}u^{-2mL}\left[\Lambda_{\left(\alpha\right)}+a^{2}u^{2}_{H}\lambda_{\left(\alpha\right)}\right].\label{eq:4.6}
\end{equation}
where $\mathcal{G}^{\left(\alpha,\alpha\right)}_{R,0}\equiv\mathcal{G}^{\left(\alpha,\alpha\right)}_{R}\big|_{a=0},\ \Lambda_{\left(\alpha\right)}$
are the Green function and solution for the ratios (\ref{eq:3.12})
on the AdS-Schwarzschild black brane background as the zeroth-order
results given in Section 2.2. So the leading order perturbation $\lambda_{\left(\alpha\right)}$
can be evaluated numerically by solving (\ref{eq:3.14}) with respect
to the metric (\ref{eq:4.3}), and the associated numerical results
for $\mathcal{G}^{\left(\alpha,\alpha\right)}_{R,1}$ are illustrated
in Figure \ref{fig:3}. Note that, the perturbative solutions remain
satisfying $\lambda_{\left(1\right)}\left(\omega,\mathrm{k}\right)=\lambda_{\left(2\right)}\left(\omega,-\mathrm{k}\right)$
and $\lambda^{*}_{\left(1\right)}\left(\omega,\mathrm{k}\right)=-\lambda_{\left(2\right)}\left(-\omega,\mathrm{k}\right)$.
While the numerical results display peaks in the imaginary part (which
is the correction to the zeroth-order Green function), there is always
a negative dip beside them. The dips in the corrections coincide with
the peaks of the zeroth-order Green functions, confirming that the
perturbations indeed reduce the total Green functions. To support
our analysis, we also numerically plot the perturbatively corrected
dispersion curve in Figure \ref{fig:4}. According to these numerical
results, we demonstrate the dependence of the fermion Green function
on the momentum direction in this anisotropic background.

\subsubsection*{Turning on the chemical potential}

\begin{figure}[t]
\begin{centering}
\includegraphics[scale=0.39]{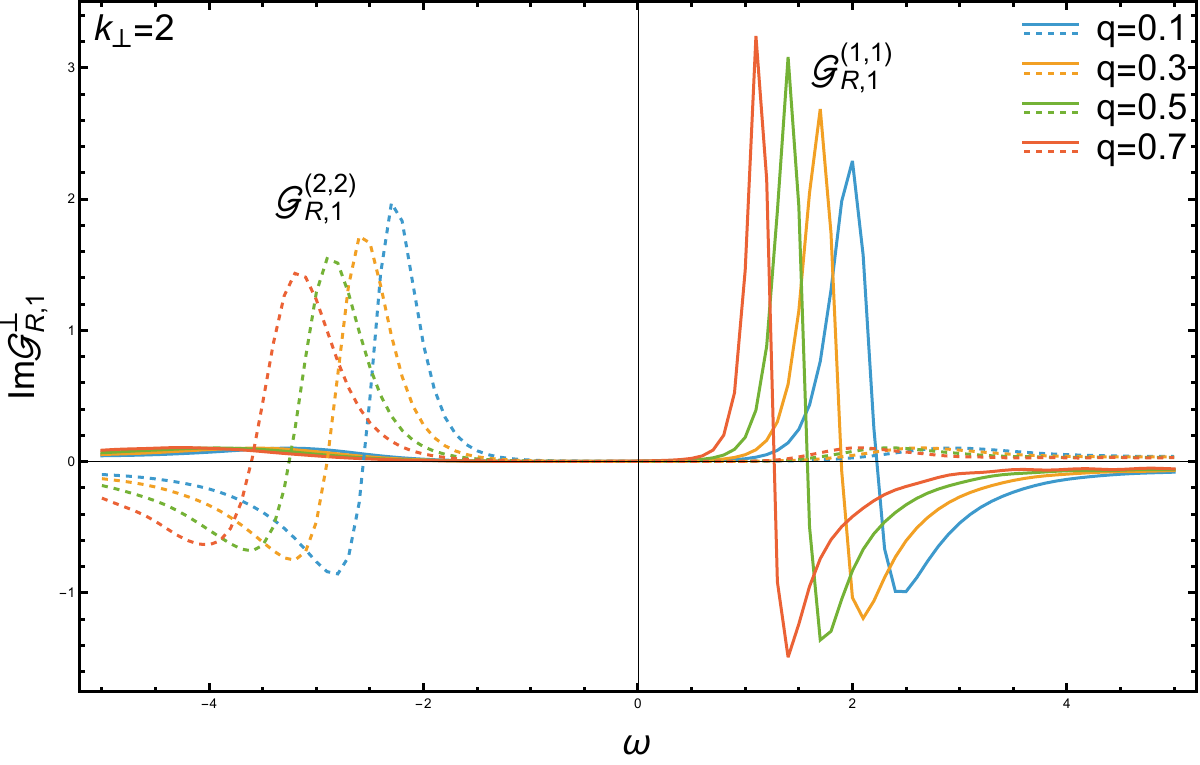}\includegraphics[scale=0.39]{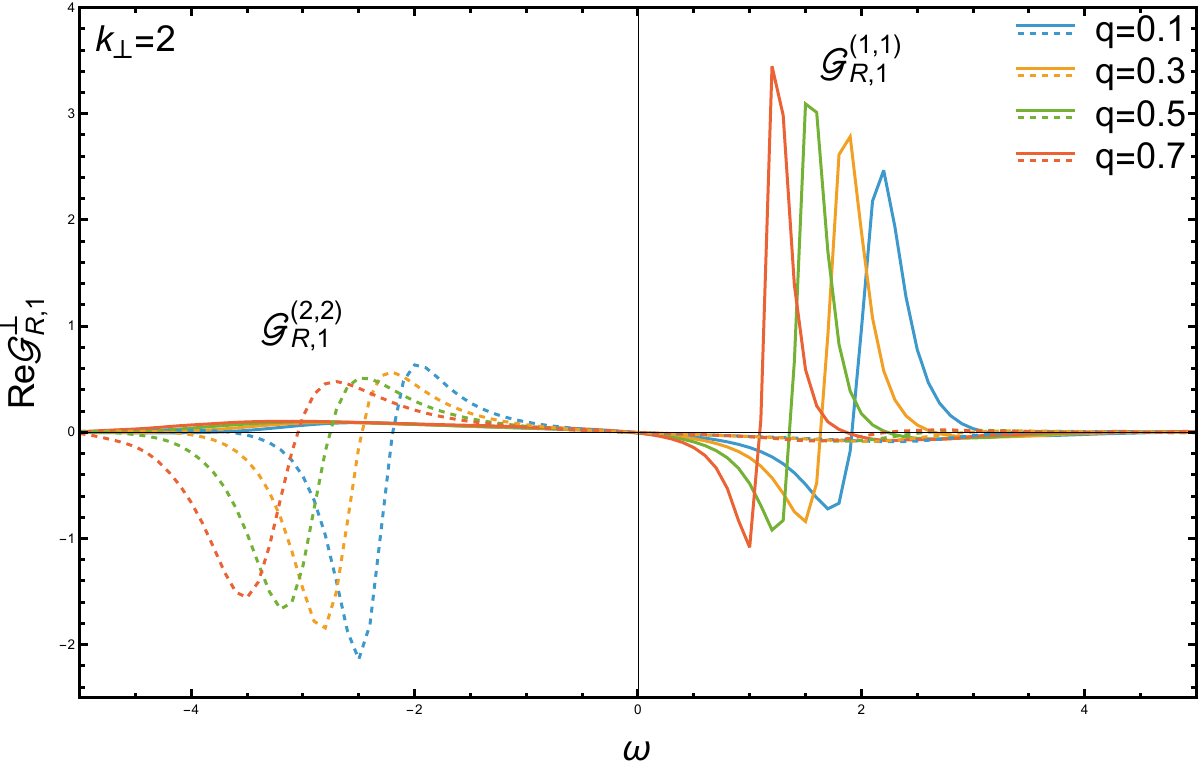} 
\par\end{centering}
\begin{centering}
\includegraphics[scale=0.39]{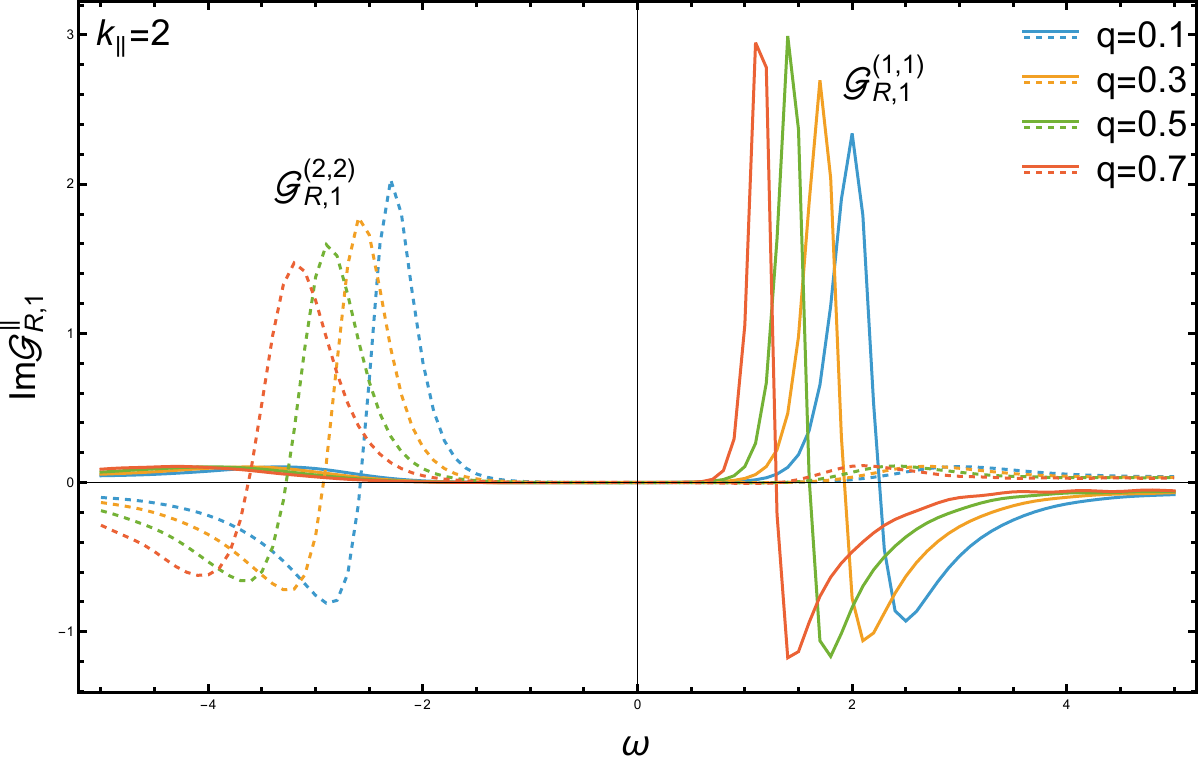}\includegraphics[scale=0.39]{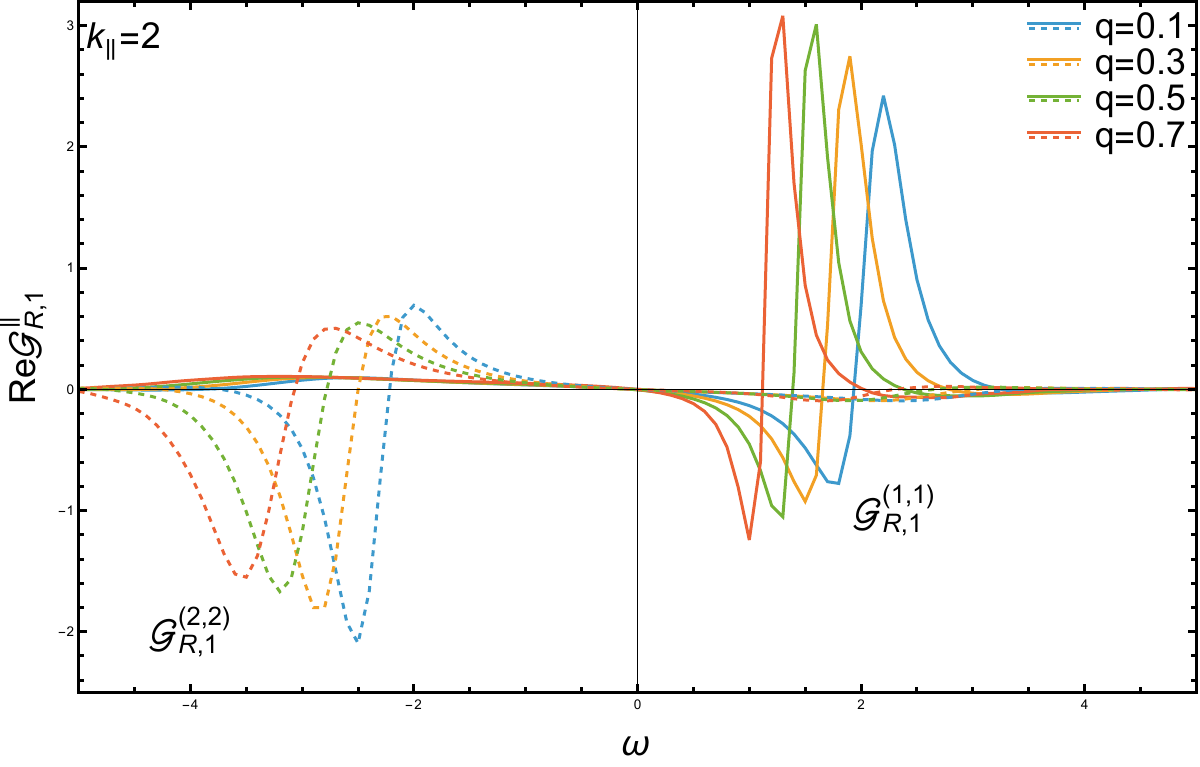} 
\par\end{centering}
\caption{Anisotropic perturbations $\mathcal{G}^{\left(\alpha,\alpha\right)}_{R,1}$
induced by axion to the fermionic correlation functions with chemical
potential and fixed momentum. The parameters are set as $u_{H}=g_{f}=L=1,m=0.01$.
\textbf{Upper:} the momentum is perpendicular to the direction of
$x^{3}$. \textbf{Lower:} the momentum is parallel to the direction
of $x^{3}$.}
\label{fig:7} 
\end{figure}

\begin{figure}[t]
\begin{centering}
\includegraphics[scale=0.39]{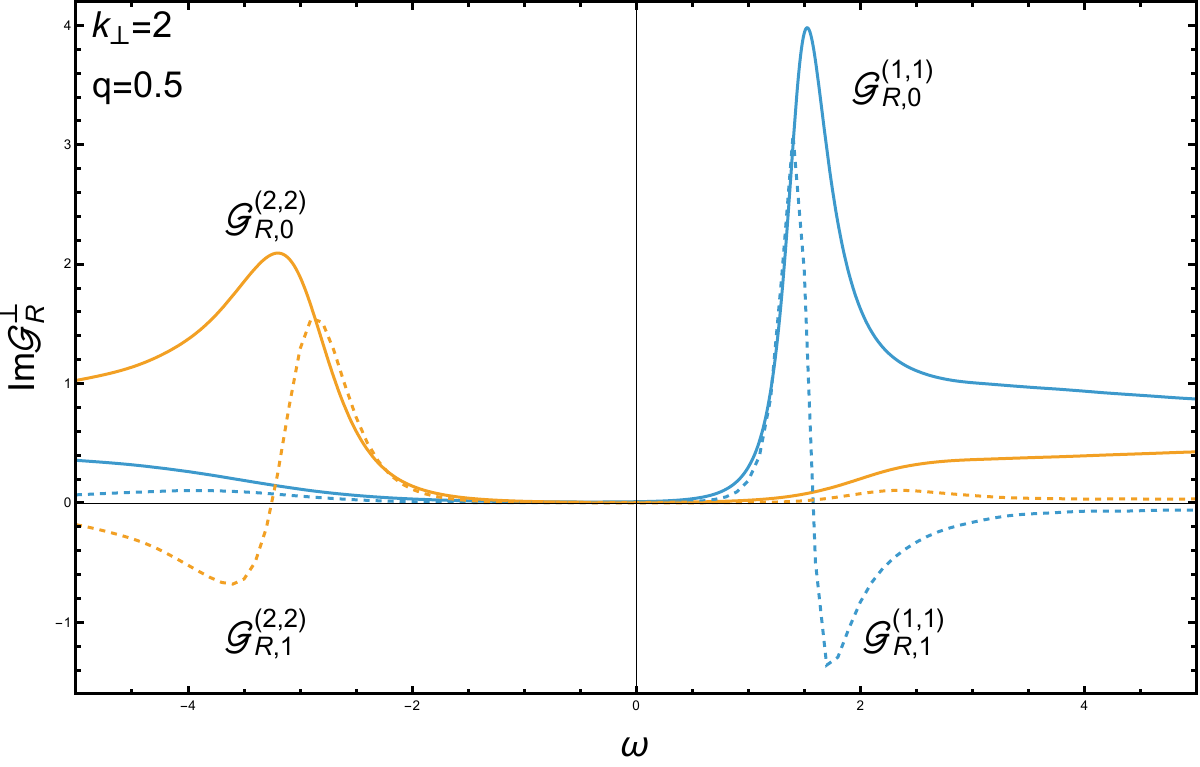}\includegraphics[scale=0.39]{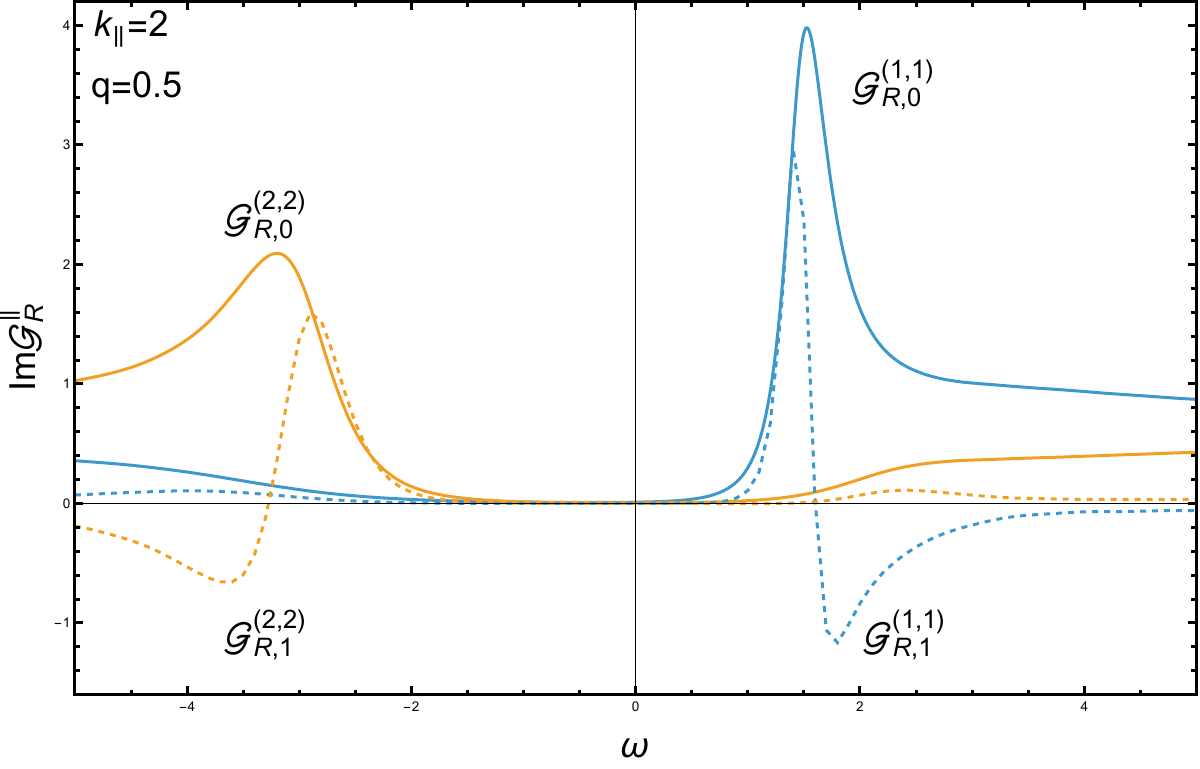} 
\par\end{centering}
\begin{centering}
\includegraphics[scale=0.39]{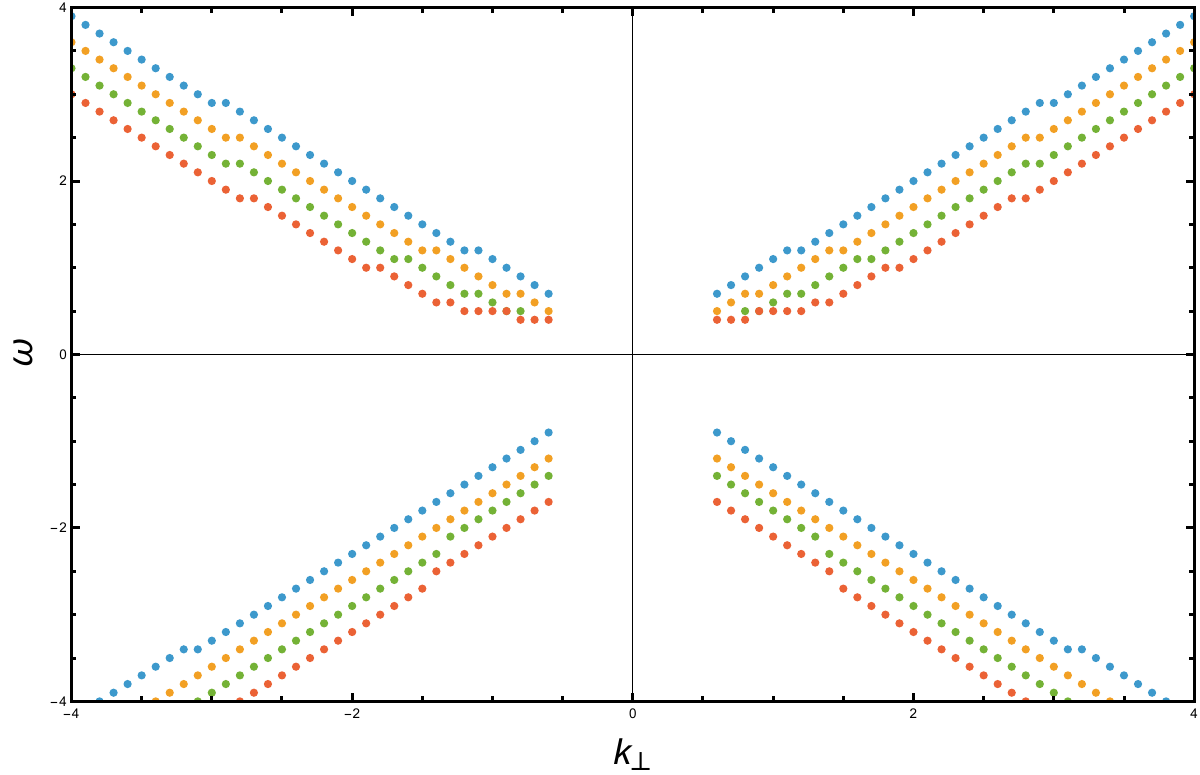}\includegraphics[scale=0.39]{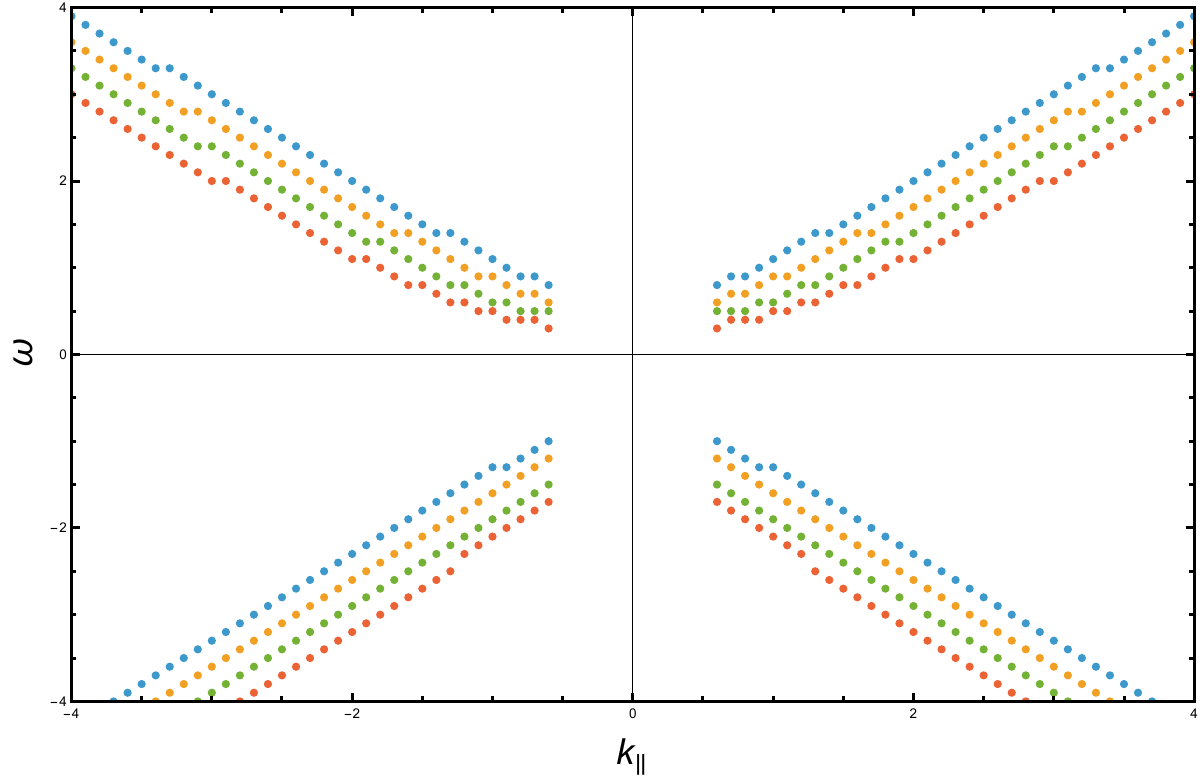} 
\par\end{centering}
\caption{The perturbative fermionic dispersion curve with axion induced anisotropy
and chemical potential. The parameters are set as $u_{H}=g_{f}=L=1,m=0.01$.
\textbf{Upper: }The zeroth Green function and its leading order corrections.
\textbf{Lower:} the dispersion curve from the leading order perturbations,
the colors blue, orange, green and red refer respectively to $q=0.1,0.3,0.5,0.7$.}
\label{fig:8} 
\end{figure}

Let us turn to the case with non-vanishing chemical potential, i.e.
$A_{0}\neq0$ and the associated perturbative metric can also be found
in the appendix. For the isotropic situation, i.e. $a=0$, the metric
with non-vanishing chemical potential taking the form in (\ref{eq:A-7})
is not exactly the metric of the AdS-Schwarzschild black brane background
(\ref{eq:2.15}), thus we have to evaluate the zeroth-order Green
function by using the metric (\ref{eq:A-7}). Recall the decomposition
(\ref{eq:4.6}) of the Green function, the zeroth-order Green function
can be computed by using (\ref{eq:2.31}) with the in-falling boundary
condition with respect to metric (\ref{eq:A-7}), so the corresponding
numerical results are given in Figure \ref{fig:5}. Note that the
parameter $q$ describes the electric charge of the black brane in
the gravity side. Due to the presence of the chemical potential, our
numerical results illustrate that the Green functions do not satisfy
$\mathcal{G}^{\left(1,1\right)*}_{R}\left(\omega,\mathrm{k}\right)=-\mathcal{G}^{\left(2,2\right)}_{R}\left(-\omega,\mathrm{k}\right)$
while the relation $\mathcal{G}^{\left(1,1\right)}_{R}\left(\omega,\mathrm{k}\right)=\mathcal{G}^{\left(2,2\right)}_{R}\left(\omega,-\mathrm{k}\right)$
remains, so the Green functions are not symmetric under $\omega\rightarrow-\omega$.
This is expected since the presence of the chemical potential breaks
down the spectral symmetry of the ``particle'' and ``anti-particle''
\cite{MLeBellac,Klimov:1981ka,Haque:2024gva}. To confirm this, we
further plot out the fermionic dispersion curve from the isotropic
Green function with chemical potential as it is displayed in Figure
\ref{fig:6}, and the numerical results indicate indeed that only
one branch of the Green functions dominates.

Afterwards, let us focus on the leading order corrections with respect
to the zeroth-order Green function. Using the perturbative method
given in Section 3.3 with respect to the functions given in (\ref{eq:A-3})
and (\ref{eq:A-4}), the leading order corrections can be numerically
evaluated with the in-falling boundary condition, and they are given
in Figure \ref{fig:7}. Accordingly, the numerical calculation illustrates
the leading order corrections remain to be unsymmetric for $\omega\rightarrow-\omega$
in the presence of the chemical potential. And it leads to the unsymmetric
fermionic spectra from the dispersion curve, which has been plotted
numerically out in Figure \ref{fig:8}. These results also illustrate
the leading order corrections dependence on the momentum direction,
and enhance the peaks presented in the zeroth order solution while,
in the mean time, they lead to a dip in the corrections to the spectral
function.

\section{The anisotropy induced by magnetic field}

In this section, we consider another holographic model in which the
anisotropy is induced by the magnetic field, then analyze the fermionic
correlation functions in this holographic system.

\subsection{The model}

Our concerned model is based on the D3-brane system which includes
a magnetic flux in the dimensional reduction. Briefly, the five-dimensional
effective gravity action for this model is given in \cite{DHoker:2009ixq,DHoker:2009mmn,Dudal:2015wfn,Zhao:2024ipr},

\begin{equation}
S_{\mathrm{5D}}=\frac{1}{2\kappa^{2}}\int d^{5}x\sqrt{-g}\left[\mathcal{R}_{\left(\mathrm{5D}\right)}-\frac{2\Lambda}{L^{2}}-\frac{1}{4}F_{MN}F^{MN}\right]+S_{\mathrm{bdry}},\label{eq:5.1}
\end{equation}
where, as before, $\Lambda=-6$ is the cosmological constant, $L$
refers to the radius of the bulk, $F_{MN}=\partial_{M}A_{N}-\partial_{N}A_{M}$
is the $U\left(1\right)$ gauge field strength, $\kappa$ is the five-dimensional
gravity coupling constant and $S_{\mathrm{bdry}}$ is the gravity
boundary terms. As we can see, this action is in fact a simplified
version of action (\ref{eq:4.2}) by setting $\phi,C_{0}=0$ with
a distinct ansatz of the $U\left(1\right)$ gauge field $A$. Therefore,
a static solution asymptotically to the AdS-Schwarzschild black brane
background with anisotropy induced by the magnetic field can be found
by using the following ansatz 
\begin{align}
ds^{2} & =\frac{L^{2}}{u^{2}}\left\{ -f_{T}dt^{2}+h_{T}\left[\left(dx^{1}\right)^{2}+\left(dx^{2}\right)^{2}\right]+q_{T}\left(dx^{3}\right)^{2}+\frac{du^{2}}{f_{T}\left(u\right)}\right\} ,\nonumber \\
A & =A_{1}dx^{1},A_{1}=-LBx^{2},\label{eq:5.2}
\end{align}
where $B$ is a constant magnetic field. By imposing the ansatz (\ref{eq:5.2})
into the equations of motion obtained from the action (\ref{eq:5.1})
and taking into account the high temperature limit i.e. $u^{2}_{H}B\ll1$
(or equivalently $T^{2}\gg B$), the functions presented in (\ref{eq:5.2})
can be solved perturbatively as \cite{DHoker:2009ixq,Dudal:2015wfn,Zhao:2024ipr},
\begin{align}
g_{tt} & \mathrel{=}{-\frac{L^{2}}{u^{2}}f_{T},g_{xx}=\frac{L^{2}}{u^{2}}h_{T},g_{zz}=\frac{L^{2}}{u^{2}}q_{T},g_{uu}=\frac{L^{2}}{u^{2}f_{T}}}\mathpunct{,}\nonumber \\
f_{T}\mathinner{\left(u\right)} & \mathrel{=}{1-\frac{u^{4}}{u^{4}_{H}}+\frac{2}{3}B^{2}u^{4}\ln\left(\frac{u}{u_{H}}\right)+\mathcal{O}\left(u^{8}_{H}B^{4}\right)}\mathpunct{,}\nonumber \\
q_{T}\mathinner{\left(u\right)} & \mathrel{=}{1+\frac{8}{3}B^{2}u^{4}_{H}\int^{u_{H}/u}_{+\infty}dy\frac{\ln y}{y\left(y^{4}-1\right)}+\mathcal{O}\left(u^{8}_{H}B^{4}\right)}\mathpunct{,}\nonumber \\
h_{T}\mathinner{\left(u\right)} & \mathrel{=}{1-\frac{4}{3}B^{2}u^{4}_{H}\int^{u_{H}/u}_{+\infty}dy\frac{\ln y}{y\left(y^{4}-1\right)}+\mathcal{O}\left(u^{8}_{H}B^{4}\right).}\label{eq:5.3}
\end{align}
In this solution, the horizon and holographic boundary are located
at $u=u_{H}$ and $u=0$ respectively. The Hawking temperature $T$
in this model is given by

\begin{equation}
T=-\frac{1}{4\pi}\frac{\partial f_{T}}{\partial u}\big|_{u=u_{H}}=\frac{1}{4\pi}\left[\frac{4}{u_{H}}-\mathinner{\frac{2}{3}}{B^{2}u^{3}_{H}}\right]+\mathcal{O}\left(B^{4}\right).\label{eq:5.4}
\end{equation}
In this section, we will focus on the perturbative solution given
in (\ref{eq:5.3}), i.e. in the high temperature limit, which would
be closer to the situations in the realistic hot plasma like QGP.

\subsection{The correlation function with magnetic field}

As the general form of the Dirac action with gauge field is given
in (\ref{eq:3.2}), it leads to\footnote{In our notation, the gauge field $A_{M}$ is dimensionless according
to (\ref{eq:5.2}) while the 5D coupling constant $g_{f}$ takes the
dimension of energy.}, 
\begin{align}
\tilde{S}_{\mathrm{bulk}}= & i\int dx^{2}du\bar{\chi}\bigg\{\sqrt{\frac{g_{xx}}{g_{uu}}}\left(\gamma\partial_{u}-m\sqrt{g_{uu}}\right)\nonumber \\
 & +i\left[-\gamma^{0}\omega\sqrt{\frac{g_{xx}}{-g_{tt}}}+\gamma^{1}\left(-i\partial_{1}+bx^{2}\right)+\gamma^{2}\left(-i\partial_{2}\right)+\gamma^{3}\sqrt{\frac{g_{xx}}{g_{zz}}}\left(-i\partial_{3}\right)\right]\bigg\}\chi,\label{eq:5.5}
\end{align}
where $b=g_{f}LB$ and we have inserted the gauge field given in (\ref{eq:5.2}).
Since the Hamiltonian associated to (\ref{eq:5.5}) includes $x^{2}$,
$-i\partial_{2}$ is not conserved i.e. it does not commutate with
the Hamiltonian. So the ansatz for $\psi$ given in (\ref{eq:2.9})
has to be changed as,

\begin{equation}
\psi=\left(\begin{array}{c}
\psi_{R}\\
\psi_{L}
\end{array}\right)=\left(-gg^{-1}_{uu}\right)^{-1/4}\int\frac{d^{4}k}{\left(2\pi\right)^{4}}e^{-i\omega t+ik_{1}x^{1}+ik_{3}x^{3}}\chi\left(u,k,x^{2}\right),\label{eq:5.6}
\end{equation}
where $\chi\left(u,k,x^{2}\right)$ is the Dirac spinor. By inserting
(\ref{eq:5.6}) into (\ref{eq:5.5}), the Dirac action becomes,

\begin{align}
\tilde{S}_{\mathrm{bulk}}= & i\int dx^{2}du\bar{\chi}\bigg\{\sqrt{\frac{g_{xx}}{g_{uu}}}\left(\gamma\partial_{u}-m\sqrt{g_{uu}}\right)\nonumber \\
 & +i\left[-\gamma^{0}\omega\sqrt{\frac{g_{xx}}{-g_{tt}}}+\gamma^{1}\left(k_{1}+bx^{2}\right)+\gamma^{2}\left(-i\partial_{2}\right)+\gamma^{3}\sqrt{\frac{g_{xx}}{g_{zz}}}k_{3}\right]\bigg\}\chi\nonumber \\
\equiv & i\int dx^{2}du\bar{\chi}\left[\sqrt{\frac{g_{xx}}{g_{uu}}}\left(\gamma\partial_{u}-m\sqrt{g_{uu}}\right)+i\gamma^{\mu}\hat{K}_{\mu}\right]\chi.\label{eq:5.7}
\end{align}
where

\begin{equation}
\hat{K}_{\mu}=\left\{ -\omega\sqrt{\frac{g_{xx}}{-g_{tt}}},k_{1}+bx^{2},-i\partial_{2},\sqrt{\frac{g_{xx}}{g_{zz}}}k_{3}\right\} .
\end{equation}
Physically, magnetic field leads to Landau levels $E_{n}$ as the
eigenvalues of the harmonic oscillator, thus we impose the condition
for Landau levels in (\ref{eq:5.7}) as, 
\begin{equation}
\left(-i\gamma^{2}\partial_{2}+\gamma^{1}bx^{2}+\gamma^{1}k_{1}\right)\chi=E_{n}\gamma^{0}\chi.\label{eq:5.9}
\end{equation}
Therefore, the effective Lagrangian (\ref{eq:5.7}) depends on the
Landau levels $E_{n}$ instead of $x^{2}$. For $b>0$, by further
defining the creation/annihilation operators for the harmonic oscillator
as,

\begin{equation}
\hat{a}\equiv\frac{1}{\sqrt{2b}}\left(by+i\hat{k}_{y}\right)=\frac{1}{\sqrt{2b}}\left(by+\partial_{y}\right),\hat{a}^{\dagger}\equiv\frac{1}{\sqrt{2b}}\left(by-i\hat{k}_{y}\right)=\frac{1}{\sqrt{2b}}\left(by-\partial_{y}\right),
\end{equation}
with 
\begin{equation}
\chi=\left(\begin{array}{c}
\chi_{R}\\
\chi_{L}
\end{array}\right),\chi_{R,L}=\left(\begin{array}{c}
\chi^{\left(1\right)}_{R,L}\\
\chi^{\left(2\right)}_{R,L}
\end{array}\right),y=x^{2}+\frac{k_{1}}{b},
\end{equation}
(\ref{eq:5.9}) becomes

\begin{align}
\sqrt{2b}\hat{a}^{\dagger}\chi^{\left(2\right)}_{L}=-E_{n}\chi^{\left(1\right)}_{L},\  & \sqrt{2b}\hat{a}\chi^{\left(1\right)}_{L}=-E_{n}\chi^{\left(2\right)}_{L},\nonumber \\
\sqrt{2b}\hat{a}^{\dagger}\chi^{\left(2\right)}_{R}=E_{n}\chi^{\left(1\right)}_{R},\  & \sqrt{2b}\hat{a}\chi^{\left(1\right)}_{R}=E_{n}\chi^{\left(2\right)}_{R},\label{eq:5.12}
\end{align}
where $\left[\hat{a},\hat{a}^{\dagger}\right]=1$. This implies the
solution for $\chi_{R,L}$ can be obtained by the combination of the
harmonic oscillator eigenstates $\phi_{n}\left(y,u,k_{1}\right)$
as,

\begin{equation}
\chi_{R}\sim\left(\begin{array}{c}
\phi_{n}\\
\phi_{n-1}
\end{array}\right),\chi_{L}\sim\left(\begin{array}{c}
\phi_{n}\\
-\phi_{n-1}
\end{array}\right),
\end{equation}
where $\phi_{n}$ satisfies

\begin{equation}
\hat{a}^{\dagger}\phi_{n}=\sqrt{\left(n+1\right)}\phi_{n+1},\ \hat{a}\phi_{n}=\sqrt{n}\phi_{n-1}.
\end{equation}
Accordingly, it implies the eigenvalues $E_{n}$, as the Landau levels,
take the form as $E_{n}=\sqrt{2nb}$. The solution for $E_{n}=-\sqrt{2nb}$
also exists and can be obtained by interchanging the rols of $\chi_{R},\chi_{L}$.
Note that, (\ref{eq:5.12}) also leads to the following second order
differential equations,

\begin{equation}
2b\hat{a}^{\dagger}\hat{a}\chi^{\left(1\right)}_{R,L}=E^{2}_{n}\chi^{\left(1\right)}_{R,L},2b\hat{a}\hat{a}^{\dagger}\chi^{\left(2\right)}_{R,L}=E^{2}_{n}\chi^{\left(2\right)}_{R,L},
\end{equation}
which also implies the eigenvalues $E_{n}$ should take the form $E_{n}=\pm\sqrt{2nb}$
as the eigenvalue of harmonic oscillator. Although we have discussed
the case of $b>0$ above, the case of $b<0$ can also be treated in
the same way, simply by interchanging the definition of the creation/annihilation
operators $\hat{a},\hat{a}^{\dagger}$, and simultaneously interchanging
roles of $\phi_{n},\phi_{n-1}$. Altogether, there must be solutions
for $\chi_{R},\chi_{L}$ with Landau levels $E_{n}=\pm\sqrt{2n\left|b\right|}$.

Afterwards, the action (\ref{eq:5.7}) with the condition (\ref{eq:5.9})
for the Landau levels becomes 
\begin{align}
\tilde{S}_{\mathrm{bulk}}= & i\int dx^{2}du\bar{\chi}\bigg\{\sqrt{\frac{g_{xx}}{g_{uu}}}\left(\gamma\partial_{u}-m\sqrt{g_{uu}}\right)\nonumber \\
 & +i\left[-\gamma^{0}\omega\sqrt{\frac{g_{xx}}{-g_{tt}}}+E_{n}\gamma^{0}+\gamma^{3}\sqrt{\frac{g_{xx}}{g_{zz}}}k_{3}\right]\bigg\}\chi,\nonumber \\
\equiv & i\int dx^{2}du\bar{\chi}\left[\sqrt{\frac{g_{xx}}{g_{uu}}}\left(\gamma\partial_{u}-m\sqrt{g_{uu}}\right)+i\gamma^{\mu}K^{B}_{\mu}\right]\chi,\label{eq:5.16}
\end{align}
where 
\begin{equation}
K^{B}_{\mu}=\left\{ -\omega\sqrt{\frac{g_{xx}}{-g_{tt}}}+E_{n},0,0,\sqrt{\frac{g_{xx}}{g_{zz}}}k_{3}\right\} .
\end{equation}
Therefore, the ratio defined in (\ref{eq:2.29}) satisfies the following
differential equation, 
\begin{align}
\sqrt{\frac{g_{xx}}{g_{uu}}}\xi^{\prime}_{\left(1\right)} & =\left(K^{B}_{0}-K^{B}_{3}\right)+\left(K^{B}_{0}+K^{B}_{3}\right)\xi^{2}_{1}+2m\sqrt{g_{xx}}\xi_{\left(1\right)},\nonumber \\
\sqrt{\frac{g_{xx}}{g_{uu}}}\xi^{\prime}_{\left(2\right)} & =-K^{B}_{0}-K^{B}_{3}+\left[K^{B}_{3}-K^{B}_{0}\right]\xi^{2}_{\left(2\right)}+2m\sqrt{g_{xx}}\xi_{\left(2\right)},\label{eq:5.18}
\end{align}
which follows the same derivation as it is demonstrated in Section
2.2. The equations in (\ref{eq:5.18}) contain terms proportional
to $u_{H}B^{1/2}$ and $u^{4}_{H}B^{2}$ due to the metric given in
(\ref{eq:5.3}). Therefore, in a perturbative treatment, the zeroth-order
Green function should be obtained by keeping only the terms up to
order $u_{H}B^{1/2}$, while its leading-order corrections follow
from further including the terms up to order $u^{4}_{H}B^{2}$. When
only the terms up to order $u_{H}B^{1/2}$ are retained in (\ref{eq:5.18}),
the metric reduces to the isotropic case. Hence, the zeroth-order
Green function at this order is obtained by solving (\ref{eq:5.18})
on the isotropic AdS-Schwarzschild black brane background. Besides,
although the Landau levels take the form $E_{n}=\pm\sqrt{2n\left|b\right|}$
indicating there are two branches, we consider the case of $E_{n}=\sqrt{2n\left|b\right|}$
in our actual numerical calculations with $\mathrm{k}=k_{3}$. Because
the Green functions $G_{R}\left(\omega,\mathrm{k},E_{n}\right)$ with
respect to the other branches can be obtained via the symmetries in
(\ref{eq:5.18}) leading to $G_{R}\left(\omega,\mathrm{k},\sqrt{2n\left|b\right|}\right)=G_{R}\left(-\omega,\mathrm{k},-\sqrt{2n\left|b\right|}\right)$
according to (\ref{eq:5.18}).

\subsection{Numerical results}

Since the perturbative gravity solution (\ref{eq:5.3}) is valid in
high temperature limit i.e $u^{2}_{H}B\ll1$ (or equivalently $T^{2}\gg B$)
according to \cite{DHoker:2009ixq,Dudal:2015wfn,Zhao:2024ipr}, our
numerical analysis in this section is performed for that high\nobreakdash-temperature
case. By keeping the above analysis in hand, we can set the perturbative
parameter in (\ref{eq:3.12}) as $\epsilon\rightarrow u^{4}_{H}B^{2}$,
thus the Green function can be written as, 
\begin{figure}[t]
\begin{centering}
\includegraphics[scale=0.39]{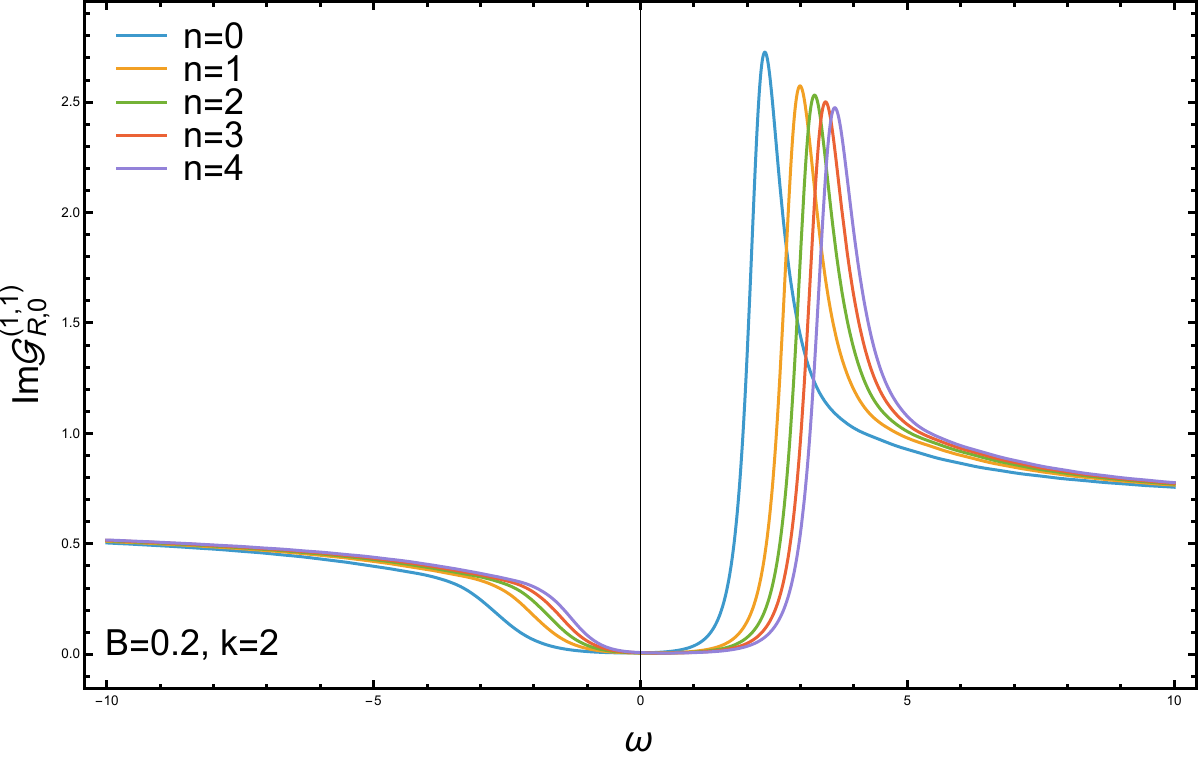}\includegraphics[scale=0.39]{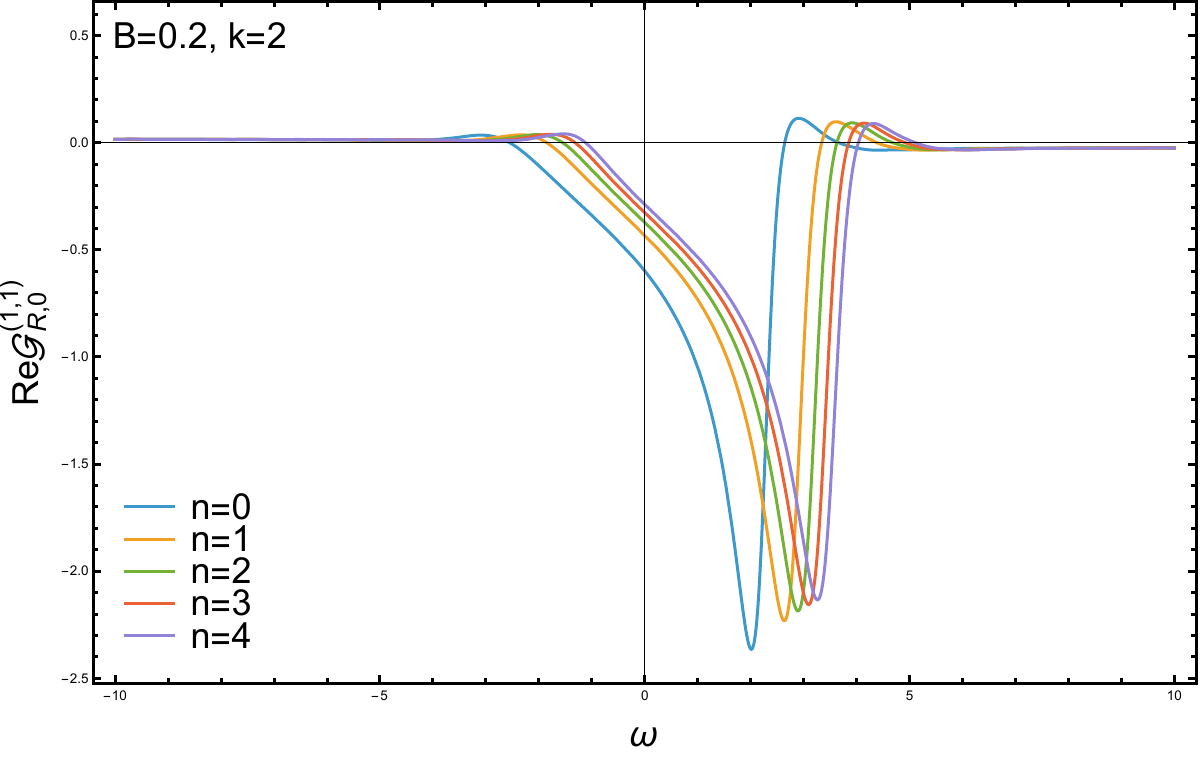} 
\par\end{centering}
\begin{centering}
\includegraphics[scale=0.39]{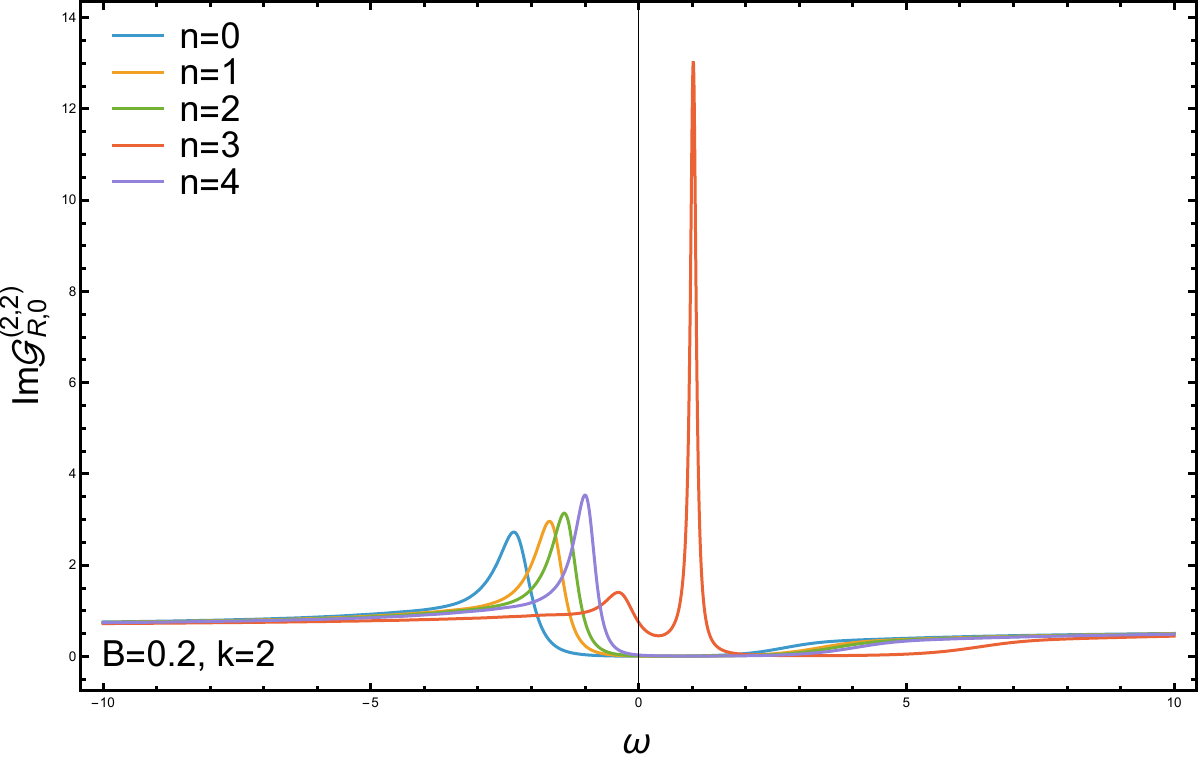}\includegraphics[scale=0.39]{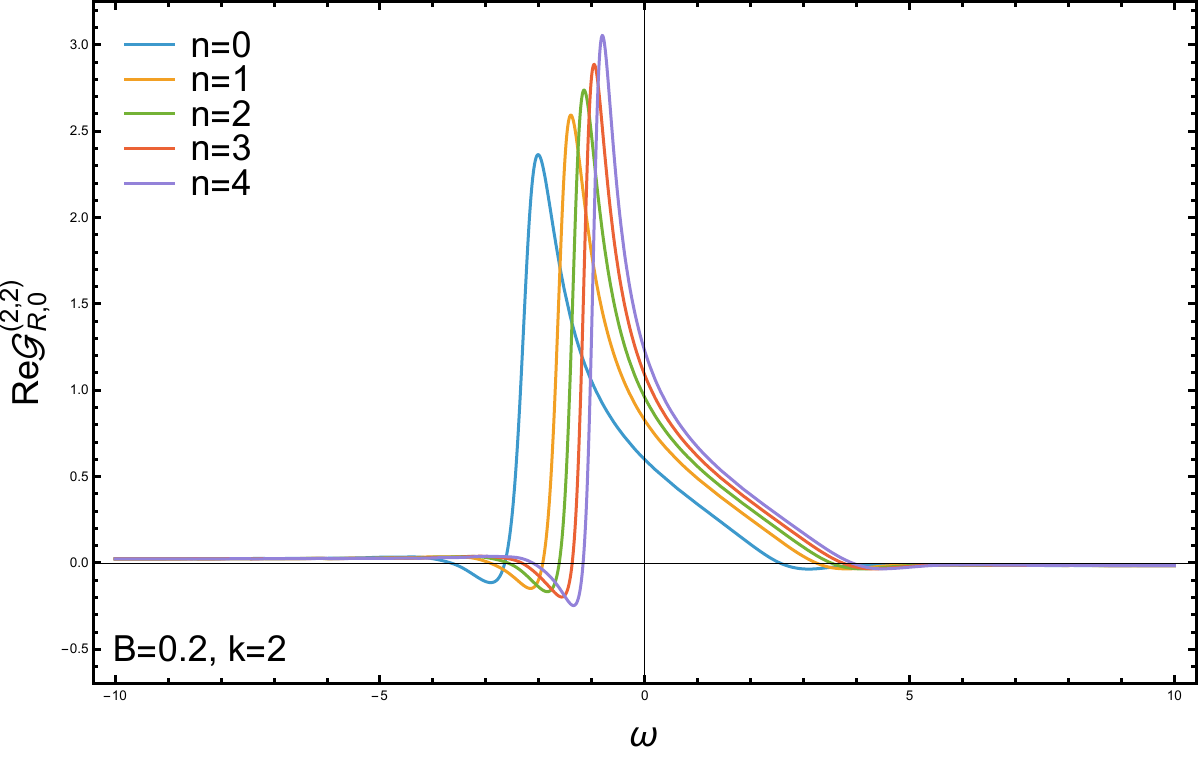} 
\par\end{centering}
\caption{The zeroth-order fermionic Green function with magnetic field. The
parameters are chosen as $u_{H}=g_{f}=L=1,m=0.01,\mathrm{k}=k_{3}$.
There could be two peaks, a dominant high peak and a secondary lower
peak, in the imaginary part of the Green functions with high Landau
levels. Here, the magnetic-field parameter shown in the plots denotes
the combination $b$. Since $g_{f}=L=1$, its numerical value coincides
with the value labelled as $B$.}
\label{fig:9} 
\end{figure}

\begin{figure}[t]
\begin{centering}
\includegraphics[scale=0.25]{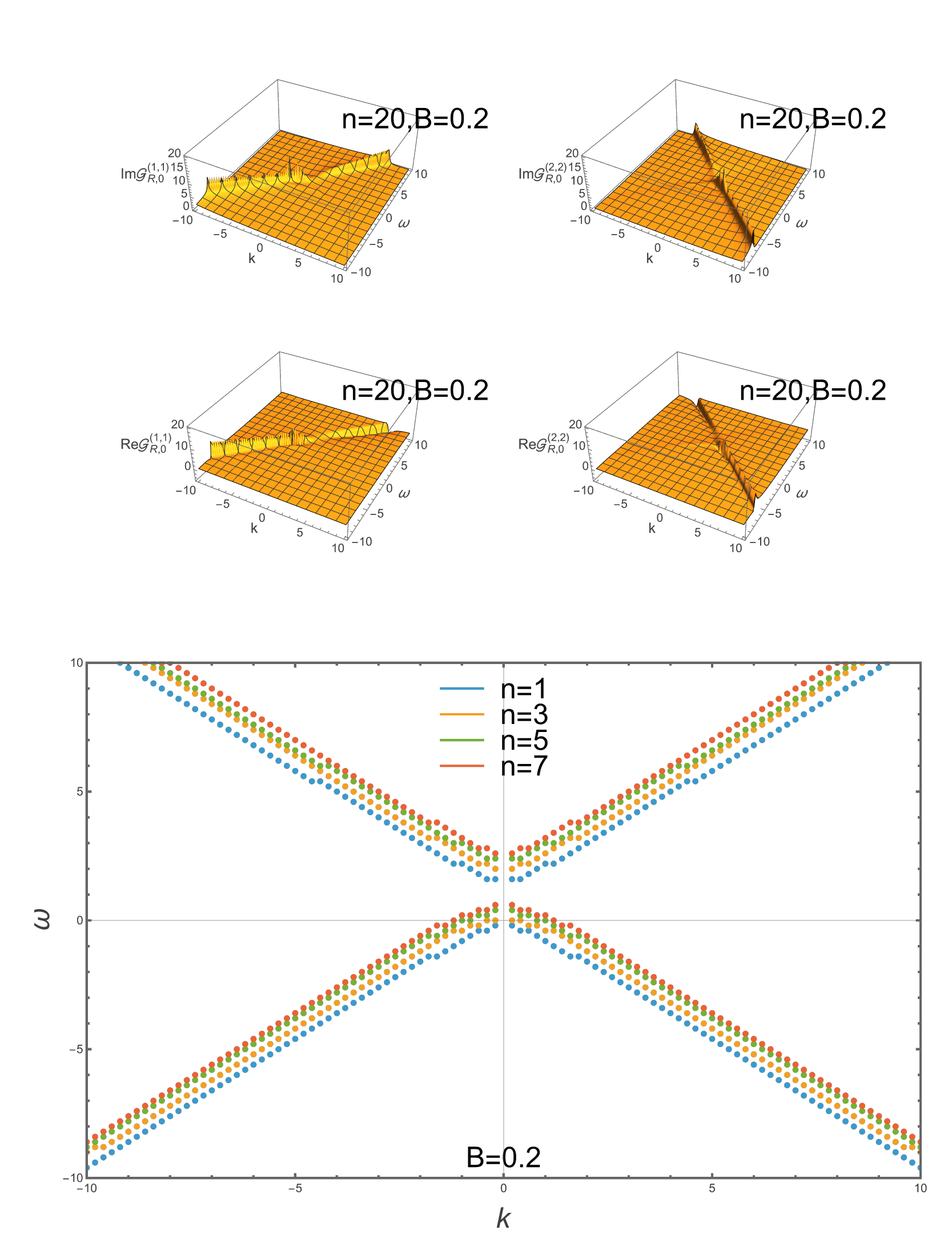}\includegraphics[scale=0.32]{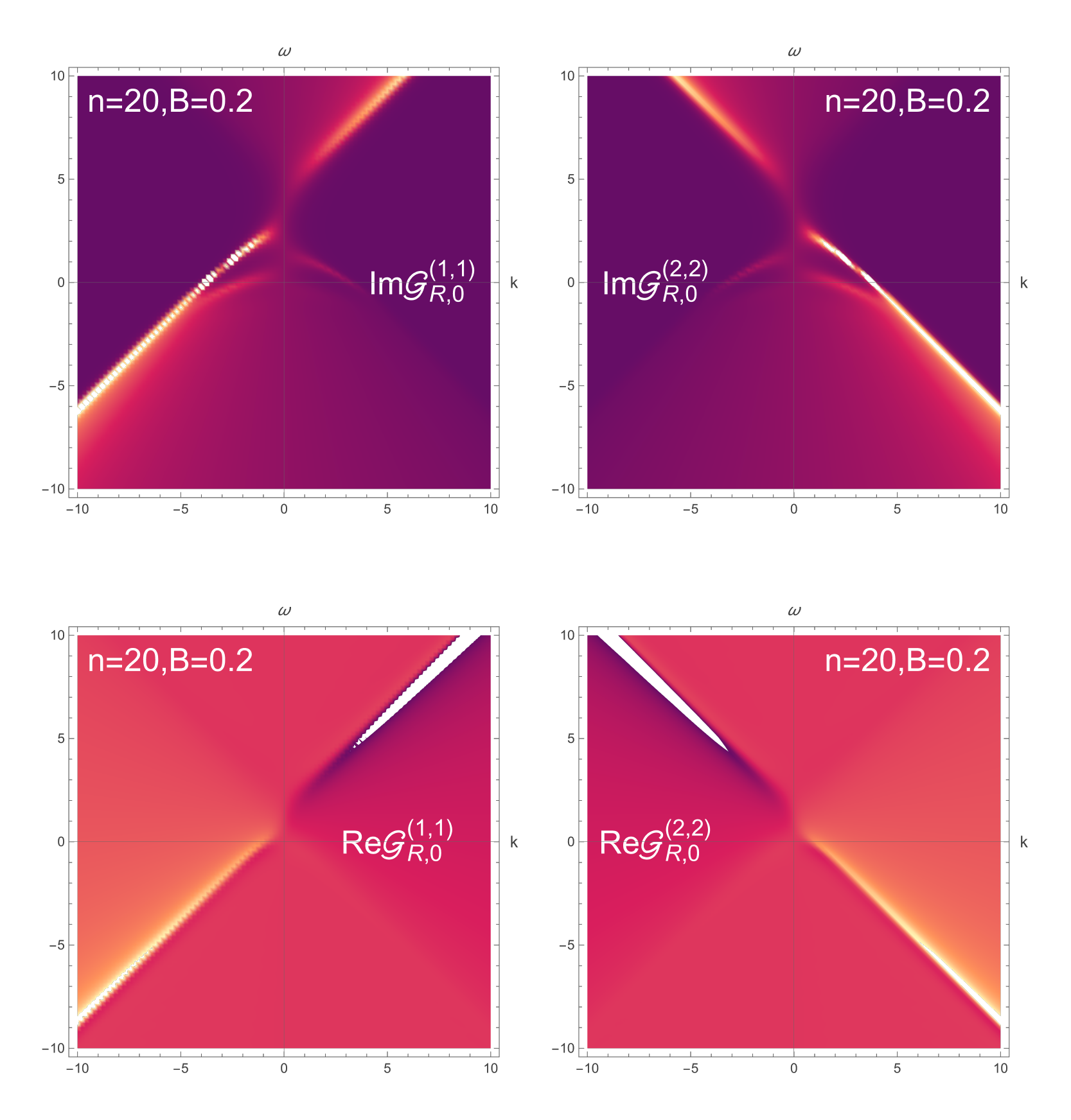} 
\par\end{centering}
\caption{The zeroth-order fermionic Green function with magnetic field and
the associated dispersion curves. The parameters are set as $u_{H}=g_{f}=L=1,m=0.01,B=0.2,\mathrm{k}=k_{3}$.
\textbf{UpperLeft:} The 3D plot of the Green function with magnetic
field. The ridge in the imaginary part of the Green function develops
into two infinitely sharp peaks. \textbf{LowerLeft:} The fermionic
dispersion curves with various Landau levels obtained from the dominant
high peaks. These dispersion curves can be obtained by choosing equivalently
the parameters as $n=1$ with $B=0.2,0.4,0.6,0.8$. \textbf{Right:
}The density plots of the fermionic Green functions, where the white
region refers to the peaks or dips. Here, the magnetic-field parameter
shown in the plots denotes the combination $g_{f}BL$. Since $g_{f}=L=1$,
its numerical value coincides with the value labelled as $B$.}
\label{fig:10} 
\end{figure}

\begin{equation}
\mathcal{G}^{\left(\alpha,\alpha\right)}_{R}=\mathcal{G}^{\left(\alpha,\alpha\right)}_{R,0}+u^{4}_{H}B^{2}\mathcal{G}^{\left(\alpha,\alpha\right)}_{R,1}=\left(-1\right)^{\alpha}\lim_{u\rightarrow0}u^{-2mL}\left[\Lambda_{\left(\alpha\right)}+u^{4}_{H}B^{2}\lambda_{\left(\alpha\right)}\right],\label{eq:5.19}
\end{equation}
where $\mathcal{G}^{\left(\alpha,\alpha\right)}_{R,0}$ is the zeroth
order Green function including order $u_{H}B^{1/2}$. Then we solve
numerically the equations in (\ref{eq:5.18}) with the in-falling
boundary condition by ruling out the terms of $\mathcal{O}\left(B^{2}\right)$
to obtain the zeroth-order Green functions, and the relevant results
are illustrated in Figure \ref{fig:9} and Figure \ref{fig:10}. Note
that, for the zeroth-order Green functions, when the physical momentum
is totally parallel to the magnetic field, it means the fermion is
fixed on the lowest Landau Level ($n=0$). In this case, the zeroth-order
Green functions return to the results obtained on the AdS-Schwarzschild
black brane background as they are given in Section 2.2. Therefore,
in this section, we focus on the case with vertical physical momentum
$k_{\perp}\neq0$ (i.e. $E_{n}\neq0$) for the zeroth-order Green
functions. 
\begin{figure}[t]
\begin{centering}
\includegraphics[scale=0.28]{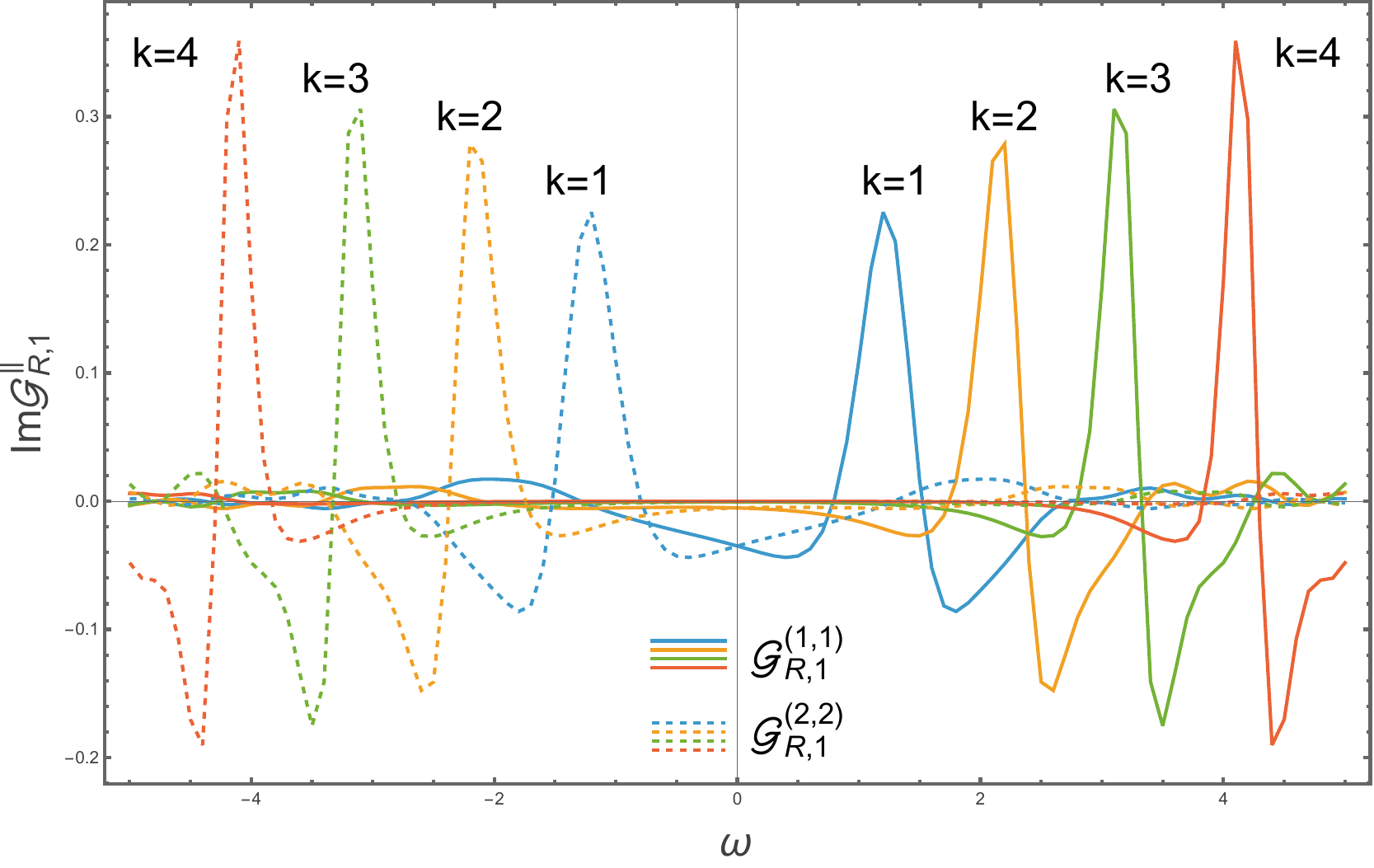}\includegraphics[scale=0.28]{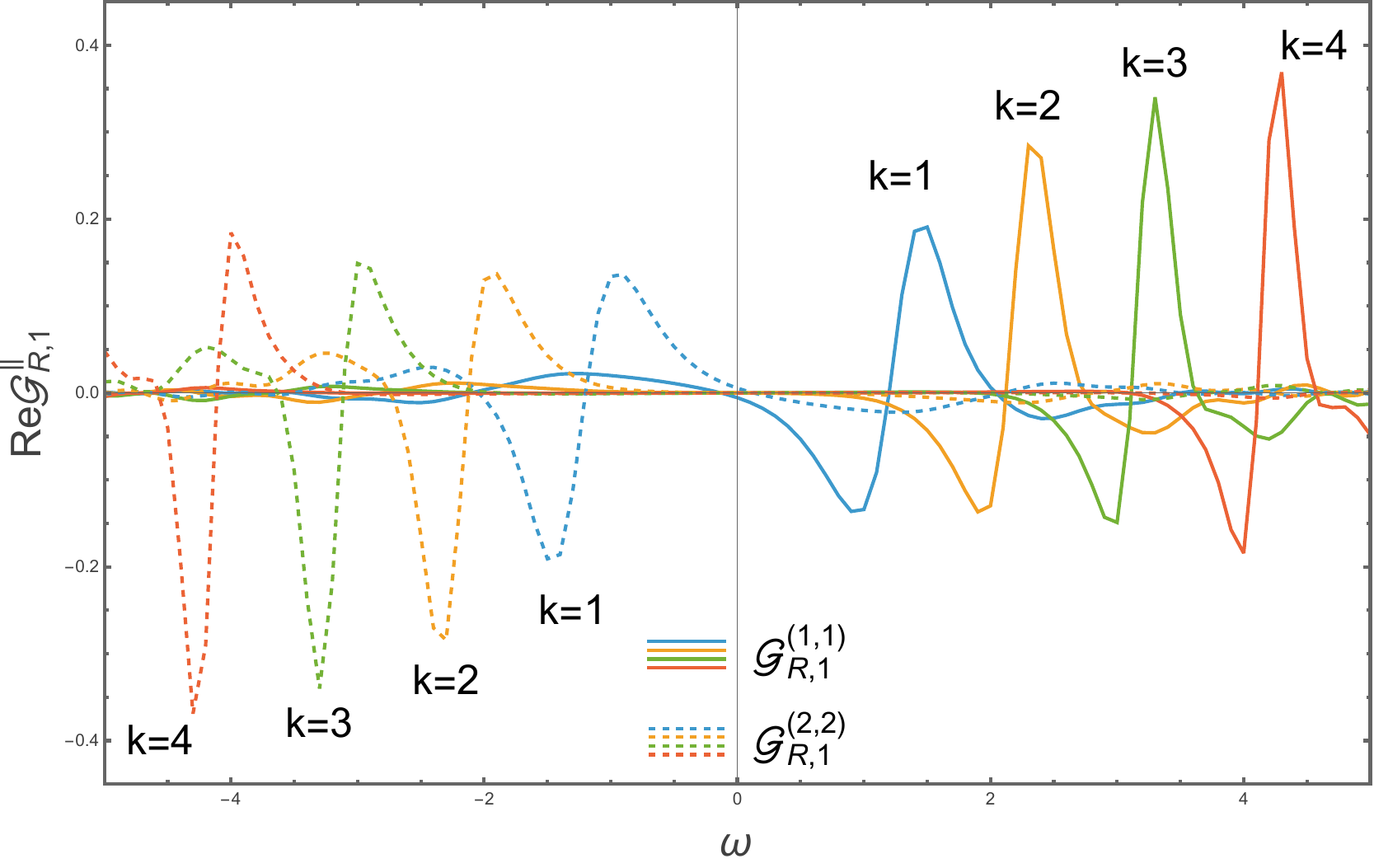} 
\par\end{centering}
\begin{centering}
\includegraphics[scale=0.36]{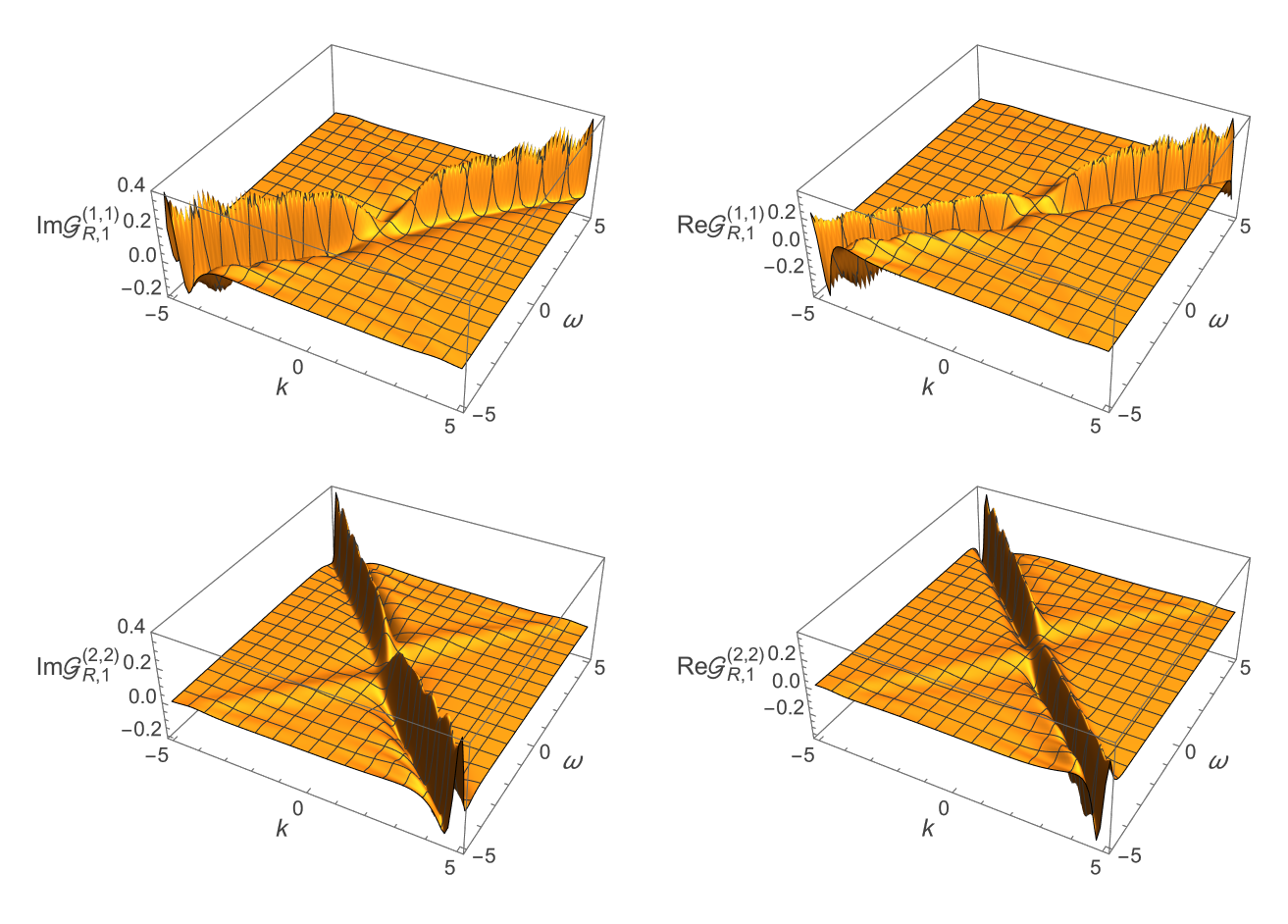}\includegraphics[scale=0.28]{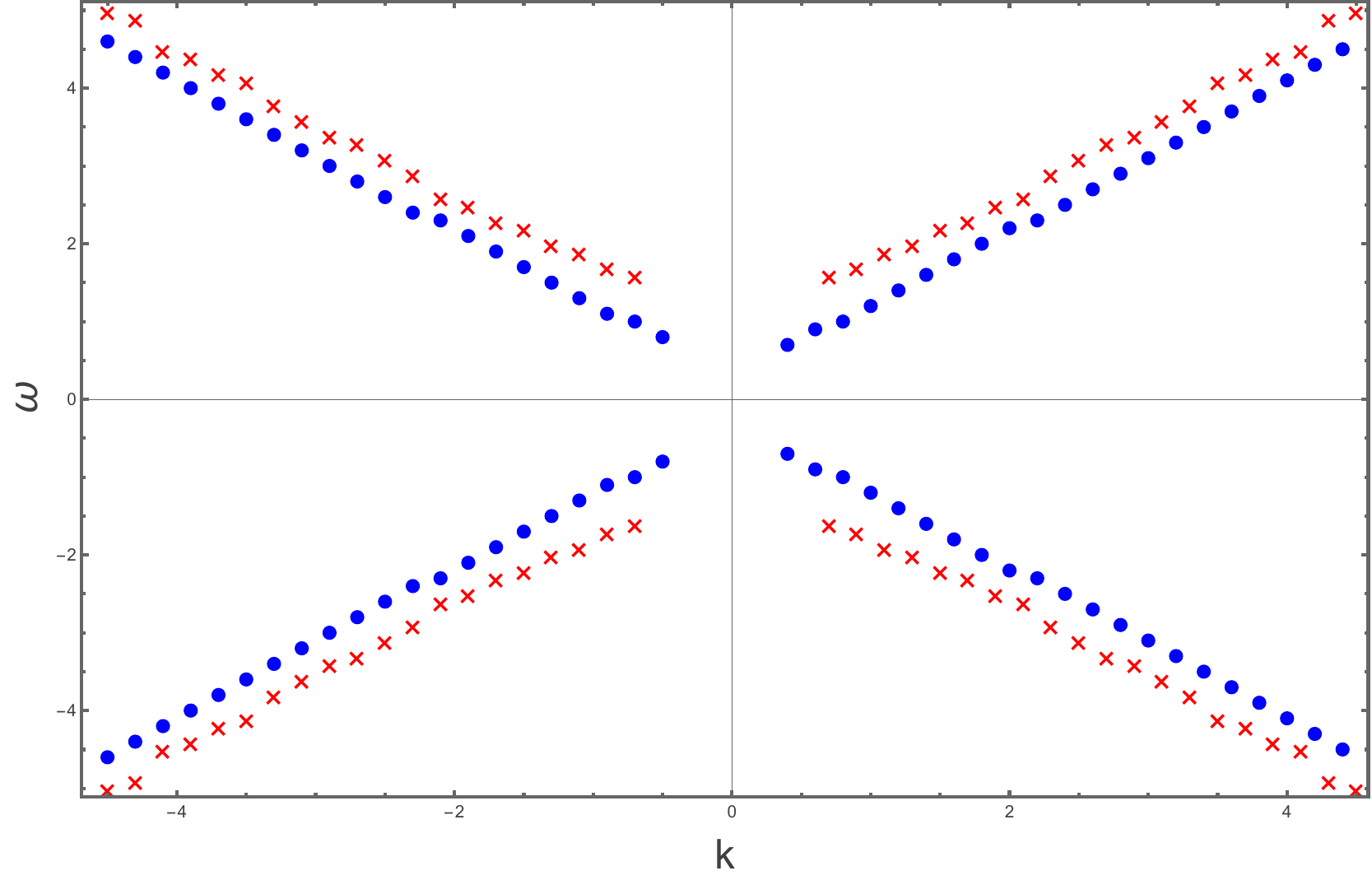} 
\par\end{centering}
\caption{Anisotropic perturbations $\mathcal{G}^{\left(\alpha,\alpha\right)}_{R,1}$
induced by magnetic field on the fermionic correlation functions.
The parameters are set as $u_{H}=L=1,m=0.01,\mathrm{k}=k_{3}$. \textbf{Upper:}
The imaginary and real part of the perturbative Green function with
fixed $\mathrm{k}$. \textbf{LowerLeft:} 3D plot of perturbative Green
function. \textbf{LowerRight:} The position of the peaks (blue dot)
and the negative dips (red cross) of the perturbative Green functions
on the $\omega,\mathrm{k}$ plane.}
\label{fig:11} 
\end{figure}

Our numerical calculations for the zeroth-order Green function reveal
the Landau levels as discrete peaks in the imaginary part, as the
spectral functions, in Figure \ref{fig:9}. Remarkably, as Dirac fermions
have two spin eigenvalues, this may explain why the spectral function
exhibits two peaks, especially with high Landau levels. And the dominant
high peak corresponds to the primary fermionic quasiparticle excitation,
significantly shifted by the Landau energy. Moreover, there are two
sharp peaks on the ridges in the imaginary part of the Green functions,
see Figure \ref{fig:10}, which may refer to the long-lived quasiparticle
resulting from the external magnetic field lifting Dirac fermion degeneracy.\textcolor{red}{{}
}To further clarify the dispersion relation, we numerically plot the
fermionic dispersion curves obtained from the dominant high peaks
in the zeroth-order Green function under the magnetic field, as a
function of Landau level number. The holographic dispersion curve
indicates that the effective mass ($\omega\big|_{k=0}$) increases
with the magnetic field. This result agrees qualitatively with the
fermionic dispersion obtained from the hard thermal loop method \cite{MLeBellac,Andersen:2014xxa}
in which effective mass for fermions is $m_{f}\propto\sqrt{T^{2}+L^{2}B^{2}}$
in the high temperature limit $T^{2}\gg B$ and without chemical potential.
However, this holographic approach does not show exactly the anisotropy
in the background since the metric without terms of $\mathcal{O}\left(u^{4}_{H}B^{2}\right)$
is isotropic.

To include the background anisotropy induced by the magnetic field,
we calculate numerically the leading order perturbation up to order
$u^{4}_{H}B^{2}$ by following the methods in Section 3.3. The numerical
results, illustrated in Figure \ref{fig:11}. By comparing the results
in Figure \ref{fig:9} with those in Figure \ref{fig:11}, we find
that the leading-order corrections in this model develop the negative
dips characterizing the distinct response behavior against the zero-th
order. Obviously, this model strictly forbids the extrapolation of
our results to large $u^{2}_{H}B$, a restriction that is dictated
by the gravity solution itself. Once the corrections of order $\mathcal{O}\left(u^{4}_{H}B^{2}\right)$
are taken into account in the gravity background (\ref{eq:5.3}),
the parameter $u^{2}_{H}B$ is restricted to a small region (i.e.
$u^{2}_{H}B\ll1$) , otherwise, the signature of the metric in (\ref{eq:5.3})
would be changed and the Hawking temperature in (\ref{eq:5.4}) becomes
negative. This reflects the limitation of the perturbative gravity
solution in (\ref{eq:5.3}) \footnote{In the gravity background (\ref{eq:5.3}), large $u^{2}_{H}B$ is
equivalent to take $T\rightarrow0$. However this model may be not
well-defined or may be singular in this case according to \cite{DHoker:2009ixq}.}. Hence, the numerical results presented in Figure \ref{fig:11} do
not encode any physical content beyond being a tiny correction to
the zeroth-order Green function, and they are not expected to lead
to a negative spectral function theoretically. In addition, the perturbative
fermionic dispersion curves are independent of the quantum numbers
$n$ of the Landau levels. So the perturbation may recover some Landau
levels displayed in the zeroth-order Green function, as it is predicted
by the hard thermal loop approximation with strong magnetic field
\cite{MLeBellac,Klimov:1981ka,Haque:2024gva}.

\section{The anisotropy induced by flavors}

In this section, we consider the holographic model with anisotropy
induced by flavors, then analyze the fermionic correlation functions
in this holographic system.

\subsection{The model}

The concerned model in this section is based on the D3-D5 intersection
\cite{Penin:2017lqt,Garbayo:2022pqp,Hoyos:2020zeg} in which the $N_{c}$
D3-branes and $N_{f}$ D5-branes refer respectively to the colors
and flavors in the dual theory. Since both the dynamics of $N_{c}$
D3-branes and $N_{f}$ D5-branes are taken into account, in the large
$N_{c}$ limit, the action for this system takes the form as,

\begin{equation}
S=S_{\mathrm{IIB}}+S_{\mathrm{D5}},\label{eq:6.1}
\end{equation}
where in the Einstein frame,

\begin{align}
S_{\mathrm{IIB}} & =\frac{1}{2\kappa^{2}_{10}}\int d^{10}x\sqrt{-g}\left(\mathcal{R}-\frac{1}{2}\partial_{M}\phi\partial^{M}\phi-\frac{1}{2\cdot3!}e^{\phi}F^{2}_{3}-\frac{1}{4\cdot5!}F^{2}_{5}\right),\nonumber \\
S_{\mathrm{D5}} & =-T_{\mathrm{D}5}N_{f}\int_{\mathrm{D5}}d^{6}xe^{\frac{\phi}{2}}\sqrt{-g_{\mathrm{D5}}}+T_{\mathrm{D}5}N_{f}\int_{\mathrm{D5}}C_{6}.
\end{align}
The action includes a Ramond-Ramond 6-form $C_{6}$ and 4-form $C_{4}$
coupling respectively to $N_{f}$ D5-branes and $N_{c}$ D3-branes
with their field strengths $F_{5}=dC_{4},F_{3}=\ ^{\star}dC_{6}$.
In ten-dimensional spacetime, the anisotropic solution for the action
can be found as,

\begin{equation}
ds^{2}=\frac{L^{2}}{u^{2}}\left[-f_{b}dt^{2}+\left(dx^{1}\right)^{2}+\left(dx^{2}\right)^{2}+e^{-2\phi}\left(dx^{3}\right)^{2}+\frac{du^{2}}{f_{b}}\right]+\frac{9}{8}L^{2}\left[ds^{2}_{\mathbb{CP}^{2}}+\frac{9}{8}\left(d\tau+B_{1}\right)^{2}\right],\label{eq:6.3}
\end{equation}
where $B_{1}$ is a 1-form\footnote{The exact formulas for dilaton $\phi$ and 1-form $B_{1}$ can be
found in \cite{Penin:2017lqt}.}. Note that this solution does not recover the $\mathrm{AdS_{5}}\times S^{5}$
by setting $N_{f}=0$. Accordingly, the action (\ref{eq:6.1}) with
solution (\ref{eq:6.3}) can be reduced to an effective five-dimensional
gravity system in which the effective five-dimensional action is given
as,

\begin{align}
S_{\mathrm{5D}}= & \frac{1}{2\kappa^{2}}\int d^{5}x\sqrt{-g_{\mathrm{5D}}}\left[\mathcal{R}_{\left(\mathrm{5D}\right)}-\frac{40}{3}\left(\partial\rho\right)^{2}-20\left(\partial\varphi\right)^{2}-\frac{1}{2}e^{4\rho+4\varphi+\phi}\left(\partial\mathcal{V}\right)^{2}-U\right]\nonumber \\
 & -\frac{1}{2\kappa^{2}}\int d^{5}x\sqrt{-g_{\mathrm{5D}}g^{-1}_{33}}\left[6Q_{f}e^{\frac{14}{3}\rho-2\varphi+\frac{\phi}{2}}\right],
\end{align}
where the potential $U$ is given as,

\begin{equation}
U=4e^{\frac{16}{3}\rho+12\varphi}-24e^{\frac{16}{3}\rho+2\varphi}+\frac{Q^{2}_{c}}{2}e^{\frac{40}{3}\rho},
\end{equation}
and the associated five-dimensional solution is given as,

\begin{align}
ds^{2} & =g_{tt}dt^{2}+g_{xx}\left[\left(dx^{1}\right)^{2}+\left(dx^{2}\right)^{2}\right]+g_{zz}\left(dx^{3}\right)^{2}+g_{uu}du^{2},\nonumber \\
g_{tt} & =-\left(\frac{9}{8}\right)^{2}\frac{L^{\frac{16}{3}}}{u^{2}}f_{b},\ g_{xx}=\left(\frac{9}{8}\right)^{2}\frac{L^{\frac{16}{3}}}{u^{2}},\ g_{zz}=\left(\frac{9}{8}\right)^{2}\left(\frac{4Q_{f}}{3}\right)^{\frac{4}{3}}\frac{L^{\frac{8}{3}}}{u^{\frac{2}{3}}},\ g_{uu}=\left(\frac{9}{8}\right)^{2}\frac{L^{\frac{16}{3}}}{u^{2}f_{b}},\nonumber \\
f_{b}\left(u\right) & =1-\left(\frac{u}{u_{H}}\right)^{\frac{10}{3}},\label{eq:6.6}
\end{align}
with the relations

\begin{align}
L^{4} & =\frac{256}{1215}Q_{c},\ Q_{c}=16\pi g_{s}\alpha^{\prime2}N_{c},\ Q_{f}=\frac{4\pi N_{f}}{9\sqrt{3}},\nonumber \\
\phi & =\frac{2}{3}\ln\left(\frac{3L^{2}}{4Q_{f}u}\right),\ \rho=\ln\left[\left(\frac{8}{9}\right)^{\frac{3}{5}}L^{-1}\right],\ \varphi=\frac{1}{10}\ln\left(\frac{9}{8}\right),\ \mathcal{V}=\sqrt{2}Q_{f}x^{3}.
\end{align}
In particular, the metric presented in (\ref{eq:6.6}) can be rescaled
as

\begin{equation}
ds^{2}\left(\frac{\alpha^{\prime}}{L^{2}}\right)^{\frac{5}{3}}\equiv\bar{ds}^{2}=-\frac{L^{2}}{u^{2}}f_{b}\left(u\right)dt^{2}+\frac{L^{2}}{u^{2}}\left[\left(dx^{1}\right)^{2}+\left(dx^{2}\right)^{2}\right]+\left(\frac{4Q_{f}}{3}\right)^{\frac{4}{3}}\frac{\left(dx^{3}\right)^{2}}{u^{\frac{2}{3}}L^{\frac{2}{3}}}+\frac{L^{2}}{u^{2}f_{b}\left(u\right)}du^{2},\label{eq:6.8}
\end{equation}
thus we will use the metric given by $\bar{ds}^{2}$ as in (\ref{eq:6.8}).
The Hawking temperature is given,

\begin{equation}
T=-\frac{1}{4\pi}\frac{\partial f_{b}}{\partial u}\big|_{u=u_{H}}=\frac{5}{6\pi u_{H}}.
\end{equation}

\subsection{Numerical results}

\begin{figure}[t]
\begin{centering}
\includegraphics[scale=0.39]{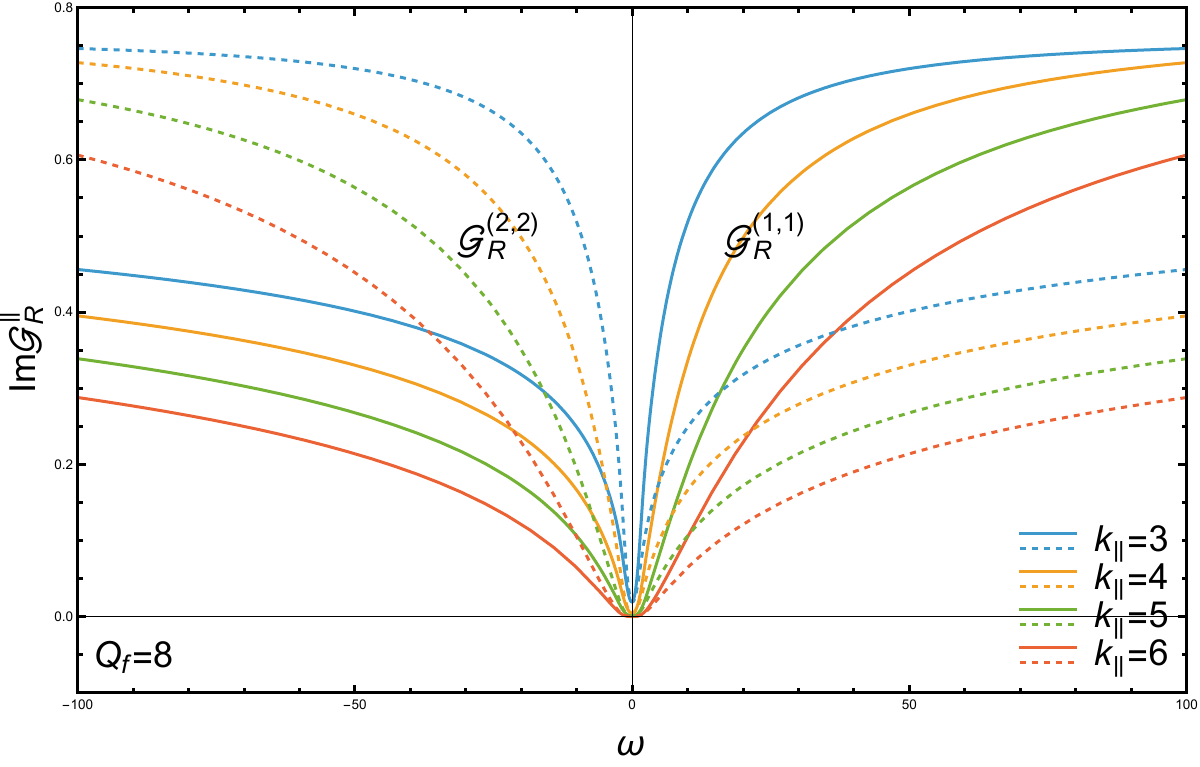}\includegraphics[scale=0.39]{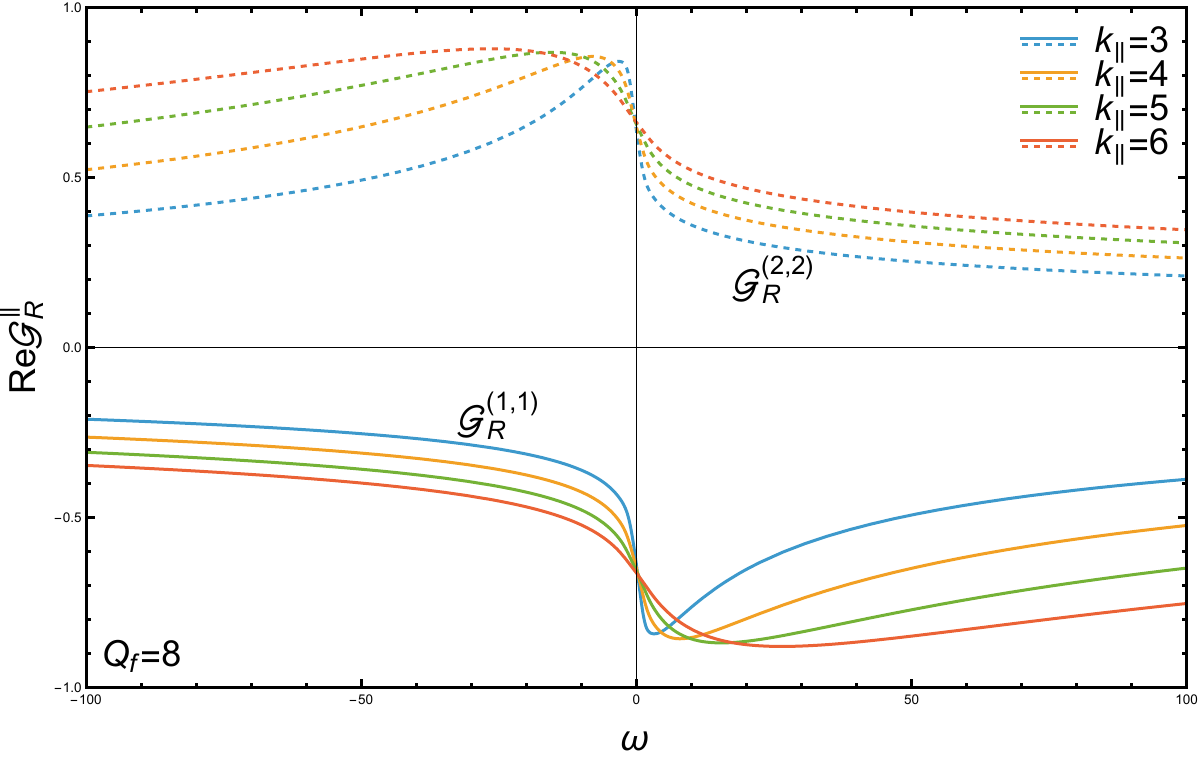} 
\par\end{centering}
\begin{centering}
\includegraphics[scale=0.39]{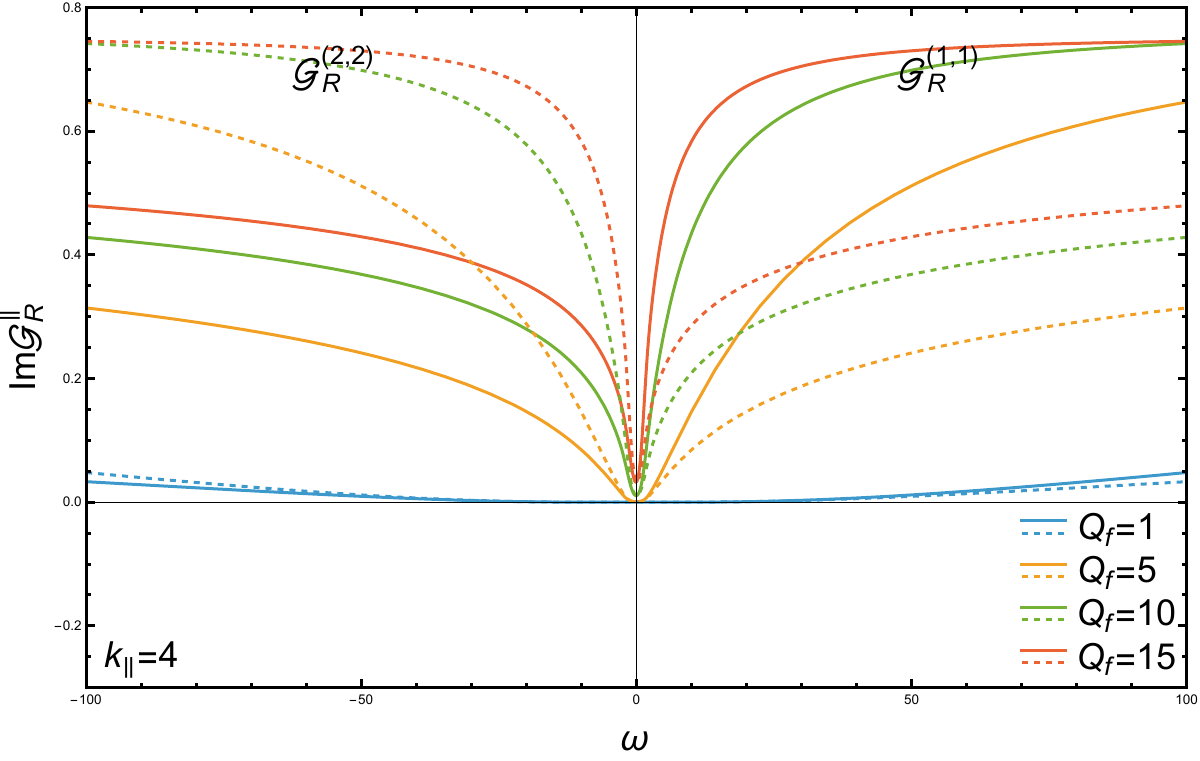}\includegraphics[scale=0.39]{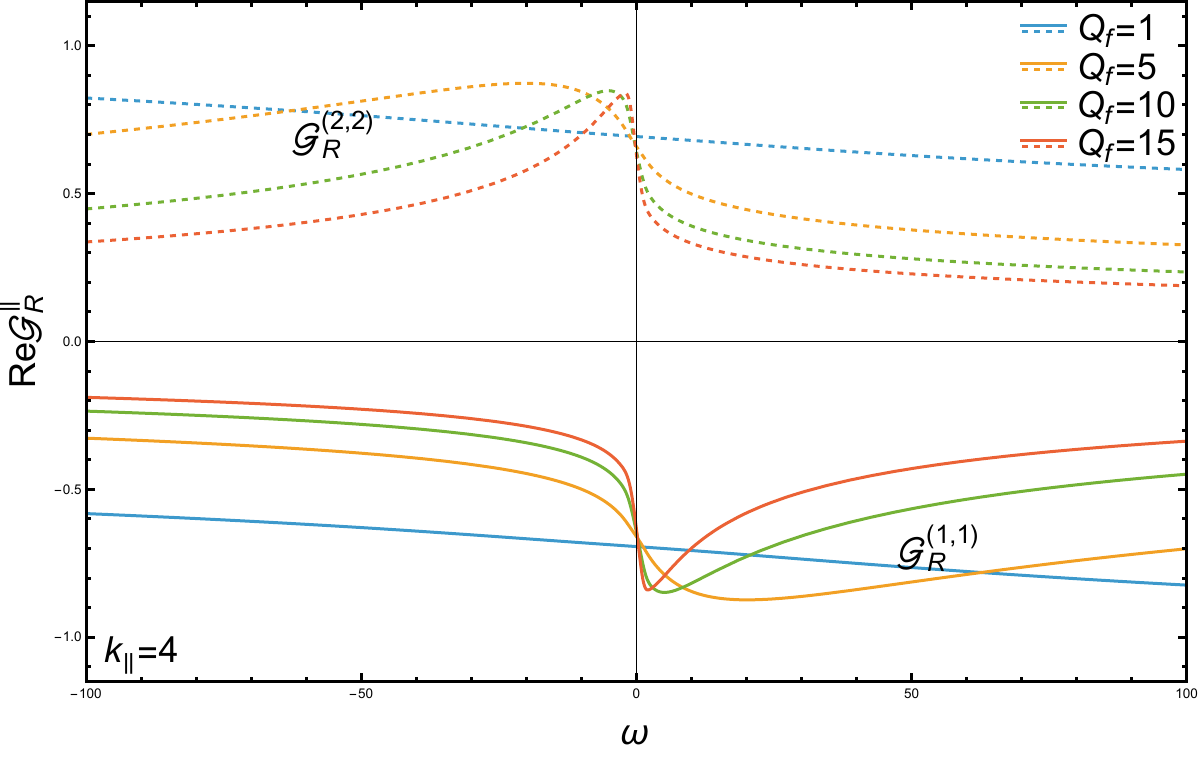} 
\par\end{centering}
\caption{The anisotropic fermionic Green function with various $\mathrm{k}_{\parallel}$
and $Q_{f}$. The parameters are set as $T=\frac{5}{6\pi}$ and $u_{H}=L=1,m=0.01$. }
\label{fig:12} 
\end{figure}

Based on the analysis, we numerically evaluate the Green functions
with anisotropy induced by flavors which are given in Figure \ref{fig:12}.
Note that we illustrate the results with respect to $k_{\Vert}$ only,
because the equations of motion for the Green functions with respect
to $k_{\perp}$ are similar to those on the AdS-Schwarzschild black
brane background, covering basically the results presented in Figure
\ref{fig:1} and \ref{fig:2}. And we do not use the perturbative
method since the flavored metric (\ref{eq:6.3}) is not a perturbation
solution with respect to flavor number in the large $N_{c}$ limit.

According to Figure \ref{fig:12}, we find there is no peak in the
imaginary part of the Green function, thus it implies the excitation
of fermion depends only on the momentum $k_{\perp}$ along the perpendicular
directions. It reveals furthermore that the fermionic spectral function
along the direction of symmetry breaking exhibits a momentum-independent
pseudogap behavior i.e. the imaginary part of the Green function develops
a minimum (a dip) at $\omega=0$, which stands in sharp contrast to
the hard thermal loop -like quasiparticle excitations observed along
the isotropic plane $\left(x^{1}-x^{2}\right)$ in this model. This
result suggests that this holographic model may host a dimensionally
reduced incoherent metallic phase (reduced to 2+1 dimensional spacetime):
the low-energy dynamics along the direction are governed entirely
by an irrelevant deformation of the infrared geometry (e.g., an anisotropic
scaling fixed point), which effectively freezes the momentum degree
of freedom and renders the fermionic excitations purely dissipative
and continuous \cite{Donos:2012js,Hartnoll:2014lpa,Davison:2014lua,Rangamani:2015hka}.
In fact, this five-dimensional model is based on the D3-D5 intersection
in which the flavored fermion in the dual theory lives on the 2+1
dimensional worldvolume of the D5-brane instead of a 3+1 dimensional
theory \cite{Penin:2017lqt,Garbayo:2022pqp,Hoyos:2020zeg}. From a
condensed matter perspective \cite{Matty,spivak2004intermediatephasesdimensionalelectron,Berg_2012},
such behavior is reminiscent of the precursor of electron localization
or Wigner crystallization along the director axis in a nematic quantum
critical point. This provides a novel holographic realization of strongly
anisotropic non-Fermi liquids.

\section{Summary and discussion}

In this work, we discuss the structure of the fermionic correlation
function in strongly coupled matter and generalize the formulas to
accommodate anisotropic gravity backgrounds, which are essential for
modelling the realistic conditions of the quark-gluon plasma (QGP)
produced in heavy-ion collisions. The key technical advancement lies
in the derivation of the Dirac equation on anisotropic backgrounds
and the perturbative method for computing the leading-order corrections
to the Green function when the anisotropy parameter is small. We then
applied this generalized method to three distinct holographic models,
each capturing a different physical origin of anisotropy.

The first model is based on the D3-D7 intersection in which the anisotropy
is induced by the axion field parameterized by the field strength
$a$. Our numerical results, obtained perturbatively in the high-temperature
limit $T\gg a$, reveal that the leading-order corrections to the
Green function exhibit peaks and negative dips in the imaginary part.
When a chemical potential is turned on, the particle-antiparticle
symmetry is broken, and the anisotropic corrections further enhance
the asymmetry in the spectral functions. The dispersion curves show
that the perturbations reduce the total spectral weight at certain
momenta. In this model, the holographic fermionic Green functions
depend on the direction of momentum.

The second model is based on the D3\nobreakdash-brane system with
a magnetic flux, where the background anisotropy is induced by the
magnetic field. The gravity solution is obtained in the high\nobreakdash-temperature
limit, i.e. $u^{2}_{H}B\ll1$ (or equivalently $T^{2}\gg B$). For
the zero-th order Green function, we find the magnetic field quantizes
the fermionic motion into Landau levels, which we incorporate into
the Dirac equation via a suitable ansatz that accounts for momentum
non\nobreakdash-conservation perpendicular to the field. Our numerical
results show discrete peaks in the Green function corresponding to
Landau levels, with an effective mass that qualitatively agrees with
the hard thermal loop approximation $m_{f}\sim\sqrt{T^{2}+B^{2}L^{2}}$.
And the leading\nobreakdash-order corrections, of order $\mathcal{O}\left(u^{4}_{H}B^{2}\right)$
induce anisotropic modifications. Notably, there are two peaks appearing
in the spectral function with high Landau levels, which may indicate
spin\nobreakdash-split in the total energy. Moreover, two infinitely
sharp peaks arise on the ridges of the spectral functions, possibly
representing long\nobreakdash-lived coherent quasiparticle poles
induced by the external field. Besides, to include the anisotropy
in the background, we numerically evaluate the leading-order corrections
to the Green function by using our perturbative method.

The third model is based on the D3-D5 intersection, in which the unquenched
flavors introduce anisotropy through the backreaction of the D5-branes.
Unlike the previous two models, this background is not a small perturbation
around isotropy, rather, it describes a genuinely anisotropic fixed
point. Our numerical calculations show a striking difference between
the perpendicular and parallel directions. Along the perpendicular
directions, the spectral function exhibits hard thermal loop -like
quasiparticle peaks. However, along the symmetry\nobreakdash-breaking
direction, the spectral function displays a momentum-independent pseudogap
i.e. minimum in the imaginary part at $\omega=0$ that persists for
all momenta. This behavior suggests the emergence of a dimensionally
reduced incoherent metallic phase \cite{Donos:2012js,Hartnoll:2014lpa,Davison:2014lua,Rangamani:2015hka},
where the low-energy dynamics along the anisotropic direction are
governed by an irrelevant deformation of the infrared geometry, effectively
freezing the momentum degree of freedom. From a condensed matter perspective,
such a pseudogap is reminiscent of the precursor to electron localization
or Wigner crystallization along the director axis in a nematic quantum
critical point, providing a novel holographic realization of strongly
anisotropic non-Fermi liquids \cite{Matty,spivak2004intermediatephasesdimensionalelectron,Berg_2012}.

Overall, our work demonstrates that holographic methods provide a
powerful and versatile framework for computing fermionic correlation
functions in strongly coupled anisotropic plasmas, a task that is
usually extremely challenging in QFT. The three models we studied
collectively illustrate that the effects of anisotropy are both rich
and physically significant: they break directional symmetry, induce
Landau levels, modify dispersion relations, and can even drive dimensional
reduction in the low-energy effective theory. These findings are particularly
relevant for the phenomenology of QGP produced in off-central heavy-ion
collisions, where strong magnetic fields and anisotropic expansion
are present, as well as for certain condensed matter systems with
nematic order.

\section*{Acknowledgements}

Si-wen Li is supported by the National Natural Science Foundation
of China (NSFC) under Grant No. 12005033, the Fundamental Research
Funds for the Central Universities under Grant No. 3132026190. Yan-qing
Zhao is supported by the National Natural Science Foundation of China
(NSFC) under Grant No. 12505151.

\section*{Appendix: The metric in the anisotropic model with axion}

In this section, we list the perturbative forms of the functions presented
in metric (\ref{eq:4.3}) according to \cite{Banks:2016fab,Giataganas:2017koz,Mateos:2011ix,A2016mno,Mateos:2011tv,Cheng:2014qia,Cheng:2014sxa}.
With vanishing chemical potential (i.e. the $U\left(1\right)$ gauge
field is set as $A_{0}=0$), the associated functions presented in
metric (\ref{eq:4.3}) can be written in the unit of $L=1$ as series
of $u_{H}a$ as,

\begin{align}
\mathcal{F}\left(u\right) & =1-\frac{u^{4}}{u^{4}_{H}}+a^{2}u^{2}_{H}\hat{\mathcal{F}}_{2}\left(u\right)+\mathcal{O}\left(a^{4}u^{4}_{H}\right),\nonumber \\
\mathcal{B}\left(u\right) & =1+a^{2}u^{2}_{H}\hat{\mathcal{B}}_{2}\left(u\right)+\mathcal{O}\left(a^{4}u^{4}_{H}\right),\nonumber \\
\phi\left(u\right) & =a^{2}u^{2}_{H}\hat{\phi}_{2}\left(u\right)+\mathcal{O}\left(a^{4}u^{4}_{H}\right),\tag{A-1}\label{eq:A-1}
\end{align}
where

\begin{align}
\hat{\mathcal{F}}_{2}\left(u\right) & =\frac{1}{24u^{4}_{H}}\left[8u^{2}\left(u^{2}_{H}-u^{2}\right)-10u^{4}\log2+\left(3u^{4}_{H}+7u^{4}\right)\log\left(1+\frac{u^{2}}{u^{2}_{H}}\right)\right],\nonumber \\
\hat{\mathcal{B}}_{2}\left(u\right) & =-\frac{1}{24}\left[\frac{10u^{2}}{u^{2}_{H}+u^{2}}+\log\left(1+\frac{u^{2}}{u^{2}_{H}}\right)\right],\nonumber \\
\hat{\phi}_{2}\left(u\right) & =-\frac{1}{4}\log\left(1+\frac{u^{2}}{u^{2}_{H}}\right).\tag{A-2}\label{eq:A-2}
\end{align}
In the above solutions with vanishing chemical potential, the metric
(\ref{eq:4.3}) recovers the AdS-Schwarzschild spacetime if the anisotropic
parameter is turned off as $a=0$.

The metric (\ref{eq:4.3}) also has a version with non-vanishing chemical
potential i.e. the $U\left(1\right)$ gauge field is set as $A_{0}\neq0$.
In this case, functions presented in metric (\ref{eq:4.3}), in the
unit of $L=1$, should be, 
\begin{align}
\mathcal{F}\left(u\right) & =1-\frac{u^{4}}{u^{4}_{H}}+\left[\frac{u^{6}}{u^{6}_{H}}-\frac{u^{4}}{u^{4}_{H}}\right]q^{2}+a^{2}u^{2}_{H}\hat{\mathcal{F}}_{2}\left(u\right)+\mathcal{O}\left(a^{4}u^{4}_{H}\right),\nonumber \\
\mathcal{B}\left(u\right) & =1+a^{2}u^{2}_{H}\hat{\mathcal{B}}_{2}\left(u\right)+\mathcal{O}\left(a^{4}u^{4}_{H}\right),\nonumber \\
\phi\left(u\right) & =a^{2}u^{2}_{H}\hat{\phi}_{2}\left(u\right)+\mathcal{O}\left(a^{4}u^{4}_{H}\right),\tag{A-3}\label{eq:A-3}
\end{align}
where

\begin{align}
\hat{\mathcal{F}}_{2}\left(u\right)= & \frac{1}{24u^{6}_{H}\sqrt{1+4q^{2}}}\bigg\{3\left(-4q^{2}u^{6}+u^{6}_{H}\right)\log\left(\frac{1+\sqrt{1+4q^{2}}+2\frac{u^{2}_{H}}{u^{2}}}{1-\sqrt{1+4q^{2}}+2\frac{u^{2}_{H}}{u^{2}}}\right)\nonumber \\
 & +u^{4}u^{2}_{H}\bigg[8\sqrt{1+4q^{2}}\left(-1+\frac{u^{2}_{H}}{u^{2}}\right)+\left(3-12q^{2}\right)\log\left(-2-2q^{2}+2\sqrt{1+4q^{2}}\right)\nonumber \\
 & +5\left(-2+q^{2}\right)\log\left(-1+2q^{2}+\sqrt{1+4q^{2}}\right)+7\left(1+q^{2}\right)\log\left(-1+2q^{2}-\sqrt{1+4q^{2}}\right)\nonumber \\
 & +7\left(1+q^{2}\right)\log\left(\frac{2q^{2}\frac{u^{2}}{u^{2}_{H}}-1+\sqrt{1+4q^{2}}}{2q^{2}\frac{u^{2}}{u^{2}_{H}}-1-\sqrt{1+4q^{2}}}\right)\bigg]\bigg\},\nonumber \\
\hat{\mathcal{B}}_{2}\left(u\right)= & \frac{1}{24}\left[\frac{10u^{2}u^{2}_{H}}{q^{2}u^{4}-u^{2}u^{2}_{H}-u^{4}}+\frac{1}{\sqrt{1+4q^{2}}}\log\left(\frac{1+\sqrt{1+4q^{2}}+2\frac{u^{2}_{H}}{u^{2}}}{1-\sqrt{1+4q^{2}}+2\frac{u^{2}_{H}}{u^{2}}}\right)\right],\nonumber \\
\hat{\phi}_{2}\left(u\right)= & \frac{u^{2}_{H}}{4\sqrt{1+4q^{2}}}\log\left(\frac{1+\sqrt{1+4q^{2}}+2\frac{u^{2}_{H}}{u^{2}}}{1-\sqrt{1+4q^{2}}+2\frac{u^{2}_{H}}{u^{2}}}\right),\nonumber \\
A_{0}\left(u\right)= & \frac{q}{8u^{3}_{H}\sqrt{3+12q^{2}}}\bigg\{24\sqrt{1+4q^{2}}\left(u^{2}_{H}-u^{2}\right)+5a^{2}u^{2}_{H}\bigg[u^{2}_{H}\log\left(\frac{3-\sqrt{1+4q^{2}}}{3+\sqrt{1+4q^{2}}}\right)\nonumber \\
 & +u^{2}\log\left(\frac{1+\sqrt{1+4q^{2}}+2\frac{u^{2}_{H}}{u^{2}}}{1-\sqrt{1+4q^{2}}+2\frac{u^{2}_{H}}{u^{2}}}\right).\tag{A-4}\label{eq:A-4}
\end{align}
Note that $q=\frac{u^{3}_{H}Q}{2\sqrt{3}}$ is the charge parameter
in which $Q$ refers to the dimensionless $U\left(1\right)$ charge
density. The physical range of $q$ is $0\leq q\leq\sqrt{2}$. The
chemical potential in this system is defined by $\mu=A_{0}|_{u=0}$,
i.e.

\begin{equation}
\mu=\frac{q}{8\sqrt{3}u_{H}}\left[24+\frac{5a^{2}u^{2}_{H}}{\sqrt{1+4q^{2}}}\log\left(\frac{3-\sqrt{1+4q^{2}}}{3+\sqrt{1+4q^{2}}}\right)\right],\tag{A-5}\label{eq:A-5}
\end{equation}
and the Hawking temperature is given as, 
\begin{equation}
T=\frac{2-q^{2}}{2\pi u_{H}}+\frac{u_{H}}{96\pi}\left[10\log2-4+5\left(3+\log2\right)q^{2}\right]a^{2}+\mathcal{O}\left(u^{4}_{H}a^{4}\right).\tag{A-6}\label{eq:A-6}
\end{equation}
In the presence of the chemical potential, the metric (\ref{eq:4.3})
takes the isotropic form under $a=0$ as,

\begin{align}
ds^{2} & =\frac{L^{2}}{u^{2}}\left[-f_{q}dt^{2}+\left(dx^{1}\right)^{2}+\left(dx^{2}\right)^{2}+\left(dx^{3}\right)^{2}+\frac{du^{2}}{f_{q}}\right],\nonumber \\
f_{q}\left(u\right) & =1-\frac{u^{4}}{u^{4}_{H}}+\left[\frac{u^{6}}{u^{6}_{H}}-\frac{u^{4}}{u^{4}_{H}}\right]q^{2}.\tag{A-7}\label{eq:A-7}
\end{align}
All the solutions given in (\ref{eq:A-3}) (\ref{eq:A-4}) returns
exactly to the solutions given in (\ref{eq:A-1}) (\ref{eq:A-2}).

 \bibliographystyle{utphys}
\bibliography{Anisotropic_fermionic_correlation_functions}

\end{document}